%% file: main.tex
\documentclass[a4paper,11pt]{article}
\usepackage[utf8]{inputenc}
\usepackage{jheppub} 
\usepackage[T1]{fontenc} 
\usepackage{amsmath,amsthm}
\usepackage{amssymb, amsfonts}
\usepackage{bbm}
\usepackage{graphicx}
\usepackage{dsfont}
\usepackage{relsize}
\usepackage{stmaryrd}
\usepackage{lmodern}
\usepackage{slantsc}
\usepackage{scalefnt}
\usepackage{wasysym}
\usepackage{xspace}
\usepackage{boxedminipage}
\usepackage{ifpdf}
\usepackage{multirow, hhline}
\usepackage{slashed}
\usepackage{xifthen}
\usepackage{tensor}
\usepackage{array, nicematrix}
\usepackage{tabu}
\usepackage{pbox}
\usepackage{multirow}
\usepackage{tikz}
\usepackage{overpic}
\usepackage{color, xcolor}
\usepackage{diagbox}
\usepackage{makecell}
\usepackage{cancel}
\usetikzlibrary{arrows, matrix, calc, scopes, decorations.markings}
\usepackage[mathcal]{euscript}
\usepackage{booktabs}
\usepackage{autobreak, subfig, diagbox, mathrsfs}
\usepackage{mathbbol}
\usepackage{tcolorbox}
\usepackage{hyperref}
\allowdisplaybreaks[4]

\newcommand{\NN}{\mathbb{N}}
\newcommand{\Z}{{\mathbb{Z}}}
\newcommand{\A}{\mathcal{A}}
\newcommand{\Hil}{\mathcal{H}}

\newcommand{\Fus}{\mathscr{F}}

\newcommand{\CC}{\mathbb{C}}
\newcommand{\Cent}{\mathcal{Z}}

\input{input.tex}

\newcommand{\df}{{\mathrm{d}}}

\newcommand{\dk}[2][1]{{\ifthenelse{\equal{#1}{1}}{\frac{\df{#2}}{2\pi}}{\frac{\df^{#1}{#2}}{(2\pi)^{#1}}}}}

\renewcommand{\Vec}{{\tt Vec}}

\newcommand{\Tsym}[6]{%
   \begingroup
     \renewcommand{\arraystretch}{1}%
     \begin{bmatrix}
       #1 & #2 & #3 \\
      #4 & #5 & #6
    \end{bmatrix}%
  \endgroup
}

\def\unitcell{
\hspace{-0.6em}
\scalebox{2.0}{
\begin{tikzpicture}[baseline=0.0ex,scale=1.5]
 \draw [line width=0.35pt] (-2pt,-2pt)--(0pt,0pt);
 \draw [line width=0.35pt] (0pt,0pt)--(2pt,-2pt);
 \draw [line width=0.35pt] (0pt,0pt)--(0pt,3pt);
 \draw [line width=0.35pt] (-2pt,5pt)--(0pt,3pt)--(2pt,5pt);
\end{tikzpicture}
}
\hspace{-0.5em}
}

\definecolor{moduleColor}{HTML}{FF8080}
\definecolor{algColor}{HTML}{0080FF}
\def\unitcellColorAlgMod#1#2{
\hspace{-0.6em}
\scalebox{2.0}{
\begin{tikzpicture}[baseline=0.0ex,scale=1.5]
 \draw [moduleColor] (-2pt,-2pt)--(0pt,0pt);
 \draw [moduleColor] (0pt,0pt)--(2pt,-2pt);
 \draw [algColor] (0pt,0pt)--(0pt,3pt);
 \draw [moduleColor] (-2pt,5pt)--(0pt,3pt)--(2pt,5pt);
 \node [text=algColor, anchor=west, scale=0.3] at (0.5pt, 1.5pt) {$#1$};
 \node [text=moduleColor, anchor=west, scale=0.3] at (2pt,-2pt) {$#2$};
\end{tikzpicture}
}
\hspace{-0.8em}
}

\def\unitcellColorPureAlg{
\hspace{-0.6em}
\scalebox{2.0}{
\begin{tikzpicture}[baseline=0.0ex,scale=1.5]
 \draw [algColor] (-2pt,-2pt)--(0pt,0pt);
 \draw [algColor] (0pt,0pt)--(2pt,-2pt);
 \draw [algColor] (0pt,0pt)--(0pt,3pt);
 \draw [algColor] (-2pt,5pt)--(0pt,3pt)--(2pt,5pt);
 \node [text=algColor, anchor=west, scale=0.3] at (0.5pt, 1.5pt) {$\A$};
 \node [text=algColor, anchor=west, scale=0.3] at (2pt,-2pt) {$\A$};
\end{tikzpicture}
}
\hspace{-0.7em}
}

\def\unitcellColorMod#1#2{
\hspace{-0.6em}
\scalebox{2.0}{
\begin{tikzpicture}[baseline=0.0ex,scale=1.5]
 \draw [moduleColor] (-2pt,-2pt)--(0pt,0pt)--(2pt,-2pt);
 \draw [black] (0pt,0pt)--(0pt,3pt);
 \draw [moduleColor] (-2pt,5pt)--(0pt,3pt)--(2pt,5pt);
 \node [text=black, scale=0.3] at (1.5pt, 1.5pt) {$#1$};
 \node [text=moduleColor, anchor=west, scale=0.3] at (2pt,-2pt) {$#2$};
\end{tikzpicture}
}
\hspace{-0.8em}
}

\def\unitcellColorModHoz#1#2{
\hspace{-0.6em}
\scalebox{2.0}{
\begin{tikzpicture}[baseline=-1.5pt,scale=1.5]
    \draw [moduleColor](-2pt,-2pt)--(0pt,0pt)--(-2pt,2pt);
    \draw [black] (0pt,0pt)--(3pt,0pt);
    \draw [moduleColor] (5pt,-2pt)--(3pt,0pt)--(5pt,2pt);
    \node [text=black, anchor=south, scale=0.3] at (1.5pt, 0pt) {$#1$};
    \node [text=moduleColor, anchor=west, scale=0.3] at (5pt,-2pt) {$#2$};
\end{tikzpicture}
}
\hspace{-0.8em}
}

\def\OrientedLocalState#1#2#3#4#5{
\hspace{-0.6em}
\scalebox{1.0}{
\begin{tikzpicture}[baseline=0.0ex,scale=3]
\draw[black,->, >=stealth, shorten >=2.2pt,line width=0.5pt](0pt,0pt)--(-2pt,-2pt);
\draw[black,->, >=stealth, shorten >=2pt,line width=0.5pt](2pt,-2pt)--(0pt,0pt);
\draw[black,->, >=stealth, shorten >=2pt,line width=0.5pt](-2pt,5pt)--(0pt,3pt);
\draw[black,->, >=stealth, shorten >=2pt,line width=0.5pt](0pt,3pt)--(2pt,5pt);
\draw[black,->, >=stealth, shorten >=2pt,line width=0.5pt](0pt,0pt)--(0pt,3pt);
\draw[black](-2pt,-2pt)--(0pt,0pt)--(2pt,-2pt);
 \draw [black] (0pt,0pt)--(0pt,3pt);
 \draw [black] (-2pt,5pt)--(0pt,3pt)--(2pt,5pt);
 \node [text=black, scale=0.8] at (1.5pt, 1.5pt) {$#1$};
 \node [text=black, anchor=east, scale=0.8] at (-2pt,5pt) {$#2$};
 \node [text=black, anchor=west, scale=0.8] at (2pt,5pt) {$#3$};
 \node [text=black, anchor=west, scale=0.8] at (2pt,-2pt) {$#5$};
 \node [text=black, anchor=east, scale=0.8] at (-2pt,-2pt) {$#4$};
\end{tikzpicture}
}
\hspace{-0.5em}
}

\newcommand{\eqs}[1]{{\texorpdfstring{${#1}$}{Lg}}}

\newcommand{\ket}[1]{{\left\vert{#1}\right\rangle}}
\newcommand{\bra}[1]{{\left\langle{#1}\right\vert}}

\newcommand{\eqn}[2][0]{\ifthenelse{\equal{#1}{0}}{\begin{equation}\begin{aligned}#2\end{aligned}\end{equation}}{\begin{equation}\begin{aligned}#2\end{aligned}\label{#1}\end{equation}}}

\usetikzlibrary{shapes.geometric, snakes}
\tikzset{>=latex}
\usetikzlibrary{arrows, matrix, calc, scopes, decorations.markings}
\usetikzlibrary{decorations.pathmorphing, positioning}
\tikzset{snake it/.style={decorate, decoration={snake,amplitude=0.2mm,segment length=1mm}}}
\tikzset{->-/.style={decoration={
			 markings,
			 mark=at position .5*\pgfdecoratedpathlength+2pt with {\arrow{>}}},postaction={decorate}}}

\tikzset{-<-/.style={decoration={
			 markings,
			 mark=at position .5*\pgfdecoratedpathlength+2pt with {\arrow{<}}},postaction={decorate}}}

\newtabulinestyle{dash=off 2pt}

\renewcommand{\arraystretch}{1.5}

{\null\,\vcenter\bgroup\scriptsize
\renewcommand{\arraystretch}{0.7}%
\arraycolsep=.13885em
\hbox\bgroup$\array{@{}#1@{}}}
{\endarray$\egroup\egroup\,\null}
\title{A 2D-CFT Factory: Critical Lattice Models from Competing Anyon Condensation Processes in SymTO/SymTFTs}

\date{\today}
\author[a,c]{Ling-Yan Hung\footnote{Corresponding author}}
\author[b]{Kaixin Ji}
\author[c]{Ce Shen}
\author[b,d,e]{Yidun Wan\footnote{Corresponding author}}

\author[b]{Yu Zhao}
\affiliation[a]{Yau Mathematical Sciences Center, Tsinghua University, Beijing 100084, China}
\affiliation[b]{State Key Laboratory of Surface Physics, Center for Astronomy and Astrophysics, Department of Physics, Center for Field Theory and Particle Physics, and Institute for Nanoelectronic devices and Quantum Computing, Fudan University, 2005 Songhu Road, Shanghai 200433, China}
\affiliation[c]{Beijing Institute of Mathematical Sciences and Applications, Beijing 101408, China}
\affiliation[d]{Shanghai Research Center for Quantum Sciences, 99 Xiupu Road, Shanghai 201315, China}
\affiliation[e]{Hefei National Laboratory, Hefei 230088, China}
\emailAdd{lyhung@tsinghua.edu.cn, kxji21@m.fudan.edu.cn, 
scbebetterme@gmail.com, ydwan@fudan.edu.cn, yuzhao20@fudan.edu.cn}

\abstract{In this paper, we introduce a ``\textbf{CFT factory}'' - a novel algorithm of methodically generating 2D lattice models that would flow to 2D conformal fixed points in the infrared. These 2D models are realised by giving critical boundary conditions to 3D topological orders (SymTOs/SymTFTs) described by string-net models, often called the strange correlators. We engineer these critical boundary conditions by introducing a commensurate amount of non-commuting anyon condensates. The non-invertible symmetries preserved at the critical point can be controlled by studying a novel ``refined condensation tree''. \textbf{Our structured method generates} an infinite family of critical lattice models, including the $A$-series minimal models, and uncovers previously unknown critical points. Notably, we find at least three novel critical points (c$\approx1.3$, $1.8$, and $2.5$ respectively) preserving the Haagerup symmetries, in addition to recovering previously reported ones. The condensation tree, together with a generalised Kramers-Wannier duality, \textbf{predicts} precisely large swathes of phase boundaries, \textbf{fixes} almost completely the global phase diagram, and \textbf{sieves} out second order phase transitions. This is not only illustrated in well-known examples (such as the 8-vertex model related to the $A_5$ category) but also further verified with precision numerics, using our improved \textbf{(non-invertible) symmetry-preserving} tensor-network RG, in novel examples involving the Haagerup symmetries. We show that critical couplings can be precisely encoded in the categorical data (Frobenius algebras and quantum dimensions in unitary fusion categories), thus establishing a powerful, \textbf{systematic} route to discovering and potentially classifying new conformal field theories.

}

\begin{document}

\maketitle

\section{Introduction}

Critical lattice models are an exceedingly economical way of encoding CFTs, so their systematic construction might be a promising avenue to classifying and solving for novel CFTs, a long-standing open problem even in 2 dimensions, although much more is known there. Ideally, any successful framework of such critical lattice models should sufficiently constrain these models based on physical properties (such as symmetries or matter content), such that the landscape of lattice models can be scanned through in a controlled way. Unfortunately, not only is it difficult to engineer physical properties of a lattice model, a generic lattice model would produce a gapped phase in the thermodynamic limit, and phase transitions between gapped phases are also typically first order. To recover a CFT, it is necessary to sieve out second order phase transitions, but this is generally an arduous problem. Attempts have been made in the last 50 years to find patterns in critical couplings, notably for example, with the notion of discrete holomorphicity \cite{Bernard:1991za,Rajabpour:2007er,Ikhlef:2015sxa,Fendley:2020fxv}; there are instances where criticality is observed in the Kramers-Wannier({KW}) self-dual models, such as the Ising model.
About 20 years ago, a series of lattice models called the “golden chain” \cite{Feiguin:2006ydp} are found
to enjoy criticality protected by non-invertible symmetries. The golden chain model and a large class of its generalisations \cite{Aasen:2016dop,Aasen:2020jwb, Vanhove:2018wlb} (also called the ``strange correlator''\footnote{The idea of a strange correlator was first introduced in \cite{You_2014} to detect SPT phases. They also pointed out that this is equivalent to studying non-trivial boundary conditions of the 2+1D topological model. }) are a realization of the topological holographic principle \cite{Ji:2019jhk,Kong:2020cie,Kong:2020jne, Gaiotto:2020iye, Freed:2022qnc,Apruzzi:2021nmk, Chatterjee:2022kxb,Freed:2018cec,Kong:2019byq, Bhardwaj:2017xup}, in which a transfer matrix with given (non-invertible) symmetries is expressible as the path-integral of a topological order---a symTO or symTFT---in one dimension higher associated with the said symmetries, with appropriate boundary conditions. This higher dimensional path-integral is often termed the ``sandwich construction''.  The search for CFTs preserving given (generalized) symmetries can thus be translated into the search for critical boundary conditions of the associated symTO. This problem remains widely open, albeit progress in re-expressing a large class of known integrable lattice models as strange correlators \cite{Aasen:2016dop,Aasen:2020jwb, Vanhove:2018wlb}, and more recent guesses of critical models with Haagerup symmetries\cite{Vanhove:2021zop,Huang:2021nvb}.

 In this paper, we tackle this decades-long problem. We establish a CFT factory by proposing a systematic method of engineering boundary conditions in any given 2+1 dimensional string-net model (SN) \cite{Levin:2004mi} of symTO to produce infinite classes of critical lattice models in 1+1 dimensions. By drawing insights from the (generalised) golden chain,connecting the physics of competing anyon condensations with criticality, we can control the minimal amount of (non-invertible) symmetries preserved in the construction, making it possible to systematically scan through and generate novel models containing given symmetries. 

As a first test, we show that an infinite collection of well-known 2D lattice models, such as
the $A$-series integrable models, can be recovered naturally in our construction. The critical couplings
now admit simple interpretations in terms of the data of the input unitary fusion category (UFC) that defines the string-net model---Frobenius algebras and their modules in the UFC. The Hu-Geer-Wu ({HGW}) model \cite{Hu:2015dga} that extends the original Levin-Wen string-net model by manifesting the internal spaces of anyons, such that anyon condensation can be studied in a finer and more local manner \cite{Zhao:2024ilc} proved to be a powerful tool in searching for these critical points. A large class of these commensurate points are also shown to be self-dual points of a generalised {KW} duality. 
Then, we construct novel critical models with Haagerup symmetries, showcasing its power to
discover new CFTs. Moreover, by means of the Frobenius algebras in the model's input UFC,
one can analytically predict large swathes of the phase boundaries between gapped phases. In the family of 8-vertex models \cite{Baxter:1971cr} (duals of the Ashkin-Teller (AT) model \cite{FJWegner_1972}) that are related to the string net model defined with the $A_5$ category \cite{Gils_2013,Aasen:2020jwb}, we analytically recover positions of critical points previously obtained via integrable methods and almost completely fix the global phase diagram and locate phase boundaries upto a small area near the tri-critical point there. In fact, in all the examples we checked, including novel Haagerup models hitherto unknown in the literature, our analytical results match accurate numerical data of the global phase diagram obtained using an upgraded version of the symmetric RG method developed in \cite{Vanhove:2018wlb,Chen:2022wvy,Bao:2024ixc}.

\begin{figure}
    \centering
    \includegraphics[width=0.4\linewidth]{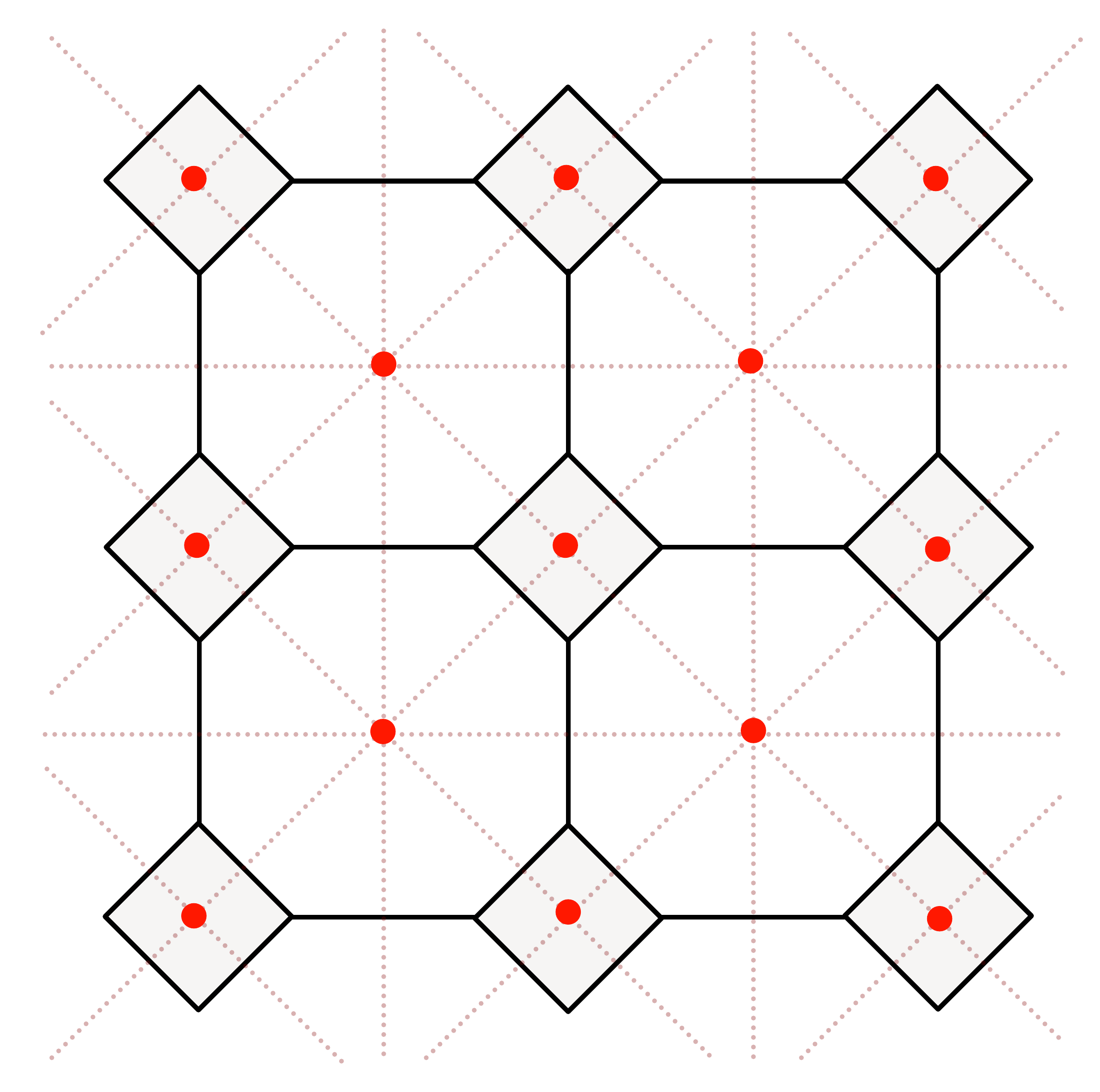}
    \caption{The truncated square lattice and its dual triangulation $\{\Delta\}$.}
    \label{fig:4-8-lattice}
\end{figure}

Our strategy can be outlined as follows. 
\begin{enumerate}
    \item We pick a wave-function $|\Psi\rangle_{\text{SN}({\Fus})}$ of a given string-net model with an input UFC ${\Fus}$. The model outputs a topological phase, whose anyon types are the simple objects of the UMTC $\Cent({\Fus})$---Drinfeld center of ${\Fus}$. We pick a truncated square lattice (Fig. \ref{fig:4-8-lattice}), with the intention of constructing a critical square lattice model, where the vertex-plaquette equality in number would be an important consideration in addition to realizing discrete translation and rotation invariance.
    
    \item As the topological holographic principle dictates, we will then have to design a suitable state $\langle \Omega_{\textrm{critical}} |$ such that \cite{Feiguin:2006ydp,Aasen:2016dop,Vanhove:2018wlb,Aasen:2020jwb}
    \begin{equation} \label{strangecorr}
        Z_{2D \, \textrm{critical}} = \langle \Omega_{\textrm{critical}} |\Psi\rangle_{\text{SN}({\Fus})}.
    \end{equation} 
    It was argued \cite{Levin_2012,Levin:2019ifu} and proved rigorously \cite{Levin2} at least in the case of the 2D Ising model, that when a pair of non-commuting topological defects (which arises as anyons in the 3D symTO) remain unbroken (or uncondensed), the 2D model is either degenerate or gapless. Therefore $\langle \Omega_{\textrm{critical}} |$ can be constructed by creating a commensurate amount of two different kinds of anyon condensates that contain mutually non-commuting anyons. This is practically conducted by dividing the lattice into unit cells and making use of properties of the Frobenius algebra objects in ${\Fus}$ characterising the condensates in the topological order. When competing anyons are commensurate in one unit cell, it turns out globally $\langle \Omega_{\textrm{critical}}|\Psi\rangle_{\text{SN}({\Fus})}$ produces the partition function of a critical lattice model.  
    
    \item By studying the competition of three competing condensates we also produce analytical prediction of the location of phase boundary. This is achieved by adjusting the amount of a third competing condensate in a way that does not disturb the equilibrium between the other two.

    \item Each choice of a pair of non-commuting competing condensates spontaneously break different amount of the bulk symmetries contained in $\Cent({\Fus})$, potentially producing CFTs with different amount of preserved symmetries. The actual minimal amount of symmetries preserved in the resultant critical model can be deduced by studying a {\it{condensation tree}}---a refinement of the Hasse tree previousely considered in the literature \cite{Chatterjee:2022tyg,Bhardwaj:2024qrf}. The refined condensation tree would also supply precise information of the phase diagram. Constructing CFTs preserving all of $\Cent({\Fus})$ can be engineered by studying the tree and picking appropriate pairs of competing anyon condensates.  The refined condensation tree also enables us to systematically {\it exclude} first order phase transitions and hence engineer second order phase transitions.

    \item We will discuss in detail a generalised {KW} duality, generalizing the electromagnetic duality symmetries found in the Ising spin model, and particularly its natural realisation in our lattice. This would prove mighty in confirming a large class of the critical couplings we found and provide detailed information of the resultant CFT. 

\end{enumerate}  

We check our theory with numerics. We refine the symmetry preserving RG proposed in \cite{Vanhove:2018wlb, Chen:2022wvy} to obtain global phase diagrams in multiple examples, confirming our theoretical analysis. Moreover, by applying our method to exotic models such as the string-net model with input Haagerup $H_3$ UFC, we discover new critical points. 

Note that the form of $\langle \Omega|$ that recovers a renormalisation group (RG) fixed point, be it topological or conformal, is not unique. The universality of fixed points implies that many UV features in the lattice would be washed out under RG. Therefore, the seed state $\langle \Omega_{\textrm{critical}}|$ that produces a CFT fixed point is not unique. The essence of our construction is to provide an algorithm to create a seed state that is destined to flow to a CFT in the infrared using only the category describing the symmetries of the CFT.  This is thus a systematic {\it CFT factory}.  

Our paper is organised as follows. Section \ref{sec:puddle} describes our ansatz for dividing the lattice into unit cells and the construction of critical points from commensurate amount of condensates within a unit cell, illustrated by examples. 
Section \ref{sec:general_theory} presents a general theory to organise the condensates via the refined condensation tree, engineer second order phase transitions and obtain locations of phase boundaries and read off symmetries of the critical points. These are again illustrated with many examples and checked with numerics. We will also describe a powerful generalisation of the KW duality. 
Section \ref{sec:Haagerup} applies the theory developed to studying the critical points from competing anyon condensations in the Doubled Haagerup SymTO. Section \ref{sec:Discussion} discusses our approach and possible generalisations. 
The appendices contain many important details. Appendix \ref{sec:HGW}  reviews the HGW model and implementation of anyon condensation there. Appendix \ref{appendix:data} provides important data of the categories used in the current paper. 
Appendix \ref{app:symRG} explains how the symmetric RG method is implemented in the precision numerical computations performed in the current paper. Last but not least,  Appendix \ref{app:summary tables} summarises in the form of tables of all the critical points obtained in the current paper.

\section{Constructing Competing Condensates in a Square Lattice  in {String-net} Models}
\label{sec:puddle}
As mentioned in the introduction, we consider two-dimensional lattice partition functions as a strange correlator of the form \eqref{strangecorr}. 

\begin{figure}
    \centering
    \includegraphics[width=0.5\linewidth]{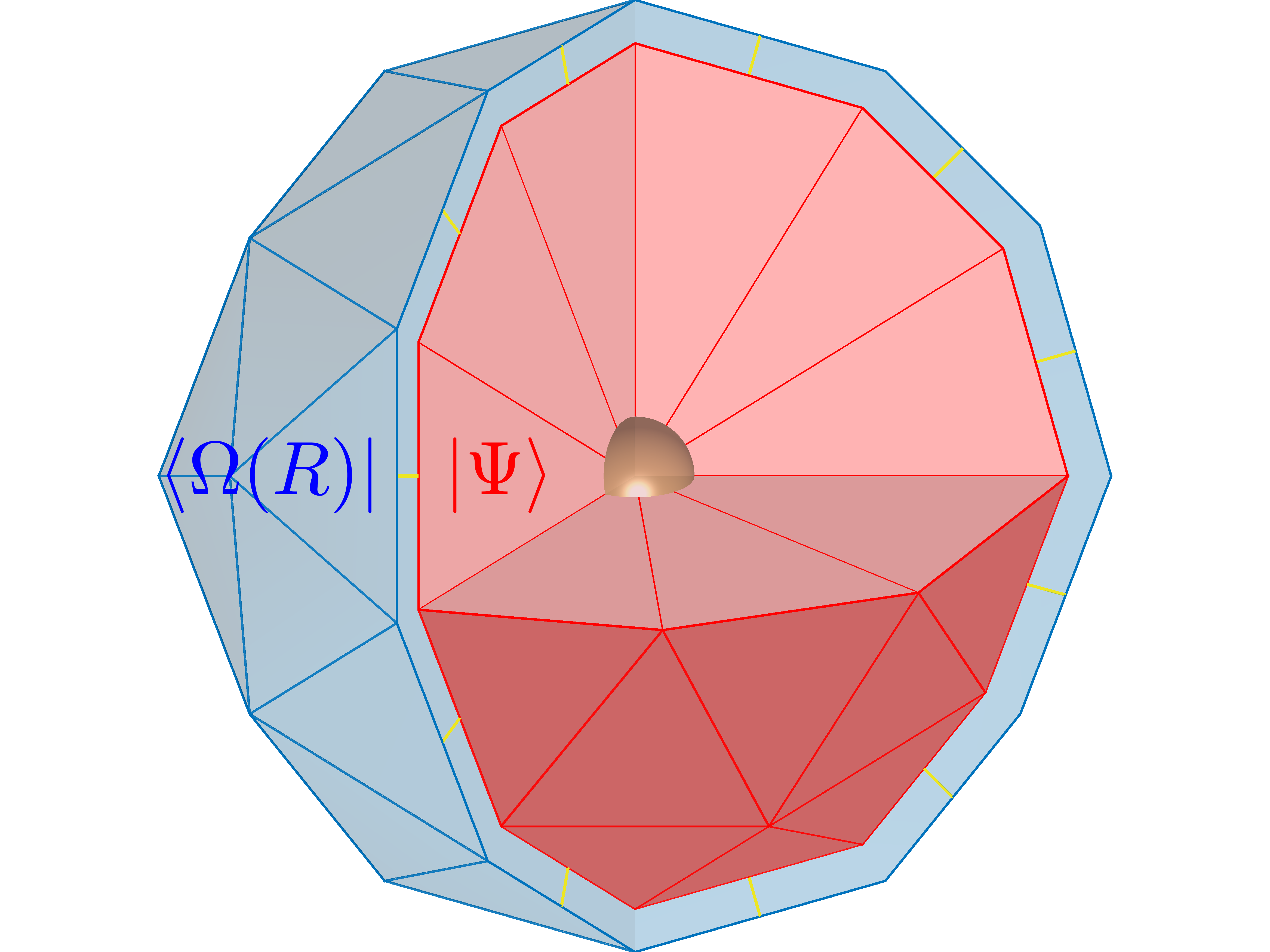}
    \caption{Geometry of the strange correlator as a 3D path-integral. The boundary of $|\Psi\rangle_{\text{SN}({\Fus})}$ is the surface on which the 2D lattice model is defined (which is chosen to be a sphere for concreteness). The state $\bra{\Omega(R)}$ essentially determines the boundary condition of the 3D solid, and it is drawn like a PEPS state to produce a local 2D partition function together with $|\Psi\rangle_{\text{SN}({\Fus})}$. (Its dependence on different couplings is denoted schematically by $R$.) The inner hole (in brown) of the solid ball is colored by a topological boundary condition in a sandwich construction. In the current construction, the boundary condition is equivalent to filling up the hole. For more general topological boundaries (as those discussed in section \ref{sec:topbc}), the hole would not be filled up. }
    \label{sandwich}
\end{figure}

This strange correlator is defined as the overlap between the ground-state wave function \(|\Psi\rangle_{\text{SN}({\Fus})}\) of the string-net model and a state \(\langle \Omega|\). The string-net ground state $|\Psi\rangle$ is expressed as a projected entangled pair states (PEPS) in terms of quantum \(6j\) symbols \cite{Aasen:2020jwb, Vanhove:2018wlb}.  This was reviewed in Appendix \ref{app:symRG}, where $|\Psi\rangle_{\text{SN}({\Fus})}$ is broken down into a collection of triangles, and each triangle is given by the quantum 6j symbol of the input category ${\Fus}$. It can be interpreted as a 3D path-integral over a solid handle-body (say a solid ball), such that its boundary (say a sphere) is the 2D surface on which the lower dimensional theory is defined. This is illustrated in figure \ref{sandwich}. The state \(\langle \Omega|\) can thus be understood as boundary conditions imposed on the 2D boundary of the 3D path-integral encapsulated in $|\Psi\rangle_{\text{SN}({\Fus})}$. The strange correlator (\ref{strangecorr}) is thus an explicit realisation of the sandwich construction. 

In this section, we would like to provide a recipe for designing  \(\langle \Omega|\), such that the resultant strange correlator describes a critical point in 1+1 D. We denote this special state by \(\langle \Omega_{\text{Critical}}|\) and will construct it by applying the idea of competing anyon condensations. We will also provide an algorithm to engineer continuous phase transitions, producing 1+1 D conformal field theories in the infrared.

\subsection{Choosing a Unit Cell and Constructing Condensate Puddles}

To construct competing condensates, the first step is to construct \(\langle \Omega|\) corresponding to a condensate on the 2D surface on which the string-net ground state $|\Psi\rangle_{\text{SN}({\Fus})}$ is defined. 
An anyon condensation process in a parent 2+1 D topological order literally corresponds to proliferation of a collection of anyons, driving a phase transition, and producing a child topological order\cite{bais2009condensate, hung2014symmetry, hung2015generalized, hu2022anyon}. This can be realized explicitly on the lattice with proliferation of anyons using anyon creation operators\cite{zhao2023characteristic, christian2023lattice, lin2024anyon, Zhao:2024ilc}, which can be constructed in the HGW model\cite{hu2018full, zhao2024noninvertible, Zhao:2024ilc}, an extension of the original Levin-Wen string-net model\cite{levin2005string}. These machineries help catch important microscopic details of the condensates solely by the input UFC of the model and play a key role in inspring the constructions below. The HGW model's full glory is not needed to understand the main text in the current paper, but we review it in detail in Appendix \ref{sec:HGW}.  After anyon condensation, the parent topological order reduces to a different phase that we refer to as a child order. The child order is described by the string-net model with a different input category $\mathcal{K}$, where one can show that $\mathcal{K}$ is a sub-category of the input category  $  {\Fus}$ of the parent model. This fact would be important in the next section on classifying phase transitions. 

When the child order is trivial, i.e., the vacuum, the condensate in the bulk is characterized by a ``Lagrangian algebra'' $L$, which is a Frobenius algebra object in $\Cent(\Fus)$. The input category $\mathcal{K}$ of the child order is correspondingly reduced to a trivial category with one object---a unitary connected symmetric Frobenius algebra object $\A$ in the parent input category ${\Fus}$. 
Mathematically, $L$ is the {\it full center} of $\A$. There are systematic ways of deducing $L$ from $\A$ reviewed in Appendix \ref{sec:HGW}.

A unitary Frobenius algebra in ${\Fus}$ is a (possibly composite) object in ${\Fus}$ i.e. 
\begin{equation}
    \A = \bigoplus_an_aa, \qquad a \in {\Fus},\qquad n_a \in \mathbb{N},
\end{equation}
equipped with an algebra multiplication expressed as a cyclically symmetric function \(f^\A: \A^3\to\CC\). Here, an $n_a$ is a non-negative integer that is the multiplicity of simple object $a\in {\Fus}$ enrolled in $\mathcal{A}$. To avoid clutter, the discussion in this section is limited to $n_a \le 1$, although it can be readily generalized.

The first step is to construct $\langle \Omega |$ representing a global condensate $L$ characterised by $\A$ in ${\Fus}$. 
This was originally achieved in \cite{Hu:2017lrs, Chen:2022wvy}. For a given trivalent lattice (the dual graph of a triangulation ), 
\begin{equation} \label{fixed_point}
        \Big\langle \unitcellColorPureAlg \Big| = \sum_{abcef\in L_\A} \Big\langle \OrientedLocalState{a}{b}{c}{e}{f} \Big| \ f^\A_{abc^\ast}f^\A_{a^\ast f e^\ast},
\end{equation}
In the state above, we assign a weight $f^\A_{abc^\ast}$ to each tri-valent vertex (dual to each triangle). It was shown that (\ref{fixed_point}) is a gapped fixed point under the symmetric RG operation and describes the topological trivial phase of the input Frobenius algebra \(\A\) \cite{Chen:2022wvy, Zhao:2024ilc}. 

In what follows, however, we would like to modify the construction. Rather than constructing a global condensate on the 2D surface, we would like to construct a small puddle of condensate in a unit cell. The global state will be reconstructed by patching together these small puddles. This is a crucial way to reduce the number of effective degrees of freedom to a small number.  Also, anticipating the construction of a conformal point, we would like to consider an effective square lattice, to make translation and some rotation symmetry easily imposed. Inspired by previous models \cite{Vanhove:2018wlb, Aasen:2016dop, Aasen:2020jwb}, we consider the square-octagon lattice, also called the truncated square lattice, as depicted in figure \ref{fig:4-8-lattice}.

The unit cell in this construction takes the shape of \unitcell . We will be conidering (almost) tensor product states. Namely, the global state $\langle \Omega|$ we consider can be constructed by specifying the state on the unit cell \unitcell, which we denote by $\langle \unitcellColorAlgMod{A}{M_A}|$ and is chosen as follows. The four external slanted edges each carries a \emph{right module} $M_{\A}$ (essentially, a representation) of a Frobenius algebra $\mathcal{A}$, and the horizontal/ vertical edge between the slanted ones carries $\mathcal{A}$ itself.

To be explicit, a right module of a Frobenius algebra \(\A\) in a UFC \(\Fus\) is a (possibly composite) object
\[
M_\A := \bigoplus_{x\in \Fus} m_x x,
\]
equipped with an algebra action encoded by a \emph{module function} $\rho_{M_\A}: \A\otimes {M_\A}^2\to\CC$. 

\begin{figure}
    \centering
    \includegraphics[width=0.4\linewidth]{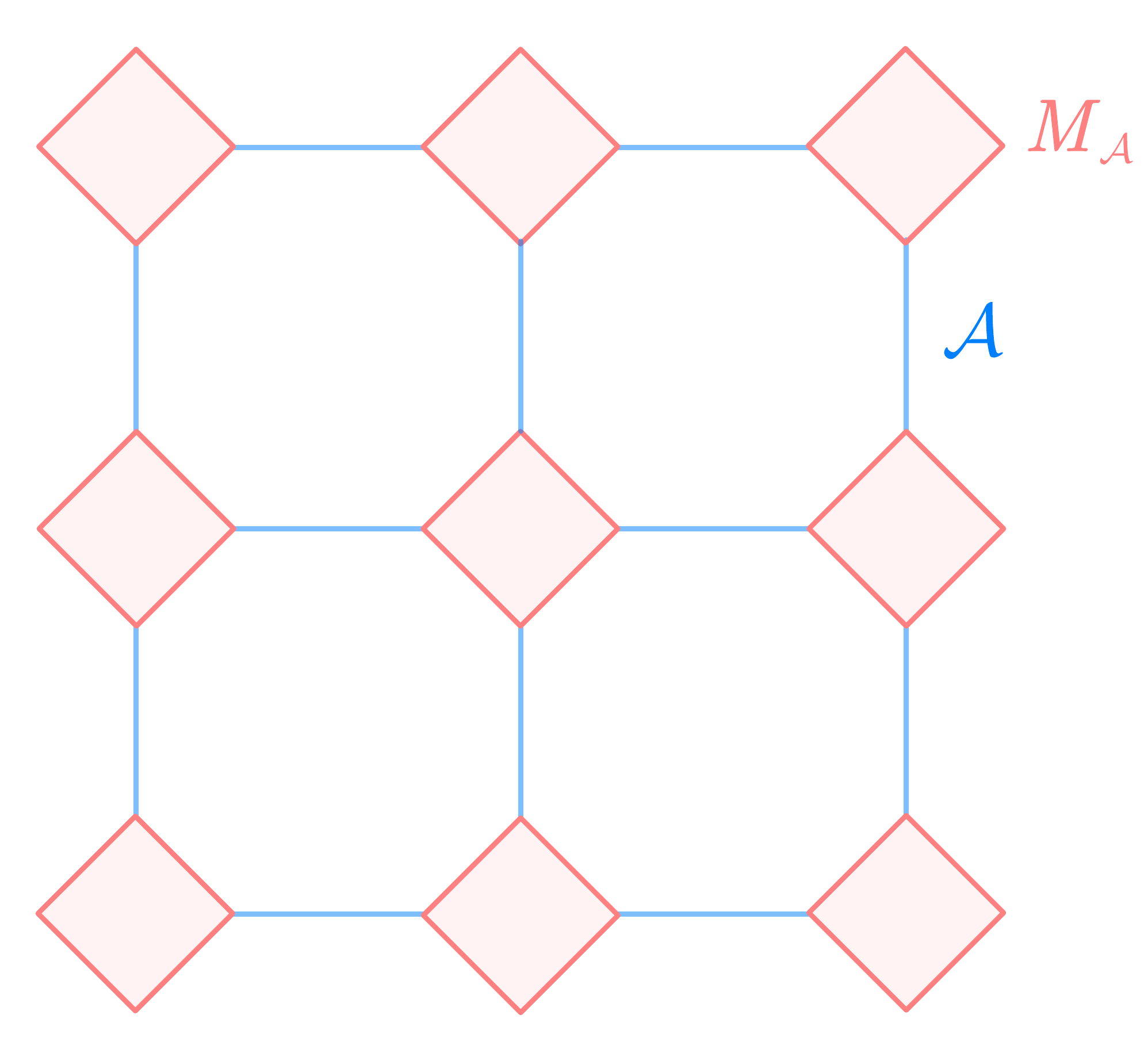}
    \caption{Ocean of condensate.}
    \label{fig:ocean_of_A}
\end{figure}

Summarising, 
\begin{equation} \label{gapped_state}
    \langle \unitcellColorAlgMod{\A}{M_\A} | = \sum_{a\in L_\A}\sum_{x,y,u,v\in L_{M_\A}} \Big\langle \OrientedLocalState{a}{x}{y}{u}{v}\Big| \ [\rho_{M_\A}]_{xy}^a([\rho_{M_\A}]_{uv}^{a})^\ast.
\end{equation}      
Globally, this arrangement is depicted in figure \ref{fig:ocean_of_A}. It is evident that the modules form small closed loops in an ocean consisting of the condensate $L$ made out of the Frobenius algebra $\mathcal{A}$. By the results in Appendix \ref{app:symRG}, one can readily show that within one RG step, where one removes the module loops using properties satisfied by modules, only the ocean of $\mathcal{A}$ remains. That is, the global state constructed from (\ref{gapped_state}) reduces to (\ref{fixed_point}). Therefore, (\ref{gapped_state}) is exactly one RG step away from a gapped fixed point, for {\it any} choice of module $M_{\mathcal{A}}$ of $\mathcal{A}$. 


\subsubsection*{Illustration: $\bra{\Omega}$ corresponding to Gapped Boundary Conditions of the Doubled Ising SymTO}

We now extend our construction of the state $\bra{\Omega}$ to the case of competing anyon condensations, illustrated by a concrete example. In the HGW string-net model, anyons reside in the plaquettes of the lattice, and their types are labeled by the simple objects of the Drinfeld center \(\Cent(\Fus)\) of the input UFC \(\Fus\) (see Appendix \ref{sec:HGW}). A key feature of the HGW model is that it manifests the internal gauge degrees of freedom of non-Abelian anyons, such that when condensing a non-Abelian anyon, a certain explicit gauge choice of the condensate can be made, analogous to e.g. the unitary gauge in the Higgs boson condensation. Specifically, anyons are realized in the model by \emph{dyons}---a pair consisting of an anyon type \(J\) and its internal gauge degree of freedom \(p\). A given anyon type \(J\), as a simple object in \(\Cent(\Fus)\), may carry multiple values of \(p\). An anyon with certain fixed internal degree of freedom $p$ value is a \textit{dyon} if $p$ is a nontrivial simple object of $\Fus$\footnote{The internal spaces of anyons should better be understood in an enlarged version of the HGW model\cite{Zhao:2024ilc}. But we shall not need this much detail here in this paper.}. Although such internal degrees of freedom \(p\) is unobservable in a topological phase due to topological invariance, they are pivotal in constructing the CFT states, which break topological invariance and thus expose them as physical local degrees of freedom that determine the physics likely to be captured by a critical CFT.

Consider the bulk symTO being the doubled Ising as an example. It is described by the string-net model with the input Ising UFC \cite{zhao2023characteristic}, which contains three simple objects \(\{1, \psi, \sigma\}\), respectively with quantum dimensions \(d_1 = d_\psi = 1, d_\sigma = \sqrt{2}\), subject to the fusion rules:
\[
\sigma \otimes \sigma = 1 \oplus \psi, \qquad \sigma \otimes \psi = \psi \otimes \sigma = \sigma, \qquad 1\otimes \psi = \psi,\qquad 1\otimes\sigma=\sigma.
\]
The output Ising topological order has \(9\) anyon types---labeled by the simple objects in the Drinfeld center of the Ising UFC---but \(10\) dyon types: 
\[(1\bar 1, 1),\qquad(1\bar\psi, \psi),\qquad(1\bar \sigma, \sigma),\qquad(\psi\bar 1, \psi),\qquad(\psi\bar\psi, 1),\]\[(\psi\bar\sigma, \sigma),\qquad(\sigma\bar 1, \sigma),\qquad(\sigma\bar\psi, \sigma),\qquad(\sigma\bar\sigma, 1),\qquad(\sigma\bar\sigma, \psi),\]
as anyon \(\sigma\bar{\sigma}\) possesses a 2-dimensional internal gauge space, spanned by simple objects \(1\) and \(\psi\)---appearing in the second entry in each pair above. 
There are two Frobenius algebras of the input Ising UFC:
\[
\A_0 = 1, \qquad f^1_{111} = 1;\qquad \A_0 = 1 \oplus \psi, \qquad f^2_{111} = f^2_{1\psi\psi} = 1.
\]
These two input Frobenius algebras are ``Morita equivalent'' as they have the same full center \(L = 1\bar 1\oplus\psi\bar\psi\oplus\sigma\bar\sigma\), which is the Lagrangian algebra of the doubled Ising phase. That is, anyon condensations associated with the input Frobenius algebras \(\A_0\) and \(\A_1\) both correspond to condensing anyons \(1\bar{1}\), \(\psi\bar{\psi}\), and \(\sigma\bar{\sigma}\), but the two condensed phase differ in the specific condensed components of anyon \(\sigma\bar{\sigma}\): \(\A_0\) corresponds to condensing \((\sigma\bar\sigma, 1)\), while \(\A_1\) condenses \((\sigma\bar\sigma, \psi)\)\cite{Zhao:2024ilc}. The set of Morita equivalent Frobenius algebras $\A$ of $\Fus$ is analogous to the goldstone mode parameter $\theta$ in the order parameter of the Cooper pair condensation; hence, it captures the microscopic detail of the corresponding anyon condensate $L$. 

Frobenius algebras $\A_0$ and $\A_1$ share a module $M = \sigma$, with its module functions defined as 
\[\qquad [\rho_{M_{\A_0}}]^1_{\sigma\sigma} = 1,\qquad [\rho_{M_{\A_1}}]^1_{\sigma\sigma} = [\rho_{M_{\A_1}}]^\psi_{\sigma\sigma} = 1.\]

We can consider $\A_0$ and $\A_1$ separately, picking the common module $M=\sigma$, and apply (\ref{gapped_state}) to each algebra to generate two different condensate puddles: 
\begin{subequations}
\begin{align} \label{isingA}
 \langle\unitcellColorAlgMod{\A_0}{\sigma}| &=   \langle{\unitcellColorMod{1}{\sigma}}|,\\ \qquad
\langle\unitcellColorAlgMod{\A_1}{\sigma}| &=   \langle{\unitcellColorMod{1}{\sigma}}| + \langle{\unitcellColorMod{\psi}{\sigma}}| .
\end{align} 
\end{subequations}
Here, we omit the directions of edges because the simple objects are self-dual and introduce the following notation:
\[{\langle\unitcellColorMod{a}{b}|}=\langle\OrientedLocalState{a}{b}{b}{b}{b}|.\] 
\subsection{Competing Condensates in a Unit Cell and Critical Points in a Square Lattice} 

In the above, we have divided the square lattice into unit cells of the form of \unitcell, and form a Lagrangian condensate $L_i$ corresponding to $\mathcal{A}_i$ in this cell. 

Now we are ready to set up competition between two condensates within a unit cell. It is simply given by 
\begin{equation} \label{pairofcondensates}
    \big\langle \unitcell_{(r_1,\mathcal{A}_1, M_{\mathcal{A}_1}), (r_2, \mathcal{A}_2, M_{\mathcal{A}_2})} \big| = r_1\langle\unitcellColorAlgMod{\A_1}{M_{\A_1}}| + r_2\langle \unitcellColorAlgMod{\A_2}{M_{\A_2}}|,
\end{equation}
where the coefficients $r_i$ are generically complex variables determining the relative weight of the condensates. In this paper however, inspired by 2D classical statistical models, we consider real and positive weights. The effect of complex weights would be explored elsewhere.
To ensure that the competition can be arranged locally and independently of other unit cells, the module $M_{\A_i}$ has to share exactly the same collection of objects, i.e., 
\begin{equation}
    M_{\A_i} = M_{\A_j},
\end{equation}
even though the algebra actions may differ. When there exist objects in either modules that are not shared by both modules, one can show that the competition between different unit cells is entangled. In which case, the equilibrium within a unit cell would be spoiled by its neighbours, making it hard to predict the form of global equilibrium.  This is explained with further examples in the appendix \ref{app:entangle}. We keep the subscript $\A_i$ in the modules to remind ourselves that the algebra actions of different algberas on the shared module are not required to be the same in general. We note that whether the algebra actions are chosen to be the same makes a difference. This will be explained and further illustrated with examples in section \ref{sec:first_order}.

More generally, one can introduce a competition of multiple condensates in each unit cell :
\begin{equation} \label{many_condensates}
    \big\langle\unitcell_{\{(r_i, \mathcal{A}_i, M_{\mathcal{A}_i})\}}\big| = \sum_i  r_{i} \langle \unitcellColorAlgMod{\A_i}{M_{\A_i}}|.
\end{equation}
Similarly, in this case, we require that 

$$    M_{\A_i} = M_{\A_j}
$$
for all $i,j$ that appears in the collection. 

By varying these coefficients, we expect to identify not only critical points between pairs of phases but also multi-critical points among $n$ phases. To make analytical prediction of critical values of $r_i$, such that we land on a critical point, we need to have a notion of the ``amount'' of condensate $L_i$ created in $\langle \unitcellColorAlgMod{\A_i}{M_{\A_i}}|$. In other words, we need to determine a normalisation for $\langle \unitcellColorAlgMod{\A_i}{M_{\A_i}}|$, which standardises the measure of how much condensate is created in the normalised state. 

The global state $\langle \Omega |$ is a tensor product over all the unit cells (in the sense that unit cells sharing the same slanted edge should have matching shared edge labels) 
\begin{equation} \label{globalfromunit}
\langle \Omega | = \otimes_{\textrm{all unit cells}} \big\langle\unitcell_{\{(r_i, \mathcal{A}_i, M_{\mathcal{A}_i})\}}\big|
\end{equation}

\subsubsection{Normalising a Condensate Puddle}
\label{sec:normalise}
\begin{figure}
    \centering
    \includegraphics[width=\linewidth]{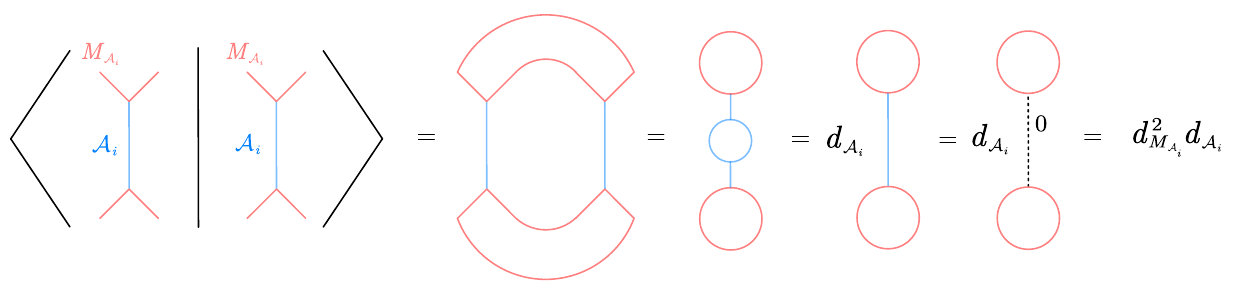}
    \caption{Calculation of vertex normalization. In the second and third step we have used the definition of module and algebra respectively. See \ref{app:alg_mod} for more detail.}
    \label{fig:vertex_normalization}
\end{figure}

To determine the correct normalisation of a condensate, we resort to a natural notion of ``inner product'' in an input UFC $\Fus$. The normalisation in a given unit cell \unitcell is given by taking the conjugate of the diagram and joining them, as depicted in figure \ref{fig:vertex_normalization}. 
This evaluates to 
\begin{equation}
N^2_{(\mathcal{A}_i, M_{\mathcal{A}_i})} \equiv   \langle \unitcellColorAlgMod{\A_i}{M_{\A_i}}| \unitcellColorAlgMod{\A_i}{M_{\A_i}}  \rangle =  d_{M_{\mathcal{A}_i}}^2 d_{\mathcal{A}_i}.
\end{equation}
In figure \ref{fig:vertex_normalization}, we used module properties to simplify the diagram into a product of two $M_{\mathcal{A}_i}$ loops and one $\mathcal{A}_i$ loop. A loop is a trace, equal to the quantum dimension of the object, giving the last equality above. In particular 
\begin{equation}
    d_{M_{\mathcal{A}_i}} \equiv \sum_{c \in M_{\mathcal{A}_i}} m_c \, d_{c},\qquad d_{\mathcal{A}_i} \equiv \sum_{a\in \mathcal{A}_i} n_a\, d_a.
\end{equation}

Given this inner product, one can normalise the state as follows,
\begin{equation} 
    \langle \widehat \A_i |_{M_{\A_i}} \equiv  \frac{\langle\unitcellColorAlgMod{\A_i}{M_{\A_i}}| }{N_{(\mathcal{\A}_i, M_{\mathcal{\A}_i})}}.
\label{Anorm}
\end{equation}
We propose that the location of the phase transition point between two competing anyon condensations $L_i$ and $L_j$ corresponding respectively to the Frobenius algebra $\mathcal{A}_i$ and $\mathcal{A}_i$ is given by the state. 
\begin{equation} \label{pair_compete}
     \big\langle \unitcell_{(\mathcal{A}_i, M_{\mathcal{A}_i}),(\mathcal{A}_j, M_{\mathcal{A}_j})}\big|_{\textrm{critical}} =\langle \widehat \A_i |_{M_{\A_i}}+\langle \widehat \A_j |_{M_{\A_j}}
\end{equation}

This is a central result of the current paper. 
Let us comment on the connection between the weights and the square lattice. When the symTO is a lattice gauge theory such as the toric code model, the anyons are either electric excitations that are created at the vertices, or magnetic excitations, that are created in the plaquettes.
The square lattice has an equal number of vertices and plaquettes (i.e. the square plaquette circumferenced by the module loop can be considered as a small island without any condensate and can be considered also as a vertex), thus justifying the location of the equilibrium point to be one of equal weight summation of the competing anyons. 
For more general lattice, we expect that weights of the condensing anyons depend on the vertice to plaquette ratio. 
It is worth noting that when $\A_i$ and $\A_j$ share the same module $M$ i.e. $M_{\A_i} = M_{\A_j} = M$ and that all the module functions agree, then the critical state is independent of the precise choice of $M$ among different shared modules. 
Generically $\A_i$ and $\A_j$ share modules if $\A_i \subset \A_j$, or vice-versa. 

\subsubsection{Illustration: Competing Condensates in the Ising String-Net and the Critical Ising Spin Model} \label{sec:ising}

We constructed the states in a unit cell corresponding to individual condensates in the Ising string net model in (\ref{isingA}).

The normalisations of the condensate puddles evaluate to

\begin{equation}
    N_{(\mathcal{A}_1, \sigma)}  = \sqrt{2}, \qquad  N_{(\mathcal{A}_2, \sigma)}  = 2.
\end{equation}
Taking into (\ref{pair_compete}), we have
\begin{equation} \label{isingomeg}
\big\langle{\hskip 0.3em\unitcell \hskip 0.3em}_{\textrm{critical}} \big|= \frac{1}{2}\left((\sqrt{2} + 1)\langle{\unitcellColorMod{1}{\sigma}}| + \langle{\unitcellColorMod{\psi}{\sigma}}| \right).
\end{equation}

The overall normalization coefficients is irrelevant. Comparing  the ratio between the two local states with results in \cite{Vanhove:2018wlb,Aasen:2016dop,Chen:2022wvy} that express the Ising spin partition functionn as a strange correlator, one finds that the strange correlator (\ref{strangecorr}) constructed from the tensor product of (\ref{isingomeg}) precisely reproduces the Ising spin model at criticality with inverse temperature
\[
\beta_c = \frac{1}{2}\ln\left( \sqrt{2}+1\right).
\] 
This equilibrium point between the two competing anyon condensates thus reproduces the celebrated critical temperature of the 2D classical Ising model and faithfully realizes the Ising CFT. We give a concise review of this result in the appendix \ref{sec:critIsing}.

We note that state (\ref{isingomeg}) is precisely the eigenstate with largest eigenvalue of the equal weight sum of projectors creating the condensates in the HGW lattice as shown in (\ref{projector}) in Appendix \ref{sec:HGW}. This is the key inspiration that led to a general construction.



\subsection{Another Series of Examples: the \eqs{A_{k+1}} Minimal Series}
We note that the Ising spin model from the Ising string net model discussed above is a special case in a series, namely the $A$-series lattice integrable models reproducing the $A$-series minimal models at criticality. They can be constructed from strange correlators using the \(A_{k+1}\) fusion category as input category of the string-net model \cite{Aasen:2020jwb, Chen:2022wvy}. Here we demonstrate that the critical state $\langle \Omega_{\textrm{critical}}|$ can be computed using (\ref{pair_compete}).
The category \(A_{k+1}\) contains \(k+1\) simple objects labeled \(0,1,2,\dots,k\), with fusion rules given by
\[
a \otimes b = |a-b| \oplus (|a-b| + 2) \oplus \cdots \oplus \min\{a + b,\, 2k - a - b\}.
\]
The quantum dimension of a simple object \(a\) is
\[
d_a = \frac{\sin \left(\frac{(a+1)\pi}{k+2} \right)}{\sin \left(\frac{\pi}{k+2} \right)}.
\]
For \(k \ge 2\), the first few quantum dimensions are \(d_0 = 1\), \(d_1 = 2\cos\tfrac{\pi}{k+2}\), and \(d_2 = d_1^2 - 1\). In particular, the Ising UFC corresponds to the special case \(k = 2\), where the simple objects \(0\), \(1\), and \(2\) map to \(1\), \(\sigma\), and \(\psi\), respectively.

The string-net model with input \(A_{k+1}\) UFC outputs the $\Cent(A_{k+1})$ {UMTC}, which contains \((k+1)^2\) anyon types, labeled by a pair \(a\bar b\), where $0\le a, b \le k$. The dyon types of anyon \(a\bar b\) are labeled by a pair \((a\bar b, p)\), where \(p\) is the internal degree of freedom, satisfying
\[0\le p\le k,\qquad |a - b|\le p\le \min(a + b, 2p - a - b),\qquad 2 | (a + b - p).\]

For all $k\ge 2$, one can check that $\A_0 = 0$ and $\A_1 = 0\oplus 2$ are two Frobenius algebra in \(A_{k+1}\) UFC. Specifically,
\[\A_0 = 0,\qquad [f_0]_{000} = 1,\qquad d_{\A_0} = 1,\]
\[\A_1 = 0\;\oplus\;2,\qquad [f_{1}]_{000} \;=\; [f_1]_{022} = 1,\qquad [f_1]_{222} = \sqrt{\sqrt{d_2}-\dfrac{1}{\sqrt{d_2}}},\qquad d_{\A_1} = d_1^{\,2}=d_2+1 .\]
Both algebras admit a common right-module \(M = 1\) containing only one simple object, with module function
\[[\rho_{\A_0}]^{0}_{11}=1,\qquad [\rho_{\A_1}]^{0}_{11}=1,\qquad [\rho_{\A_1}]^{2}_{11}=\sqrt[4]{d_2}.
\]

Applying again the prescription in (\ref{pair_compete}) gives (up to overall normalization)
\begin{equation} \label{Akcrit}
\big\langle{\hskip 0.3em\unitcell \hskip 0.3em}_{\textrm{critical}} \ \big| \quad =\quad \langle{\unitcellColorMod{0}{1}}|\quad +\quad \frac{\sqrt{2 \cos \left(\frac{2 \pi}{k+2}\right)+1}}{2 \cos \left(\frac{\pi}{k+2}\right)+1}\quad\langle{\unitcellColorMod{2}{1}}|\ ,
\end{equation}
which matches precisely with the entire Andrews-Baxter-Forrester (ABF) lattice realisation of the $A_{k+1}$ series of minimal models \cite{Andrews:1984af} for all $k$. The critical couplings, when expressed in terms of the strange correlator as above, can be found in \cite{Chen:2022wvy}, which completely agree with the results here.  These critical points are all second-order and known to approach the minimal model CFT with central charge $c=1-\frac{6}{(k+1)(k+2)}$ in the thermodynamic limit.


The above state can be physically interpreted as injecting an \emph{equal} amount of two competing anyon condensates \(L_1\) and \(L_2\) into each unit cell.  The inability for either party to condense results in criticality. Here, \(L_1\) and \(L_2\) are Lagrangian algebras
\[
L_1 = L_2 = 0\bar{0} \oplus 1\bar{1} \oplus 2\bar{2} \oplus \cdots \oplus k\bar{k}.
\]
in MTC \(\Cent(A_{k+1})\) and are the full center of the Morita equivalent input Frobenius algebras \(\A_0 = 0 \) and \(\A_1 = 0\oplus 2\), respectively. The naive identity \(L_1 = L_2\) highlights a key limitation of using output Lagrangian algebras alone to characterize the competing anyon condensation processes. Although the condensed anyon \emph{types} of \(L_1\) and \(L_2\) coincide, the condensed \emph{dyonic} sectors in the two phase transitions differ:

\(\A_0\) corresponds to condensing all dyons \((a\bar{a}, 0), 0\le a\le k\) with trivial internal degreee of freedom \(0\), whereas \(\A_1\) corresponds to condensing dyonic sectors with both internal degrees of freedom \(0\) and \(2\).

We illustrated how the $A$-series ABF models can be reformulated in terms of equal mixtures of competing condensates. We note that when the module $M$ coloring the slanted edge in a unit cell consists of a single object, these examples naturally coincide with the anyon chain models.

On the other hand, while it is generally expected that the interplay between two anyon condensations can drive the system to a phase transition point, it is not immediately clear why such a transition should be second-order rather than first-order. The competition among multiple anyon condensations plays a central role in determining the critical behavior of the system. Therefore, to identify genuine second-order phase transitions—and, more ambitiously, to obtain a comprehensive understanding of the global phase diagram—it is essential to systematically explore the relationships between all possible condensates. This is discussed in the next section. 

\subsubsection{Generalised Kramers-Wannier Duality}
The Kramers-Wannier ({KW}) duality is a duality transformation in statistical mechanics that relates the ordered and disordered phases of a system. The phase transition point between these two phases coincide with the self-dual point, where the system often exhibits critical behavior. For example, in the classical Ising model, the KW duality relates the high-temperature and low-temperature phases. The critical temperature is found to coincide with the self-dual point. In $\Z_N$ lattice models, KW duality refers to a Fourier-type transform of couplings \cite{Alcaraz:1980bb}. In these cases, the KW duality is essentially the electromagnetic duality, which exchanges the electric and magnetic charges of the global $\Z_N$ symmetry of the lattice model. (i.e. $N=2$ for the Ising model.)
Moreover, the KW duality turns out to be closely related to Fourier analysis in the context of planar algebra \cite{jiangBlockMapsFourier2017,luClassificationExchangeRelation2024,liuFunctionalIntegralConstruction2024}, via the recently developed Alterfold theory\cite{liuAlterfoldTopologicalQuantum2023,liu3AlterfoldsQuantumInvariants2023,liuAlterfoldTheoryTopological2024}.

\begin{figure}
    \centering
    \includegraphics[width=0.95\linewidth]{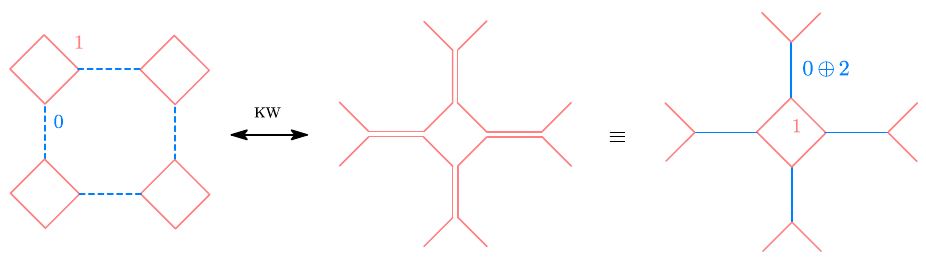}
    \caption{Generalised KW duality for the $A_{k+1}$ category. The common module is $1$ and the competing Frobenius algebras are $\mathcal{A}_1 = 0$ and $\mathcal{A}_2 = 0\oplus 2$. Under KW duality the $1$-squares are mapped to $1$-octagons, and the positions of squares and octagons are exchanged. The horizontal and vertical octagon edges are exchanged under KW. The double $1$-line on the octagon edge is equivalent to algebra $\mathcal{A}_2$ ending on the module $M=1$. So we see that the two gapped phases given by $\mathcal{A}_1$ and $\mathcal{A}_2$ are KW dual to each other.}
    \label{fig:Ak_KW}
\end{figure}

In the context of our construction, it turns out that these KW dualities known in the literature can be understood as a transformation that maps the original strange correlator on the square-octagon lattice to a dual strange correlator on the dual lattice. One of the advantages of using the square-octagon lattice is that its dual lattice has the same shape. The dual lattice is obtained by exchanging squares and octagons in the original lattice. This exchange is achieved by making an F-move on every octagon edge, as illustrated in figure \ref{fig:Ak_KW}. 

Under the given setup, the unit cell is $\langle{\unitcellColorMod{0}{1}}|+x\langle{\unitcellColorMod{2}{1}}|$ (up to normalization factors). Choosing \(x=0\) realises the child string-net model corresponding to algebra $\mathcal{A}_1$, while \(x=\sqrt{d_2}\) realises that corresponding to $\mathcal{A}_2$. 
In fact, the underlying system with $x=\sqrt{d_2}$ and $x=0$ is exchanged under the {KW} duality, as shown in figure \ref{fig:Ak_KW}. The double line on the octagon edge is a standard mathematical notation for the tensor product. Throughout this paper, a double line on an octagon edge denotes the insertion of the object $X \otimes X^*$. To be clear, we have
\begin{center}
\includegraphics[width=0.7\linewidth]{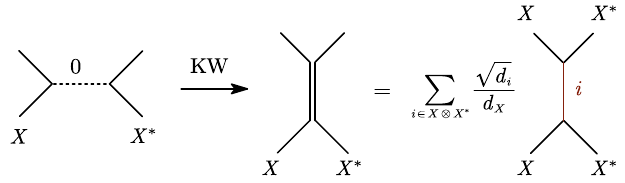}
\end{center}
The right-hand side coincides precisely with the state \eqref{gapped_state} for the algebra $X \otimes X^*$ and module $X$, up to a normalization.\footnote{From this relation, one immediately obtains the module coefficients $[\rho]^a_{xy}$ for the algebra $\mathcal{A} = X \otimes X^*$ with module $M = X$.  Moreover, the construction generalises straightforwardly to the case $\mathcal{A} = X \otimes X^*$ with $ M = X \otimes i$ for any simple object $i$, thereby encompassing all algebra–module pairs for algebras in the trivial Morita class.  By an analogous strategy, one determines a broad family of modules for algebras in nontrivial Morita classes purely from the algebra coefficients $f_{abc}$.  Only a small number of modules for some nontrivial minimal algebras (see (\ref{eq:XAX})) elude this method and must instead be computed by solving the corresponding polynomial consistency equations.}

The self-dual point $x = x_c$ is obtained by requiring that $\langle{\unitcellColorMod{0}{1}}|+x_c\langle{\unitcellColorMod{2}{1}}|$ is invariant under an F-move.
The solution is simply
\begin{equation} \label{eq:critAK}
    x_c = \frac{d_1-1}{\sqrt{d_2}}
\end{equation}
which exactly matches (\ref{Akcrit}) obtained by the equal sum of normalized states.
This transformation was also discussed in \cite{Aasen:2020jwb, Vanhove:2018wlb}. The self-dual point was discussed in the case of the Ising spin model, which correctly produces the critical temperature as above. The critical coupling for general $k$ as shown in (\ref{eq:critAK}) was obtained using the above method in \cite{Chen:2022wvy}. 
We discuss it systematically here because we find that this duality formulated in the above manner is a general feature in this kind of square-octagon lattice. In simple cases it simply coincides with electromagnetic duality, but for more general symTO which is not obviously related to group symmetries, the duality continues to play an important role in many of the cases.
This duality, combining with properties of Morita equivalent Frobenius algebra, can be further generalised to produce detailed information of the critical point and more generally the phase diagram. These will be illustrated in more examples below.  


\section{General Theory: Phase Diagrams and Critical Points via a {\it Refined} Tree of Anyon Condensation}  \label{sec:general_theory}
In this section we would like to understand the relationships between condensates, in order to sieve out first order phase transitions. For given second order phase transitions, it is also important to know what (minimal amount of) symmetries are preserved. 
Moreover, we would like to explain the global structure of phase diagrams, and predict precise locations of phase boundaries at least where it is close to the critical point obtained from equal mixtures. 

One very important ingredient organising the condensate is the notion of a {\it refined} anyon condensation tree.   


\subsection{{\it Refined} Condensation Tree}

Anyon condensation is mathematically described by a twist-free commutative, separable Frobenius algebra (CSFA) in the symTO characterised by the {UMTC} $\Cent({\Fus})$.  One twist-free CSFA\footnote{We use an italic letter $A$ for a CSFA in the output data $\Cent(\Fus)$ of the HGW model, compaired with the curly letter $\A$ for a Frobenius algebra in an input UFC $\Fus$.} $A_i$ can be a sub-algebra of another bigger one $A_2$ i.e. $A_1 \subset A_2$. When this occurs, it implies that $A_2$ can be condensed in two steps. First condensing anyons in $A_1$, before condensing some other unconfined anyons over the $A_1$ condensate to obtain the aggregate $A_2$ condensate.
Therefore, the sub-algebra $A_1$ can be considered as a parent of $A_2$.
Such considerations produce a hierachy of condensates \cite{Ji:2019jhk,Chatterjee:2022kxb, Bhardwaj:2024qrf} that can be conveniently arranged as a tree, which was dubbed the Hasse diagram \cite{Bhardwaj:2024qrf}, with each node corresponding to a condensable algebra, two nodes connected by an edge denotes one being the sub-algebra of the other. 
Lagrangian algebras are those twist-free CSFAs corresponding to maximal condensates---such that the resultant phase is topologically trivial---would appear as bottom nodes of the tree. That is, we start from the trivial twist-free CSFA including only the identity object, which is a sub-algebra to all other algebras, as the top node of the tree and grow the tree downwards.
In this section, we will introduce a refined version of the condensation tree, which is necessary to understand the phase diagram of these lattice models. 

The key refinement of our condensation tree over Hasse diagrams previousely considered \cite{Bhardwaj:2024qrf} is the inclusion of Morita equivalent Frobenius algebras as physically distinct condensate. While these algebras correspond to the same set of condensed anyons in the output theory, they produce physically distinct gapped phases\cite{Chen:2022wvy,Zhao:2024ilc,zhao2025}, such that non-trivial phase transitions between them can in fact occur, as is evident in the previous section. Our construction of explicit lattice models relies on constructing Frobenius algebra $\mathcal{A}_i$ in the ``input'' UFC ${\Fus}$. 
The simplest example is the symTO corresponding to the doubled Ising model that we have taken as our first example in the previous section. 
Recall that there are three objects denoted $1,\sigma, \psi$ in the input Ising UFC, and there are two connected Frobenius algebra $\mathcal{A}_0$ and $\mathcal{A}_1$ as in (\ref{A0A1A2}). As mentioned, they are Morita equivalent and correspond to the same Lagrangian algebra.  (Choosing a Frobenius algbera is analogous to gauge-fixing a Goldstone mode in the Higgs mechanism.\cite{Zhao:2024ilc,zhao2025}) 
 Yet, the Ising CFT is the critical CFT at the phase transition between them, implying that they are physically different, and that focusing solely on the anyon content in the output condensate $L_i$ is inadequate. The physical distinction between $\mathcal{A}_0$ and $\mathcal{A}_1$ lies in the microscopic details of the condensate. For an anyon $X \in L$ with quantum dimension $d_X \ge 2$, which is the case for $X= \sigma \bar \sigma $, can participate in $L$ in more than one way. As to $\sigma \bar \sigma$, it can be understood as the direct sum of the electric charge and magnetic flux $ e \oplus m$ in the $\Z_2$ toric code topological order \cite{Bais:2008ni, Hung:2013nla,  Lu:2013jqa, Hung:2013qpa, Barkeshli:2014cna, wang2020electric, Zhao:2024ilc} after gauging the $em$-exchange symmetry, and that $\mathcal{A}_0$ describes $e$ condensation and $\mathcal{A}_1$ describes $m$ condensation\cite{Hu:2017lrs,Hu:2017faw, Chen:2022wvy}.   

 Summarising, Morita equivalent Frobenius algebras describe physically distinguishable phases, enabling phase transitions between Morita equivalent condensates. The most natural way of distinguishing these condensates is to directly consider different Frobenius algebras in the input category ${\Fus}$, rather than focusing merely on the output condensate $L_i$. 

 Intermediate phases where some non-maximal set of anyons condensed can be described by sub-category\footnote{A simple object of $\mathcal{K}$ may be a nonsimple object of $\Fus$.} $\mathcal{K} $ of the input category ${\Fus}$\cite{Zhao:2024ilc}. This is natural since a non-trivial condensate would take the bulk symTO to a child topological order, and the corresponding degrees of freedom on the lattice is expected to be reduced to some subset of ${\Fus}$. When the condensate $L_i$ in the symTO is Lagrangian, the corresponding input string-net model is projected effectively down to one-dimensional, reflecting the fact that the corresponding condensed phase is topologically trivial. In the condensed phase, each edge is projected to the chosen Frobenius algebra $\mathcal{A}_i$ corresponding to $L_i$. For instance, if we select the trivial Frobenius algebra $\mathcal{A}_0 = 1 $, the resulting child model has only a single physical state, where every edge is labeled by the trivial degree of freedom \( 1 \).

Similar to the Morita equivalence of Frobenius algebras, multiple sub-categories of ${\Fus}$ could correspond to the same set of condensable anyons in $\Cent(\Fus)$, but they contain more microscopic details that are physically distinguishable. Therefore, we introduce the refined condensation tree as follows.

Each node is labeled by a subcategory \(\mathcal{K}\) of \({\Fus}\). Note that the identity object of a subcategory \(\mathcal{K}\) is, in general, a composite object in \({\Fus}\) unless \(\mathcal{K}\) is a full subcategory. The requirement that \(\mathcal{K}\) possesses an identity object is precisely the requirement that there exists a separable Frobenius algebra---after all, the defining property of a separable Frobenius algebra is that the composite object \(\mathcal{A}\) is strictly associative under fusion, and that ``bubbles'' (i.e., trivial summands) can be freely included or removed. Thus, the identity object of \(\mathcal{K}\) is given by a Frobenius algebra \(\mathcal{A}\), and the remaining objects in \(\mathcal{K}\) are precisely the bimodules over \(\mathcal{A}\).

Two nodes are connected by a line if the identity object of \(\mathcal{K}_i\), given by a Frobenius algebra \(\mathcal{A}_i\), is a subalgebra of the identity object \(\mathcal{A}_j\) of \(\mathcal{K}_j\), and \(\mathcal{A}_j\) can be obtained from \(\mathcal{A}_i\) as a Frobenius algebra in the subcategory \(\mathcal{K}_i\).

The above statements is essentially rephrasing the relation between two nodes in the original Hasse diagram -- namely that a node is connected to another if the latter phase can be obtained from the former by anyon condensation. The refined condensation tree contains more nodes in general. 

Here are a few examples. 
\begin{figure}
    \centering
   \begin{tikzpicture}
[grow = down,
sibling distance = 50mm,
    level distance = 30mm,
     edge from parent/.style = {draw, -latex},
  every node/.style = {align=center}
]
	\node{$\mathcal{K} = \Fus = A_3$ , \\ identity = $\mathcal{A}_0 = 0$ \\
    $\Cent(A_3)$ = Doubled Ising TO}
		child {
        node {$\mathcal{K} = \textrm{Vec}\Z_2 = \{0,2\}$\,\, \\ identity = $\mathcal{A}_0 = 0 $ \\
        $\Cent(\mathcal{K}) = \Cent(\Z_2)$ = toric code}
            child{node {$\mathcal{K}= \{0\}=\mathbb{I}$, \\
            identity = $\mathcal{A}_0$ \\
            $L = 1\bar 1 \oplus \psi\bar\psi \oplus \sigma\bar \sigma$ \\ electric condensate\\ $1\oplus e$}
            edge from parent node[left]{Condense $(\sigma\bar\sigma,\psi)= e$}}
            child{node{$\mathcal{K}= \{0\oplus2\} =\mathbb{I}$, \\
            identity = $\mathcal{A}_1$ \\
            $L = 1\bar 1 \oplus \psi\bar\psi \oplus \sigma\bar\sigma$ \\ magnetic condensate \\ $1\oplus m$}
            edge from parent node[right]{Condense $(\sigma\bar\sigma, 1) = m $}}  
            edge from parent node [right] {Condense $\psi\bar \psi$}};
\end{tikzpicture}
 \caption{Condensation tree for the doubled Ising TO. Each link is graced with the output bulk anyons that condensed. Each node is a subcategory and also the corresponding output condensed phase.}
    \label{fig:Ising tree}
\end{figure}
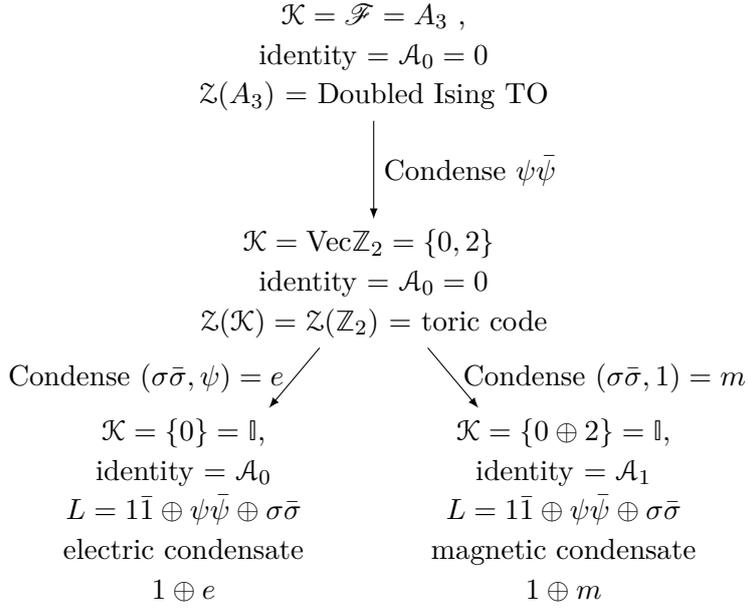
First, the condensation tree of the Ising {string-net} model is shown in figure \ref{fig:Ising tree}. 

A more interesting example is the condensation tree for $k=4$. A result from the ADE classification is that $A_5$ has two distinct modular invariants \cite{cappelli1987ade, hung2015generalized}, this is equivalent to the fact that $A_5$ has two Morita equivalence classes of Frobenius algebras, with $\mathcal{A}_0 = 0$ and $\mathcal{A}_1 = 0\oplus 4$ as the two representatives \cite{ostrik2002module}. 


A Morita equivalence class of Frobenius algebras in a UFC $\Fus$ has the following structure: Within each Morita class $i$, there exists a representative \emph{minimal} algebra $\mathcal{A}_i$, and all other algebras in the same class $i$ take the form
\begin{equation}\label{eq:XAX}
X \otimes \mathcal{A}_i \otimes X^*
\end{equation}
where $X$ is \emph{any} object of $\Fus$. However, $X \otimes \mathcal{A}_i \otimes X^*$ is connected if and only if $X$ is a simple object that preserves the connectedness of the algebra; otherwise, for certain choices of simple object $X$, the resulting algebra may be nonconnected.

The algebras $\mathcal{A}_0 $ and $\mathcal{A}_1$ are the minimal algebras in the two Morita classes of $A_5$. Using the rule above we can show that apart from the two algebras there are three other connected Frobenius algebras. $0\oplus2$ and $0\oplus2\oplus4$ are Morita equivalent to $\mathcal{A}_0$, while $0\oplus2\oplus2\oplus4$ is Morita equivalent to $\mathcal{A}_1$.

The two Morita classes correspond to the two Lagrangian algebras in the center $\Cent(A_5)$.

\begin{figure}[!ht]
    \centering
\begin{tikzpicture}[
        scale=1.,                
        transform shape,          
        grow = down,
        level distance = 20mm,    
        edge from parent/.style = {draw, -latex, thin}, 
        every node/.style = {
            align=center,
            font=\footnotesize    
        },
        level 1/.style={
            sibling distance=36mm 
        },
        level 2/.style={
            sibling distance=36mm
        },
        level 3/.style={
            sibling distance=18.4mm 
        },
        level 4/.style={
            sibling distance=18.4mm
        }
    ]

\node { $A_5$\\
$\Cent(A_5)$}
child{ node(n1){$ \{0,2,4\}=\textrm{Rep}S_3 $\\
$\Cent(S_3)$}
child{node(n2){$\{0,4\} $\\$=\textrm{Rep}\Z_2 $ \\
$\Cent(\Z_2)$}
     child{node(m1){$\{\A_0 = 0\}$\\ $L_0$ \\$A\oplus F \oplus D$}
     edge from parent node[left]{$(D,0)$}}
     child{node(m2){$\{\A_2= 0+4\}$ \\ $L_1$ \\ $A\oplus B\oplus 2F $} 
     edge from parent node[right]{$(F,4)$}}
edge from parent node[left]{$(F,0)$}}
child{node(n3){$\{0+2+4, 0-2+4\} $\\ $= \textrm{Vec}\Z_3$\\
$\Cent(\Z_2)$}
    child{node(m3){$\{\A_3= 0+2+4\}$\\ $L_0$ \\ $A\oplus F\oplus D$}
    edge from parent node[left]{$(D,2)$}}
edge from parent node[right]{$(F,4)$}}
child{node(n4){$\{0+4,2_1, 2_2\} $\\
$= \textrm{Rep}\Z_3$ \\
$\Cent(\Z_3)$}
    child{node(m4){$\{\A_4 = 0+ 2_1+2_2 + 4\}$\\$L_1$ \\ $A\oplus B\oplus 2C$} 
    edge from parent node[left]{$2(C,2)$}}
edge from parent node[right]{$(B,4)$}    }
child{node(n5){$\{0+2, 2+4\} $\\ $ = \textrm{Vec}\Z_2$ \\
$\Cent(\Z_2)$}
    child{node(m5){$\{\A_1 = 0+2\}$\\ $L_0$ \\ $A\oplus C \oplus D$} 
    edge from parent node[left]{$(D,0)$}}
edge from parent node[right]{$(C,2)$}
}
edge from parent node[right]{$4\bar 4$}
};
\draw[->] (n3)--(m2);
\draw[->] (n4)--(m2);
\draw[->] (n5)--(m4);
\node at (-3.5, -5) {$(B, 4)$};
\node at (-.5, -5) {$(F, 0)$};
\node at (3., -5) {$(B, 4)$};
    \end{tikzpicture}
    \caption{The condensation tree for the Doubled $A_{5}$ symTO.
    In the second layer of arrows we indicate the intermediate condensing anyons using anyon labels in $\Cent(S_3)$ to avoid clutter. In the last layer, we also include the condensate as anyons in the $\Cent(S_3)$ phase. As we move down the tree, simple anyons in the partially condensed symTO corresponding to the lower nodes are composites in the parent nodes. Hence, the condensing anyons expressed in terms of $\Cent(S_3)$ anyon that label the bottom layer of arrows are only representatives of a group of anyons in $\Cent(S_3)$ that are identified in the child $\Z_2$ and $\Z_3$ nodes. 
    The second index in the bracket indicates the microscopic degrees of freedom that take part in the condensation. 
    The meanings of these symbols are explained in the previous section. }
    \label{fig:A5}
\end{figure}
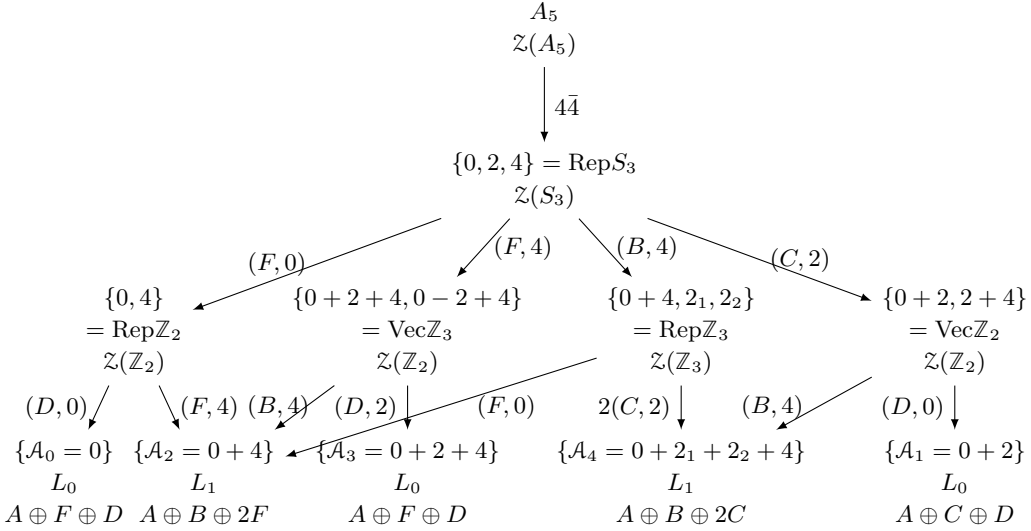
In the refined condensation tree for the symTO $\Cent(A_5)$, 
\begin{equation}
    L_0 = \oplus_{i=0}^4 i\bar i, \qquad L_1 = (0\oplus 4)  \overline{(0\oplus 4)} \,\oplus \,2 \,\, 2\bar 2.
\end{equation}
At the bottom of figure \ref{fig:A5}, we express each Lagrangian algebra also as comprised of anyons in the intermediate condensed phase $\Cent(S_3)$. The latter has 8 different anyons, and they are related to the anyons of $\Cent(A_5)$ as shown in the following table.  

\begin{table}[]
\begin{center}\begin{tabular}{||c|c|c||}\hline\hline
$\Cent({\tt Rep}S_3)$ Anyon Type & $\Cent({\tt Rep}S_3)$ Dyon Type & $Z(A_5)$ Dyon Types \\ \hline\hline
$A$ & $(A, 0)$ & $(0\bar 0, 0), (4\bar 4, 0)$ \\ \hline
$B$ & $(B, 4)$ & $(0\bar 4, 4), (4\bar 0, 4)$ \\ \hline
$C$ & $(C, 2)$ & $(2\bar 2, 2)$ \\ \hline
\multirow{2}*{$D$} & $(D, 0)$ & $(1\bar 1, 0), (3\bar 3, 0)$ \\ \cline{2-3}
& $(D, 2)$ & $(1\bar 1, 2), (3\bar 3, 2)$ \\ \hline
\multirow{2}*{$E$} & $(E, 2)$ & $(1\bar 3, 2), (3\bar 1, 2)$ \\ \cline{2-3}
& $(E, 4)$ & $(1\bar 3, 4), (3\bar 1, 4)$ \\ \hline
\multirow{2}*{$F$} & $(F, 0)$ & $(2\bar 2, 0)$ \\ \cline{2-3}
& $(F, 4)$ & $(2\bar 2, 4)$ \\ \hline
$G$ & $(G, 4)$ & $(0\bar 2, 2), (4\bar 2, 2)$ \\ \hline
$H$ & $(H, 4)$ & $(2\bar 0, 2), (2\bar 4, 2)$ \\ \hline\hline
\end{tabular}\end{center}
\caption{ Relation between $\Cent({\tt Rep}S_3)$ anyons and $\Cent(A_5)$ anyons. }
    \label{tab:S3vsA5}
\end{table}
In what follows, we will use these examples to explain how global structures of the phase diagrams and minimal amount of symmetries preserved at a critical point can be read-off from the refined condensation tree.

\subsection{Engineering Second Order Phase Transitions and Excluding First Order Transitions via Shared Modules}
As aforementioned, when competing anyons are at a tie, it is expected that the resultant phase is either gapless or degenerate. The latter corresponds to first order phase transitions from the perspective of the 2D classical statistical lattice model. Usually at a first order phase transition between two phases, the free energies of the respective phases become equal, and that is the classical interpretation of degeneracies in the 1+1 D quantum model. 

In our construction of the lattice model, we focus on a unit cell with 5 edges. The dimension of phase space is thus reduced to roughly speaking $|{\Fus}|^5$, where $|{\Fus}|$ denotes the number of simple objects of $\Fus$. For generic state  $\langle \Omega |$ with our unit cell, it is far from any RG fixed point. The interpolation (\ref{many_condensates}) at first sight involves a small number of couplings compared with $|{\Fus}|^5$. Under RG, however, all other linear combination of the fusion channels in the unit cell could be invoked. i.e. the dimension of possible couplings is much greater than the number of gapped phases nearby described by Frobenius algebras. In such cases, there would be plenty of room for first order phase transitions. 

In principle, as in the golden chain and other cases considered in \cite{Feiguin:2006ydp, Vanhove:2018wlb, Aasen:2016dop, Aasen:2020jwb}, the strange correlator construction imposes (non-invertible) topological symmetries in the 2D theory. Nonetheless, it is generally unclear how to ensure that the symmetry is not broken spontaneously in the quantum spin-chain by $\langle \Omega|$.

To exclude first order phase transitions or degeneracies, the key is to reduce the dimension of phase space as much as possible. To be more concrete, we need to reduce the number of new fusion channels that are generated under RG in much of the space of couplings. For a coupling describing a specific gapped phase, it would quickly converge to the gapped fixed point state (\ref{fixed_point}) under RG, i.e., a coupling corresponding to a gapped phase is essentially located within {\it the basin of attraction} under RG to a fixed point. In such a case, new fusion channels are not generated. In other words, it is desirable to choose (\ref{many_condensates}), such that we are restricted to fewer attraction basins for the space of couplings we explore.

One could count the dimension of couplings introduced as follows. 
The coefficient of each state $\sum_{x,y,u,v \in M}\Big\langle \OrientedLocalState{a}{x}{y}{u}{v}\Big| \ [\rho_{M_\A}]_{xy}^a([\rho_{M_\A}]_{uv}^{a})^\ast$ for fixed $a$ where $a \in \A_i$ constitutes an independent coupling in this reduced phase space. $N_{\A}$ is the number of objects in $\A$. Recall that to decouple different unit cells, we required that all $M_{\A_i}$ considered in a competition of condensates share the same set of module objects. As such, we can simply denote this common module by $M$; however, for the most generic case, the algebra actions of $\A_i$ and $\A_j$ on $M$ may differ. In such a case, in each Frobenius algebra $\A_i$, any basis element introduces an independent linear combination of the fusion channel $\sum_{x,y,u,v \in M}\Big\langle \OrientedLocalState{a}{x}{y}{u}{v}\Big| \ [\rho_{M_\A}]_{xy}^a([\rho_{M_\A}]_{uv}^{a})^\ast$. For a certain basis element $a \in \A_i \cap \A_j$, when it acts identically as an element of $\A_i$ and as an element of $\A_j$ on $M$, a single linear combination of fusion channel is attached to this element $a$. We refer to such a shared basis element $a$ a \textit{perfectly shared element}, and if the common elements of all algebras are perfectly shared on $M$, we term $M$ a \textit{perfect common module} for these algebras. Therefore, we can count the dimension  $D_\textrm{unit}$  of vector space in a unit cell as follows: 
\begin{equation}
    D_\textrm{unit} = \sum_{i} N_{\A_i}  - \sum_{\textrm{perfectly shared} \ a} (N^{\textrm{common}}_{a} - 1),
\end{equation}
where $N^{\textrm{common}}_{a}$ is the number of algebras $\A_i$ having a perfectly shared element $a$. For example, if $a$ is perfectly shared by two algebras acting on $M$, $N^{\textrm{common}}_{a}=2$. In particular, the identity object on any $M$ is always a perfectly shared element of all Frobenius algebras of a UFC. So for $n$ algebras sharing a module, $N^{\textrm{common}}_{\textrm{identity}}=n$. Since the overall normalisation of the state $\langle \Omega|$ is physically irrelevant in the strange correlator,  $D_\textrm{unit}-1$ gives the dimension $D_\textrm{couplings}$ of phase space introduced in the ansatz, i.e.,
\begin{equation} \label{eq:Dc}
    D_\textrm{couplings} = D_\textrm{unit} -1.
\end{equation}

For two competing algebras only, the number of degrees of freedom introduced in a unit cell is reduced to 
\begin{equation} \label{eq:Dunit}
    D_\textrm{unit}(\A_i, \A_j) = N_{\A_i} + N_{\A_j}  - N_{\textrm{objects $a$ with shared $\rho$}},
\end{equation}
where $N_{\textrm{objects with shared module}}$ directly counts the number of perfectly shared element $a\in \A_i \cap \A_j$. 

There is a special case in which the $D_\textrm{unit}$ is reduced most drastically, That is to choose a perfect common module $M$ of all competing Frobenius algebras $\A_i$. This can be typically achieved when the competing algebras are ordered as sub-algebras of one another: $\cdots \subset \A_i\subset \A_j \subset \A_k$. Suppose $\A_i\subset \A_j$, they must share a common ancestor in the refined condensation tree, where the identity object of the sub-input-category $\mathcal{K}\subset \Fus$ is given by the smaller Frobenius algebra $\A_i$. In the said special case, 
\begin{equation}
    D_\textrm{unit}( \cdots \subset \A_i\subset \A_j \subset \A_k) = N_{\A_k}.
\end{equation}
That is, the number of degrees of freedom is simply given by the number of objects in the largest algebra among all algebras sharing the same perfect common module $M$. 

As already noted in section \ref{sec:normalise}, in this announced special case, the choice of perfect common module $M$ is immaterial. One can show that within one step of symmetric RG (as reviewed in appendix \ref{sec:symRG}.), $\langle\Omega(M)|$ constructed from interpolations of normalised Frobenius algebras with module $M$ in each unit cell reduces to the same state $\langle\Omega|$ for all choices of $M$, up to overall normalisations depending on the quantum dimension of $M$. 

Now for a choice of perfect common module $M$ shared only by
$n_M$ different Frobenius algebra among all competing ones, we note that all other gapped phases corresponding to Frobenius algebras not sharing $M$ would not occur under RG. The RG attraction basins would be restricted to those gapped phases that share this $M$. This is an effective method to reduce the possibilities of gapped phases showing up in the reduced phase space. 

If $D_c := D_\textrm{couplings}=D_\textrm{unit}-1$ is larger than the number of $n_M$ of algebras $\A_i$ sharing the module $M$ (i.e. $1\le i\le n_M$, and $n_M$ is the number of gapped phases allowed in the reduced phase space), there is naviely still room for degeneracies and thus room for first order phase transitions. We thus arrive at the following \textit{sufficient condition} for second order phase transitions:

{\bf All phase transitions in a phase diagram are forced to be second order if }
\begin{equation} \label{eq:2ndorder}
 n_M > D_c . 
\end{equation}
When (\ref{eq:2ndorder}) is satisfied, the number of competing anyon condensates is greater than the number of independent degrees of freedom within a unit cell that can be varied. Therefore consider tuning a pair of condensates to be commensurate, changing the weight of any of the $D_c-1$ independent coefficients correspond to moving in a direction towards another condensate that cannot be condensed at the same time as the pair anyway, leaving no room for degeneracy. 

Now let us illustrate situations where there are first order phase transitions. 
\subsubsection{First Order Transitions -- with \eqs{{\Z}_N}  as Illustrations}
\label{sec:first_order}
The Abelian fusion category  $\Vec{\Z_N}, N \ge 2$ is the representation category of the cyclic group $\Z_N$. Basically, we can treat $\Z_N$ as the input category in the {string-net} model.  Its simple objects are the group elements of $\Z_N$, and are conveniently labeled by $0,1,\dots ,N-1$. The fusion rule is inherited from the group multiplication of $\Z_N$:
\[i\otimes j \;=\; i+j \pmod{N}, \qquad 0\le i,j < N ,\]
and every simple object has quantum dimension $d_i = 1$. The output topological order is a $\Z_N$ quantum double.

The case of $N=2$ as an example is in fact the $\Z_2$ toric code model, and for generic $N$ they are simply the $\Z_N$ quantum double/ Dijkgraaf-Witten models. 

Fusion category $\Vec(\Z_N)$ contains at least two connected special symmetric Frobenius algebras:
\begin{align*}
  \A_e &\;=\; 0, & f^{\A_e}_{000} &= 1, & d_{\A_e}&=1,\\
  \A_m &\;=\; \bigoplus_{a=0}^{N-1} a, & f^{\A_m}_{abc} &= \delta_{a\otimes b,c}, & d_{\A_m}&=N .
\end{align*}
These two Frobenius algebras correspond to two Lagrangian condensates. Namely, they correspond to the all-electric condensate and all-magnetic condensate respectively. Clearly, $\A_e \subset \A_m = \Z_N$, and their simplest common module is $M= \Z_N$: 
\[M_{\A_e} = M_{\A_m} = M \;=\; \bigoplus_{x=0}^{N-1} x, \qquad d_{M}=N ,\]
whose module functions are
\[\bigl[\rho_{M_{\A_e}}\bigr]^{0}_{xy} \;=\; \delta_{x,y}, \qquad \bigl[\rho_{M_{\A_m}}\bigr]^{a}_{xy} \;=\; \delta_{a\otimes x,y}, \qquad x,y,a\in{\Z}_N .
\]
Here $\delta_{p,q}$ denotes the Kronecker delta.
In fact, each subgroup of $\Z_N$ forms a Frobenius algebra in $\Z_N$, and they clearly are sub-algebra of the largest Frobenius algebra $\A_m = \Z_N$ itself. Therefore, $M$ can always be chosen as a perfect common module shared by all these Frobenius algebra.

We define normalised condensate as in (\ref{Anorm}), which gives
\[\bra{\widehat{\A_e}} \ =\ \frac{1}{N}\ \sum_{0 \le x,u < N} \OrientedLocalState{0}{x}{x}{u}{u},\quad\ \ \bra{\widehat{\A_m}} \ =\ \frac{1}{N\sqrt{N}}\ \sum_{0 \le a,x,u < N}  \OrientedLocalState{a}{x}{x\otimes a}{u}{u \otimes (N-a)}.\]

The ``equilibrium'' between electric and magnetic condensates folloiwng (\ref{pair_compete}) is thus given by

\begin{equation} \label{ZN_em}
  \bra{\Omega_\text{critical}} \propto (\sqrt{N}+1) \sum_{0 \le x,u < N} \OrientedLocalState{0}{x}{x}{u}{u} + \sum_{1 \le a < n} \sum_{0 \le x,u < n} \OrientedLocalState{a}{x}{x \otimes a}{u}{u \otimes (n-a)}.
\end{equation}

The transfer matrix following from this boundary condition is exactly the critical point of the standard $1+1$-dimensional $N$-state Potts model. 

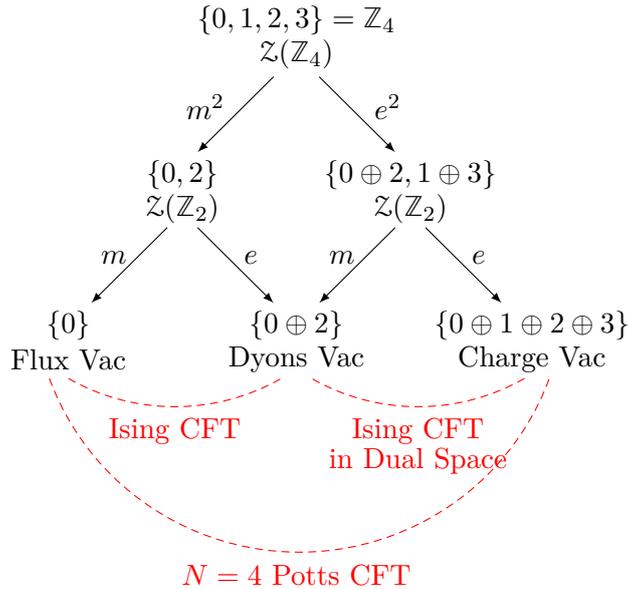
\begin{figure}
\begin{center}
\begin{tikzpicture}
\node at (0, 0.75) {$\{0,1,2,3\}=\Z_4$};
\node at (0, 0.3) {$\Cent(\Z_4)$};
\draw [->] (-.3, 0.) -- (-1.3, -1);
\draw [->] (.3, 0.) -- (1.3, -1);
\node at (-1.2, -0.4) {$m^2$};
\node at (1.2, -0.4) {$e^2$};

\node at (-1.5, -1.3) {$\{0, 2\}$};
\node at (-1.5, -1.75) {$\Cent(\Z_2)$};
\node at (1.5, -1.3) {$\{0\oplus 2, 1\oplus 3 \} $};
\node at (1.5, -1.75) {$\Cent(\Z_2)$};
\draw [->] (-1.7, -2.) -- (-2.7, -3);
\draw [->] (-1.3, -2.) -- (-.3, -3);
\draw [->] (1.3, -2.) -- (.3, -3);
\draw [->] (1.7, -2.) -- (2.7, -3);
\node at (-2.4, -2.4) {$m$};
\node at (-.6, -2.4) {$e$};
\node at (.6, -2.4) {$m$};
\node at (2.4, -2.4) {$e$};

\node at (-3., -3.3) {$\{0\}$};
\node at (-3., -3.75){Flux Vac};
\node at (0, -3.3) {$\{0 \oplus 2\}$} ;
\node at (0, -3.75) {Dyons Vac};
\node at (3.1, -3.3) {$\{0\oplus 1\oplus 2\oplus 3\}$};
\node at (3.1, -3.75) {Charge Vac} ;

\draw [red, densely dashed] (-3, -4) arc (-120:-60:2.8);
\draw [red, densely dashed] (3, -4) arc (-60:-120:2.8);
\node [red] at (-1.6, -4.7) {Ising CFT};
\node [red] at (1.6, -4.7) {Ising CFT};
\node [red] at (1.6, -5.1) {in Dual Space};
\draw [red, densely dashed] (-3.25, -4) arc (-160:-20:3.5);
\node [red] at (0, -6.6) {$N = 4$ Potts CFT};


\end{tikzpicture}
\end{center}
\caption{Refined ondensation tree of the $\Z_4$ quantum double.}
\label{Z4_condensation}
\end{figure}

At all $N$, the model with couplings defined using (\ref{ZN_em}) lie precisely at the KW self-dual line as explored in detail in \cite{Alcaraz:1980bb}. It is known that for $N \le 4$, the critical point of the Potts model is a second-order critical point that can be described by a CFT. For $N > 4$, however, the critical point becomes first-order and does not correspond to a CFT. This actually aligns with our analysis of necessary condition for first-order phase transitions:
\begin{itemize}
\item For $N = 2$, this is identical to the Ising phase transitions. The condensation diagram is equivalent to the lower half of figure \ref{fig:Ising tree}.  Within the unit cell there are essentially only two degrees of freedom, corresponding to the central edge taking $a= 0,1$. One can also express them in terms of the two Frobenius algebras $\A_e$ and $\A_m$. Therefore, the description of the phase space of the unit cell is 1-dimensional (since the overall normalisation of the linear combination does not matter) and there are exactly two competing gapped phases $\A_{e,m}$ in this space, there is no room for degeneracy. It is therefore necessarily a second order phase transition, as indeed observed. 

\item For $N=3$, the refined condensation tree is similar to the $N=2$ case. The only condensates are given by $\A_e$ and $\A_m$. In principle, the effective degrees of freedom in the unit cell is 3, corresponding to the three possibilities of the central edge $a=0,1,2$. There are only two condensates $\A_e$ and $\A_m$ in the vicinity. In this case, there is in principle a potential first order phase transition. But this did not happen---the equilibrium point (\ref{ZN_em}) that respects the symmetry between $a=1,2$ remains a continuous second order phase transition. This is in fact the critical point of the 3-state Potts model, corresponding to the minimcal CFT at $c= 4/5$. We note that the minimal CFT carries a larger symmetry than $\Z_3$, much like the Ising model carries more symmetries than $\Z_2$. Getting a second order phase transition despite the appearance of some room for first order phase transition is however not a contradiction.

\item For $N = 4$: Applying the same analysis as above, we note that there are four degrees of freedom in the unit cell, corresponding to $a=0,1,2,3$. In the case of $N=4$ however, there are additional condensates, corresponding to the Frobenius algebra $\A = 0 \oplus 2$. i.e. This algebra follows from the $\Z_2$ subgroup of $\Z_4$. This condensate corresponds to a Lagrangian condensate in which one condenses  $e^2 \oplus m^2 \oplus e^2m^2$ (i.e. two units of electric charge and magnetic fluxes, and their bound states). Figure \ref{Z4_condensation} depicts the refined condensation tree.  In a unit cell with 3 independent degrees of freedom in a space with 3 competing phases might leave room a 1 dimensional degeneracy and first order phase transition. But this did not happen. Instead it is known that the critical point for $N=4$ state Potts produces a $c=1$ CFT in the infrared corresponding to a $\Z_2$ orbifold of a $U(1)_8$ theory \cite{Dijkgraaf:1989hb}. 

\item For $N \ge 5$: The effective independent number of degrees of freedom in the unit cell is $5-1=4$, while the number of condensates is 2. 
The number of degrees of freedom thus exceeds the number of possible Frobenius algebras, and this is the first instance in the $N$-state Potts model series in which this equilibrium point (\ref{ZN_em}) produces a first order phase transition. 
For $N \ge 5$, the critical states correspond to a first-order critical point. This is consistent with the fact that the number of degrees of freedom always exceeds the number of Frobenius algebra (the latter is given by the number of subgroups of $\Z_N$). 
\end{itemize}

The $\Z_N$ models showcase the appearance of first order phase transitions. Where the dimension of phase space is large enough for degeneracies, it is necessary to further fine-tune the couplings to land on a continuous phase transition. In the case of $\Z_N$ models, it is well known that there is an extra class of gapless phases corresponding to the $\Z_N$ parafermions that were determined by imposing integrability \cite{Alcaraz:1986hs,Alcaraz:1980bb,Fateev:1985mm}, and later through the notion of ``discrete-holomorphicity''\cite{Bernard:1991za,Rajabpour:2007er,Ikhlef:2015sxa,Fendley:2020fxv}. It is not clear what discrete-holomorphicity mean from the perspecive of the anyon condensates. (See however \cite{Fendley:2020fxv} for a connection to data in a braided tensor category in some lattice models expressible in a specific form. Although how this notion is applied in more general models is unclear.) It is however clear that one has to introduce non-trivial relative weights between $a \in \Z_N$ and that {\it cannot} be explained within $\Z_N$. We note that the corresponding CFT fixed point carries a lot more emergent symmetries than $\Z_N$. ($\Z_N$ itself is not a modular tensor category, and it is not expected to capture the full symmetries of the CFT in any event. In particular for rational theories, one expects the full symmetries to be captured by a modular tensor category \cite{Moore:1989vd}.) Therefore these weights could likely admit natural interpretations in a larger category that actually captures the full symmetries of the purported gapless phases. 

It is known that the first order phase transition of the 5-state Potts model is a {\it weak first order} transition, and that it is in close proximity to a complex CFT \cite{Kaplan:2010zz} \footnote{We thank Yunfeng Jiang for pointing this out to us.}. 
Implementing the symmetric RG algorithm given in the appendix \ref{app:symRG} to the $\Z_5$ quantum doubles, the computed central charge fluctuates around $c\approx1.1$, which is very close to the real part of the known complex central charge \cite{Gorbenko:2018ncu, Gorbenko:2018dtm}. 
We are thus led to conjecture that {\bf a fluctuating central charge in the vicinity of the critical point is in fact a manifestation of the weak first order phase transitions in the proximity of a complex CFT}. Such fluctuations of the central charge is visible along the first-order phase boundary, as one will see in other examples later.

In the $\Z_N$ model, we can already see that Eq. \ref{eq:2ndorder} is a sufficient but not necessary condition for second order phase transitions. Consider $N=3$, by our counting, the commensurate state of $\Z_3$ could potentially give a first-order phase transition, but it is in fact a second order transition corresponding to the 3-state Potts model. Our condition for second order transition is thus overly strong, and could potentially leave out many CFTs. This also happens in later examples where we consider the Haagerup models, in which novel continuous transitions are confirmed numerically even though we have not cut down the phase space sufficiently. We believe our condition can be relaxed in the future. 

Let us analyse more cases below. 

\subsubsection{Only Two Frobenius Algebras Sharing a Perfect Module \eqs{M} -- Second Order Phase Transition Point -- with the A-series Models as Illustrations }


When only two Frboenius algebras $\mathcal{A}_{i,j}$ share a chosen module $M$, we expect a second order phase transition when there is only one independent parameter left in the interpolation (\ref{pairofcondensates}). In such case, the interpolation remains within the attraction basin of either the $\mathcal{A}_i$ or $\mathcal{A}_j$ fixed points as the ratio $r_2/r_1$ varies, leaving no room for degeneracy. For instance, in the refined condensation tree of the Ising model, there are only two distinct modular actions, trivially satisfying this condition. Indeed, the critical state (\ref{isingomeg}), derived from the competition of condensates (\ref{pair_compete}), corresponds exactly to the continuous phase transition of the critical Ising spin model.
This extends to the entire $A_{k+1}$ series of minimal lattice models revisited in the previous section. In these cases, the chosen module $M=1$ is perfectly shared between $\mathcal{A}_0= 0$ and $\mathcal{A}_1 = 0\oplus 2$ for all $k$. Therefore, we have $n_M =2$, and $n_M > D_c= 1$. The phase transtion point (\ref{Akcrit}) is second order for all $k$.

\subsubsection{Exactly Three Algebra Sharing a perfect Module \eqs{M} -- Phase Boundaries Precisely Predicted -- with \eqs{A_5} as an Illustration}

Consider a module shared by exactly three Frobenius algebras $\mathcal{A}_{i,j,k}$. 
Then we can consider interpolation of three algebra. Therefore, there are two independent couplings in (\ref{many_condensates}). 
First, we note that these three algebras describe three different Lagrangian condensates.  It is bound to be the case that for any pair of Lagrangian condensates, some condensed anyons in a condensate appear as confined anyons in another. They cannot condense at the same time. 
One can thus first consider critical points corresponding to equilibrium states of any two of the three Frobenius algebra using (\ref{pair_compete}).
Each of these three pair-competition produces a second order phase transition. This is because we have a two dimensional phase space. One direction is fixed by balancing two of the three competing condensates, and there is one other direction corresponding to turning on the third condensate. Since the third cannot condense along with the competing pair, dialing the strength of the third condensate is not a degenerate direction.
This suggests that not only is the pair-wise balanced point a continuous transition, one can deduce that by dailing the strength of the third condensate while preserving the relative strengths of the balanced condensates, it traces out a second-order phase boundary, at least in the vicinity of the region where the third condensate is weak. 

\subsubsection*{Three competing algebras in $A_{5}$}

In $A_5$, the collection of objects is $\{0,1,2,3,4\}$.  We consider the common module $M = 2$.
This choice corresponds to the 8-vertex model \cite{Aasen:2020jwb} \footnote{It was pointed out to us by Yuan Miao that the model realised here the parameter space spans the 6-vertex model rather than the full phase-space of the 8-vertex model, despite the terminology adopted in \cite{Aasen:2020jwb} which we followed here.}. We note that $2\otimes 2 = 0\oplus 2 \oplus 4$.  Therefore the unit cell is three dimensional : $\langle {\unitcellColorMod{i}{2}}|, \, i \in \{0,2,4\}$. There are 4 algebras containing $\{0,2,4\}$. From figure \ref{fig:A5} it is evident that all the Frobenius algebras contain basis elements from the set $\{0,2,4\}$. It would appear at first sight that all the phases could appear under RG as we change the relative weights of the basis states. However, one can check that only three algebras share the same (right) module $M = 2$. 
The details of these algebras and their action on $M=2$ are given below. 
\begin{itemize}
\item $\A_0 = 0$, such that
$$d_{A_0} = 1,\qquad f^{\A_0}_{000} = 1,\qquad [\rho_{M_{\A_0}}]^0_{22} = 1.$$

\item $\A_2 = 0\oplus 4$, such that
$$d_{\A_2} = 2,\qquad f^{\A_2}_{000} = f^{\A_2}_{044} = 1,\qquad [\rho_{M_{\A_2}}]^0_{22} = [\rho_{M_{\A_2}}]^4_{22} = 1.$$

\item $\A_3 = 0\oplus 2\oplus 4$, such that
$$d_{\A_3} = 4,\qquad f^{\A_3}_{000} = f^{\A_3}_{022} = f^{\A_3}_{044} = f^{\A_3}_{224} = 1,\qquad f^{\A_3}_{222} = 0,$$
$$[\rho_{M_{\A_3}}]^0_{22} = 1,\qquad [\rho_{M_{\A_3}}]^2_{22} = \sqrt[4]{2},\qquad [\rho_{M_{\A_3}}]^{4}_{22} = 1$$

\end{itemize}
Indeed one can see that $\A_0\subset \A_2 \subset \A_3$. As mentioned previousely, this structure is reflected in the condensation tree in figure \ref{fig:A5}. When one algebra is a non-trivial sub-algebra of the other, they should share a common parent node corresponding to some non-trivial sub-category $\mathcal{K} \subset {\Fus}$, which is confirmed here. 

Importantly, in this case $n_M =3$ and $D_c = N_{\A_3} -1 = 2$, thus satisfying $n_M>D_c$. The phase space is two dimensional with 3 gapped phases. The phase boundaries between any pair of condensates should all be second order all the way, including the tri-critical point. This expectation is confirmed perfectly in this model, as we will discuss in further detail below. 

From the discussion above, these three algebras should form three separate RG attraction basins, while the other Frobenius algebras are excluded from the phase diagram. These expectations will be confirmed in our numerical checks below. 

To parametrise the interpolation of the three algebras in a unit cell of the boundary state, we use the following orthonormal basis states

\begin{equation} \label{eq:A5 para}
    \langle (x,y,z)| \equiv x \langle{\unitcellColorMod{0}{2}}|+y\langle {\unitcellColorMod{2}{2}}|+z \langle{\unitcellColorMod{4}{2}}|.
\end{equation}

After computing the normalisation factors, one can read off from the above data
\begin{align}
    \langle \widehat{\A}_0|&=\langle (1,0,0)|, \nonumber \\
    \langle \widehat{\A}_2|&=\langle (\frac{1}{\sqrt 2},0,\frac{1}{\sqrt 2})| , \\
    \langle \widehat{\A}_3|&=\langle (\frac{1}{2},\frac{1}{\sqrt 2},\frac{1}{2})|, \nonumber
\end{align}
where we have taken the shorthand $\langle \hat{\A}_i| \equiv \langle \hat{\A}_i|_{M=2}$. This shorthand will be used whenever the module in the subscript is unambiguous.

We first confirm numerically that the three pair-wise equilibrium states
\begin{equation}\label{A5-pairwise}
    \langle C_{ij}|=\frac{1}{2}(\langle \widehat{\A}_{i}|+\langle \widehat{\A}_{j}|)
\end{equation}
all correspond to second-order phase transition points. This and all subsequent numerics follow the categorical-symmetry-preserving RG illustrated in appendix \ref{app:symRG}, which improves and generalises the algorithm brought up in \cite{Chen:2022wvy}.

We want to further track the phase boundary line between two phases $\A_i$ and $\A_j$, or at least in the vicinity of the critical points (\ref{A5-pairwise}). We certainly would like to move closer to the other phase $\A_k$ while preserving the balance between $\A_i$ and $\A_j$. The most natural guess of the phase boundary is the following set of states parametrised by $p$
\begin{subequations}\label{eq: crit line}
\begin{align}
\langle C_{ij}(p)|=\langle C_{ij}|+p\left(\langle \widehat{\A}_k|- \frac{\langle \widehat{\A}_k|B_{ij}\rangle} {\langle B_{ij}|B_{ij}\rangle} \langle B_{ij}|\right )
\end{align}
where
\begin{align}
    \langle B_{ij}|=\langle \widehat{\A}_i|-\langle \widehat{\A}_j|
\end{align}
\end{subequations}
such that 
\begin{equation}
\label{eq: crit line3}
    \langle \widehat{\A}_i| C_{ij}\rangle_p=\langle \widehat{\A}_j| C_{ij}\rangle_p.
\end{equation}

These three lines intersect at a single point (up to an overall factor)

\begin{equation}
    \langle C_{ijk}|=\sum_{abc={ijk,jki,kij}} \langle B_{ab}|B_{ac}\rangle\langle B_{bc}|B_{bc}\rangle\langle \A_a|.
\end{equation}
As we are going to see this point appears in fact to be very close to the actual tri-critical point in examples we have checked. We believe it is a good estimate of the location of the tri-critical point. Only in special cases where there are other features (such as generalised KW duality) could we refine the location of the tri-critical point. This would be the case in $A_5$ discussed below. 

The inner product is defined analogously to figure \ref{fig:vertex_normalization}. For the current case it reads, 
\begin{equation} \label{A0A1A2}
\langle \hat \A_3| \hat \A_2\rangle = \frac{1}{\sqrt{2}} = \langle \hat \A_0| \hat \A_2\rangle =  \sqrt{2}\langle \hat \A_3| \hat \A_0\rangle,
\end{equation}
which is a  result of
 \begin{equation}
\langle \hat \A_i | \hat \A_j \rangle =  \sqrt{\frac{d_{\A_i}}{d_{
A_j}}} \,\,\textrm{for}\,\, \A_i \subset \A_j. 
\end{equation}


We can now explicitly solve (\ref{eq: crit line}) for the current case. Since an overall factor of $\langle \Omega|$ is immaterial, we rescale the coefficient $x$ in front of $\langle \hat\A_0| $ to unity and replace the parameter $p$ by $y$ for clarity. The solution reads 
\begin{align}\label{eq:crit line A5}
    \langle C_{02}(y)|&=\langle 1,y,\sqrt{2}-1| ,\nonumber\\
    \langle C_{23}(y)|&=\langle 1,y,(2+\sqrt{2})y-1|,\nonumber \\
    \langle C_{03}(y)|&=\langle 1,y,-\sqrt{2}y+1|. 
\end{align}

\subsubsection*{The three-phase diagram and CFTs of $A_5$}
\begin{figure}
    \centering
    \includegraphics[width=0.7\linewidth]{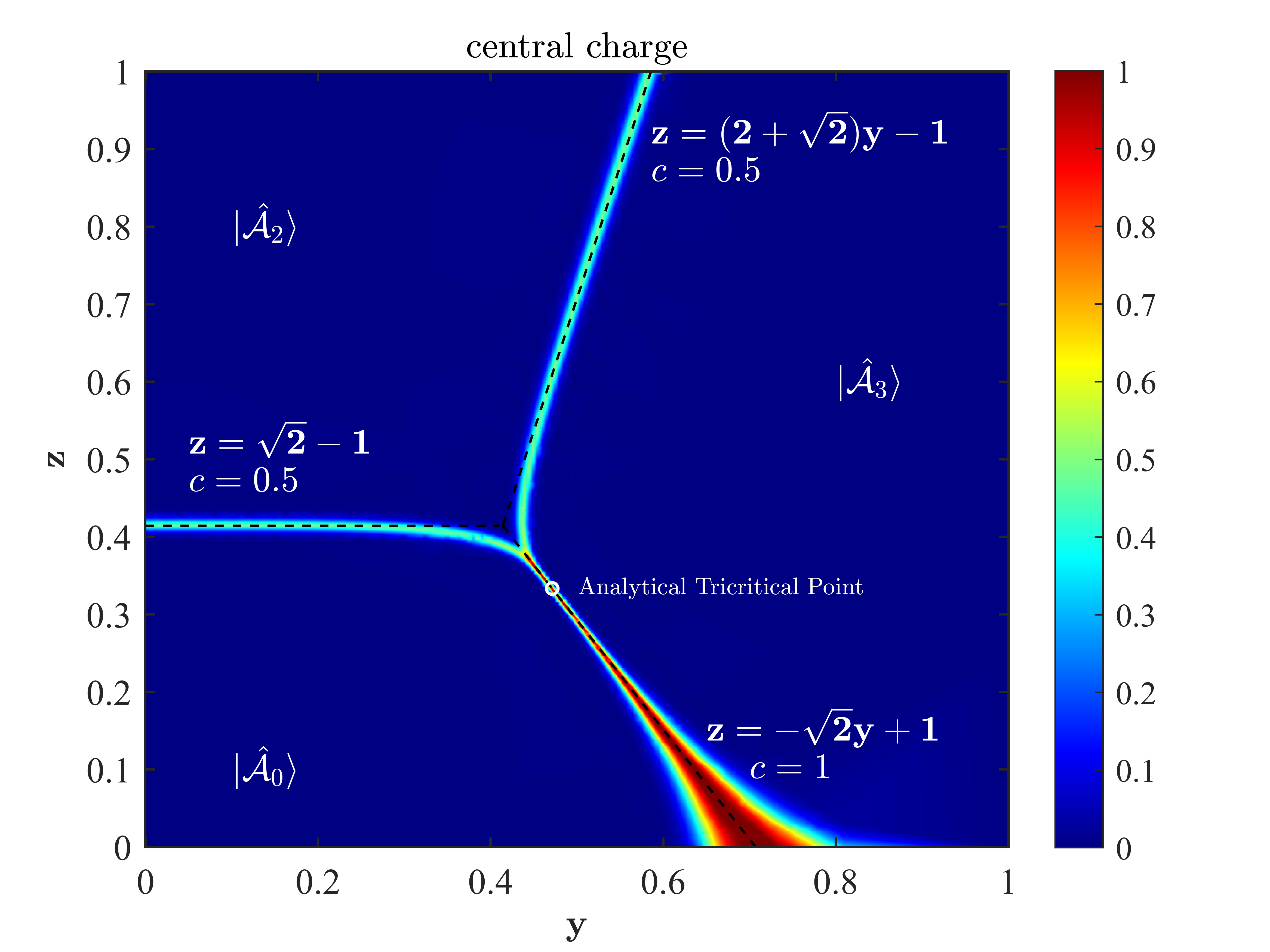}
    \caption{Phase diagram of competing Frobenius algebras $\A_{0,2,3}$ with the slanted edge in the unit cell colored by the common module $M=2$ in $A_5$. Numerically determined central charge corresponding to the strange correlator from the state $\langle {1,y,z|}$ as in (\ref{eq:A5 para}) are indicated. Each phase is marked by one of the nomalised states $\A_i$. The three phases are seperated by critical lines with central charge $c>0$. Dotted lines are the predicted critical states given explicitly by (\ref{eq:crit line A5}), which follow from (\ref{eq: crit line}). Numerical results show that the phase boundaries are predominantly a straight line in this parametrisation of the phase space, with its location aligning closely with the theoretical prediction (\ref{eq:crit line A5}). Phase boundaries with smaller central charges bend towards the phase boundary with larger central charges. Thickness of phase boundaries reflects the relative stability under perturbations. The tricritical point determined by mapping the model to the analytical result of the Ashkin-Teller model is marked in the diagram. It appears to be slightly away from the point where the phase boundaries meet.
    This mismatch is a numerical artifact, also observed in direct simulations of the Ashkin-Teller model (see Appendix \ref{sec:AT}, Fig.\ref{fig:AT_phase_diagram}). Near the tricritical point, the two Ising lines approach each other very closely and become nearly indistinguishable within the resolution of the numerical RG or tensor network methods we initially used to produce this picture. As a result, the numerically observed bifurcation (splitting) point appears slightly away from the true tricritical point.}
    \label{fig:A5 mod 2}
\end{figure}

The whole phase diagram is plotted in figure \ref{fig:A5 mod 2} in the orthonormal basis state. The above conjectured boundaries $|C_{02}\rangle_y$, $|C_{23}\rangle_y$ and $|C_{03}\rangle_y$ are plotted as dotted lines. They match well with the numerical second order phase boundaries all the way until a small region near the tri-critical point. $|C_{02}\rangle_y$ and $|C_{03}\rangle_y$ are symmetric about $|C_{23}\rangle_y$, using (\ref{A0A1A2}). The three actual second order phase boundaries meet at a point that is very close to the actual tri-critical point determined numerically.

Remarkably, this phase diagram is formally identical to that of the two-dimensional isotropic Ashkin–Teller model, in which two critical Ising lines and one $c=1$ critical line meet at the self-dual tri-critical point\cite{kohmotoHamiltonianStudies21981,marizZ4ModelCriticality1985,shen2025exploring}. By mapping the phase diagram of Ashkin–Teller model to the $A_5$ model, we found that the tri-critical point is located at $\bra{C_{03}}$, the equilibrium state between $\A_0$ and $\A_3$. 

\begin{figure}
    \centering
    \includegraphics[width=0.7\linewidth]{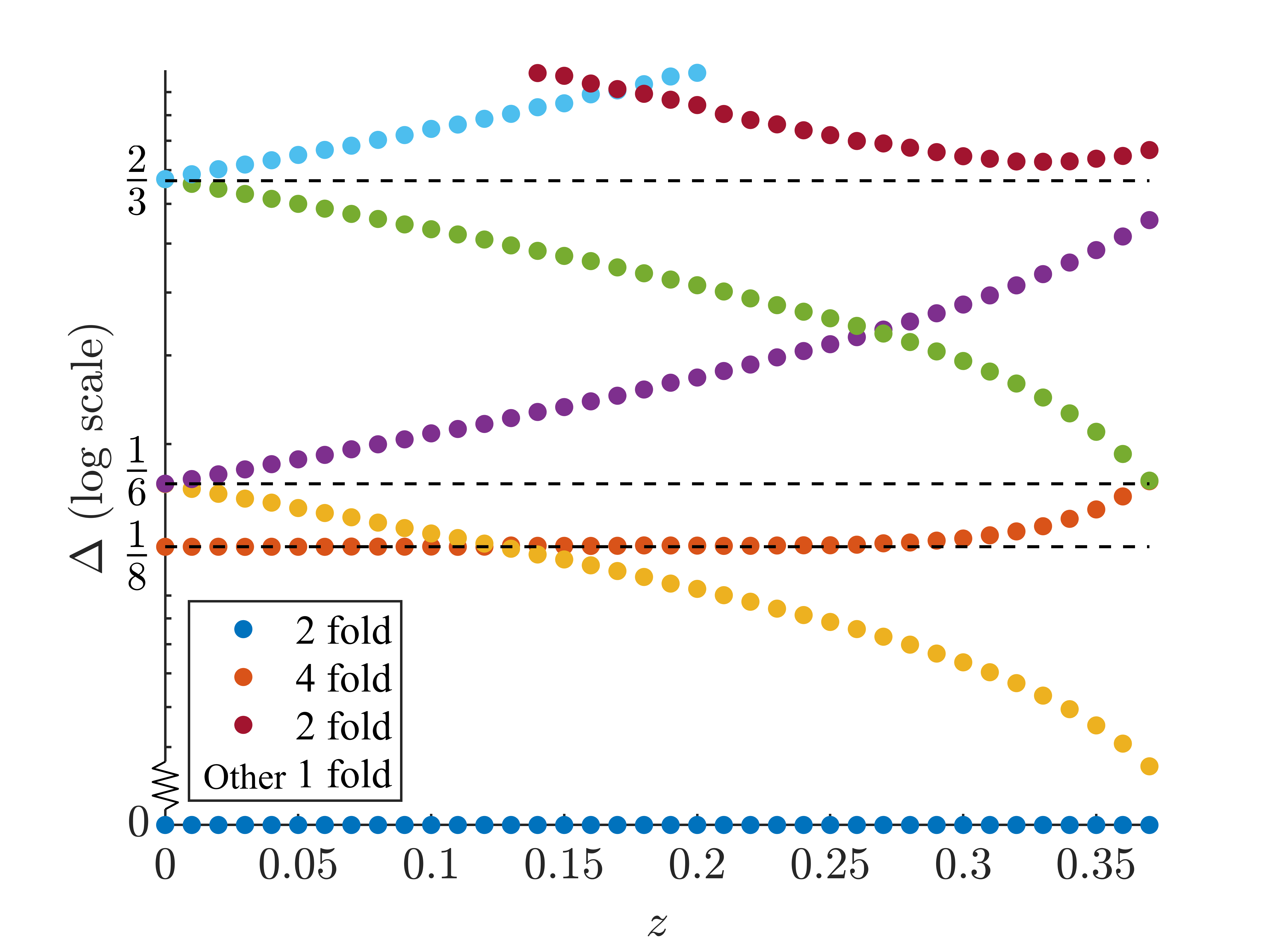}
    \caption{Low lying scalling dimensions $\Delta$ along the $c=1$ CFT between $\A_0$ and $\A_3$ in figure \ref{fig:A5 mod 2}. Ground states with $\Delta=0$ are 2-fold. Orange primaries that start at $\Delta=\frac{1}{8}$ are 4-fold. Yellow and purple primaries start being 2-fold and then split. Product of their scalling dimensions however remain unchanged, which can be easily seen in the log scale of $\Delta$. So do the green and light blue primaries.}
    \label{fig:A5 spec}
\end{figure}

We numerically calculate the spectrum of these CFTs. The two $c=0.5$ critical curves are both three copies of the critical Ising CFT. The $c=1$ critical line has varying spectrum along the line, which is 2 copies of Ising orbifolds. The low-lying spectrum of the $c=1$ line in figure \ref{fig:A5 mod 2} is plotted in figure \ref{fig:A5 spec}. 

\begin{figure}
    \centering
    \includegraphics[width=0.7\linewidth]{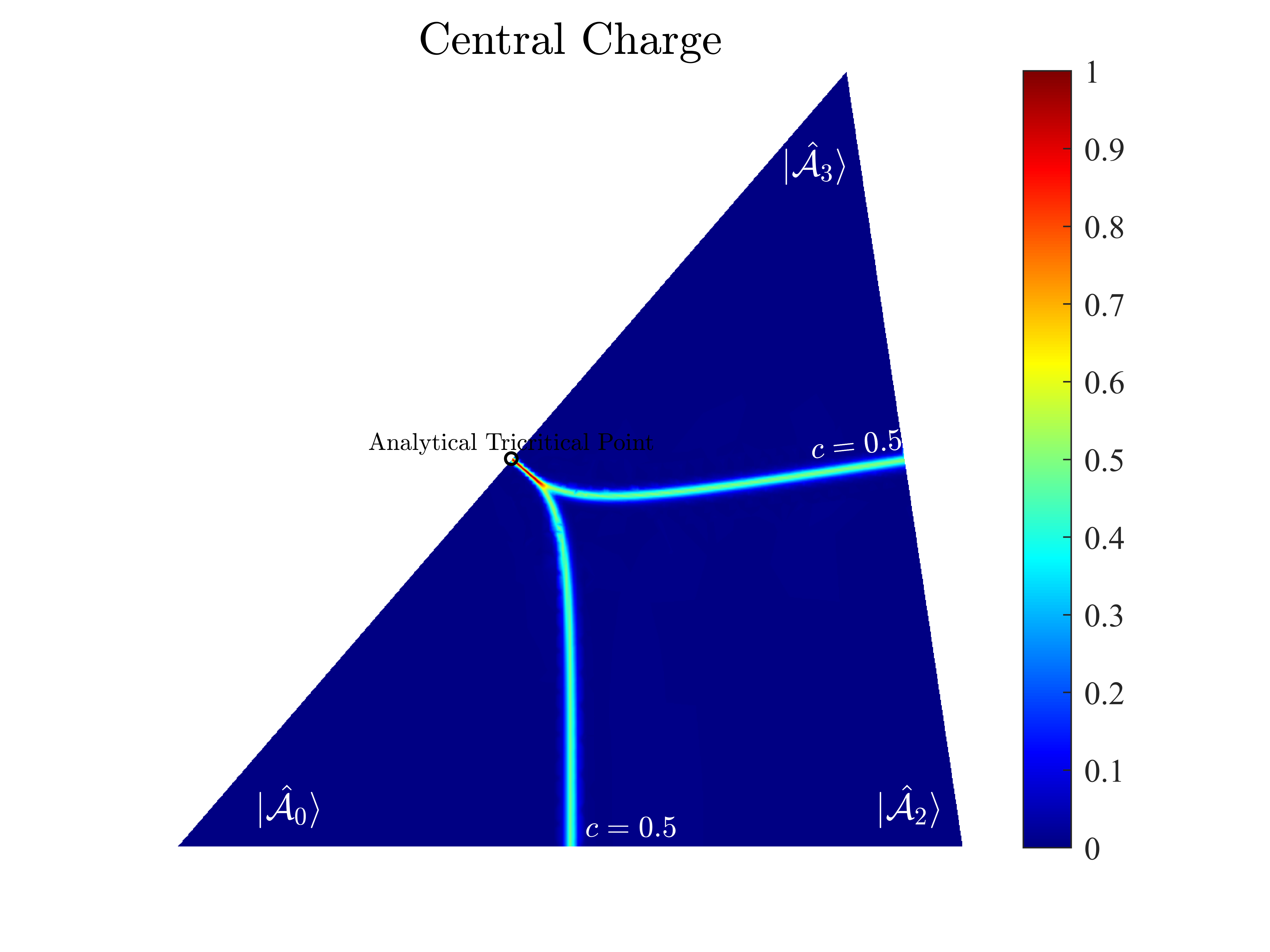}
    \caption{Ternary phase diagram of Frobenius algebras $\A_{0,2,3}$ with module $M=2$ in category $A_5$. Each vertex marks one of the three nomalised states as labelled in this figure, and any point within the triangle represents a weighted sum of these states. The three phases are seperated by the critical lines with central charge $c>0$. These critical lines coincide well with the three perpendicular bisector of the triangle all the way up to a small region close to the tricritical point. The analytically known tricritical point of the AT model is marked here, which {\it coincides} with the point obtained from pair-competition between $\A_0$ and $\A_3$ i.e. $\langle C_{03}(\alpha =0))|$ defined in (\ref{eq:A5_critical_bdry}). That this is a tri-critical point is being argued using a generalised KW duality, as discussed in section \ref{sec:KW}.  }
    \label{fig:A5 mod 2 tri}
\end{figure}

We cut out a triangular part of figure \ref{fig:A5 mod 2} and map it to the vector space spanned by $\left ( \langle \hat \A_2|-\langle \hat \A_0| \right )$ and $\left (\langle \hat \A_3|-\langle \hat \A_0|\right )$ as in figure \ref{fig:A5 mod 2 tri}. The three vertexes correspond to the RG fixed point condensate states ${\langle\widehat{\A}_{0,2,3}|}$. The three gapless phase boundaries are the perpendicular bisectors of this triangle (with inner product between vectors defined as above). This matches exactly with (\ref{eq: crit line3}). Hereafter, we plot all three-phase competetions in a ternary phase diagram. As one will see, \textbf{the phase boundaries match (\ref{eq: crit line}) all the way until a small region close to the tri-critical point.}

The tri-critical point is very close to the circumcenter of the triangle, which is the intersection of the three perpendiculars. The two critical curves preserving fewer symmetries (i.e. protected by lower nodes in the condensation tree), namely the two $c=0.5$ Ising CFT, bend towards the $c=1$ critical line which preserve more symmetries. We will witness similar effects in other examples too.


Let us try to understand the physics of these phase boundaries in light of the condensation tree. We rewrite (\ref{eq:crit line A5}) as the interpolation between $\langle\hat\A_i|$. The three phase boundaries reduce to the following: 
\begin{align}
\langle C_{02} \, (\alpha)| &\equiv \left(\langle \hat\A_0 | + \langle \hat\A_2|\right) + \alpha \left(\langle \hat\A_3| -    \langle \hat \A_3 | \hat\A_2\rangle\langle \hat\A_2| \right) \nonumber \\
\langle C_{23} \, (\alpha)| &\equiv \left(\langle \hat\A_3 | + \langle \hat\A_2|\right) + \alpha \left(\langle \hat\A_0| -    \langle \hat \A_0 | \hat\A_2\rangle\langle \hat\A_2| \right), \nonumber \\
\langle C_{03}\, (\alpha)| &\equiv \left(\langle \hat\A_3 | + \langle \hat\A_0|\right) + \alpha \langle \hat\A_2|,
\label{eq:A5_critical_bdry}
\end{align} 
where we have bundled the interpolation parameter defined previousely to a new parameter $\alpha$ to avoid clutter. As $\alpha$ is varied one moves along each of the phase boundaries.

Expressed in this form, its physical meaning becomes manifest. Without loss of generality, consider the phase boundary between $\A_0$ and $\A_2$.
As one observes the condensation tree, it is evident that $\A_0$ and $\A_2$ are competing at the lowest level, where without further fine-tuning, we expect that $(F,0)$ has already condensed. The competition between them is essentially between $(D,0)$ and $(F,4)$, which corresponds to the competition between magnetic and electric condensate in the common parent $\Cent({\Z}_2)$ node. 

Similarly $\A_2$ and $\A_3$ are also competing at the lowest level, where $(F,4)$ has condensed and the competition is between $(F,0)$ and $(D,2)$.  $\A_0$ and $\A_3$ however are competing at a higher level, where one selects between $(F,0)$ and $(F,4)$. 

Nevertheless, these two dyons can in fact condense at the same time to give $\A_2$. This suggests that increasing the weight of $\A_3$ could disrupt the balance between $\A_2$ and $\A_0$, because $\A_3$ is supporting the condensation of $(F,4)$ and so favors $\A_2$ over $\A_0$. 

Therefore we expect that one has to subtract weights on $\A_2$ as we move towards $\A_3$ along the equilibrium point between $\A_0$ and $\A_2$. This is indeed played out exactly in (\ref{eq: crit line}) automatically, which becomes manifest when expressed in the form of (\ref{eq:A5_critical_bdry}).

We note that $\A_0$ and $\A_3$ are completely symmetric in this phase diagram. Therefore the phase boundary between $\A_3$ and $\A_2$ can be explained in exactly the same manner. 

Finally, we consider moving along the phase boundary between $\A_0$ and $\A_3$. We should move simply along $\A_2$ because $\A_2$ adds weights to $\A_0$ and $\A_3$ in a completely symmetric way, as is evident from equation (\ref{A0A1A2}). Changing the weights of  $\A_2$ does not affect the equilibrium between $\A_0$ and $\A_3$, which is clearly realised in the third equation in (\ref{eq:A5_critical_bdry}).


\subsection{Symmetries Preserved at Continuous Phase Transitions}
In this subsection, we would like to read-off the minimal set of symmetries expected to be preserved at continuous phase transitions between competing Frobenius algebra. 

It is evident from a refined condensation tree that there is a notion of distance between Frobenius algebras. Down the tree, nodes denote symTO with more symmetries spontaneously broken by anyon condensation. The minimal amount of symmetry preserved in the critical state obtained from the competing pair of Frobenius algebras---assuming that we managed to choose appropriate modules $M$ to sieve out first order phase transitions -- {\bf is determined by the first common ancestor node of the pair. } The topological excitations carried by the symTO of the ancestor node are the minimal set of symmetries that are preserved by the gapless phase. However, the resulting CFT may exhibit emergent symmetries beyond those of the ancestor node, but it cannot have fewer.

Let us revisit the examples we have considered to illustrate these ideas. 
\subsubsection*{Critical theory from the doubled Ising symTO}
   In the case of the Ising spin model (or the $A_3$ model), the first common ancestor of $\A_0$ and $\A_2$ is the node corresponding to $\Cent({\Z}_2)$, as shown in figure \ref{fig:Ising tree}. Therefore the minimal set of topological symmetries is characterised by the toric code order, which has 3 non-trivial topological excitations corresponding to a ${\Z}_2$ electric charge, a magnetic flux and their bound state. The actual symmetry of the Ising CFT is enlarged to the full Ising category, describable by the top node. 

  \subsubsection*{Critical theories from the $A_5$ symTO} 
  \begin{enumerate}
      \item The first common ancestor of  $\A_0$ and $\A_2$ is $Z({\Z}_2)$. Their competition is identical to the case of the Ising spin model and we expect the critical point to be the Ising CFT.  Similarly, the competition of $\A_2$ and $\A_3$ is also expected to be Ising. 

      \item The first common ancestor of $\A_2$ and $\A_3$ is $\Cent({\Z}_3)$. As we discussed above, the electric/magnetic competition in $\Cent({\Z}_3)$ in a unit cell produces the 3-state Potts model. This is indeed inherited in the competition between $\A_2$ and $\A_3$. 

    \item The competition between $\A_0$ and $\A_3$ should produce a CFT that sees at least symmetries in $\Cent(S_3)$ which is their common ancestor. This is indeed confirmed. The critical point constructed by their equilibirum point is given by a free boson CFT with $c=1$.

    \item The competition between $\A_0$ and $\A_4$ should also produce a CFT that preserves at least the symmetry imposed by $\Cent(S_3)$. In fact, this CFT is the minimal model in the $A$ series at $k=4$, with $c=4/5$. Its complete set of topological symmetries are in fact characterised by $\Cent(A_5)$. 
  \end{enumerate}

These examples completely confirm our expectations by means of our refined condensation trees. 

We will discuss more examples in later sections as further illustrations. 

\subsection{Generalised {KW} Duality}\label{sec:KW}

The generalised KW duality can even be applied to compute tri-critical points, via a double-layer construction. We consider interpolating among $\A_{0,2,3}$ in $A_5$ with $M=2$ as an example. The unitcell is three dimensional with basis $| {\unitcellColorMod{i}{2}}\rangle, \, i \in \{0,2,4\}$. The KW duality is performed by doing an F-move on every octagon edge. This operation is represented by:
\begin{align*}
| {\unitcellColorModHoz{0}{2}}\rangle &\xrightarrow{F} \frac{1}{2}\left( | {\unitcellColorMod{0}{2}}\rangle + \sqrt{2}| {\unitcellColorMod{2}{2}}\rangle + | {\unitcellColorMod{4}{2}}\rangle \right), \\
| {\unitcellColorModHoz{2}{2}}\rangle &\xrightarrow{F} \frac{1}{\sqrt{2}}\left( | {\unitcellColorMod{0}{2}}\rangle - | {\unitcellColorMod{4}{2}}\rangle \right), \\
| {\unitcellColorModHoz{4}{2}}\rangle &\xrightarrow{F} \frac{1}{2}\left( | {\unitcellColorMod{0}{2}}\rangle - \sqrt{2}| {\unitcellColorMod{2}{2}}\rangle + |{\unitcellColorMod{4}{2}}\rangle \right) .
\end{align*}
As explained in figure \ref{fig:Ak_KW}, under an F-move the vertical edges are mapped to horizontal edges. After doing F-move we have implicitly shifted the lattice by a unit cell (or equivalently, rotated the lattice by $90$ degrees), shifting our focus on the vertical edges shown on the right-hand side.

In what follows, we denote the basis $| {\unitcellColorMod{i}{2}}\rangle$ by $\ket i$. Under the normalization where $\ket 0$ has a coefficient of $1$, the parameter space is two-dimensional, with a generic point represented as $\ket{0} + y\ket{2} + z\ket{4}$. The KW duality acts as a transformation on this parameter space, mapping it onto itself. The KW self-dual points correspond to the fixed points of this transformation. In this case, the self-dual points lie along a line in the parameter space, given by the equation:
\begin{equation}
    z = -\sqrt{2} y + 1.
\end{equation}

The phase boundary between $\mathcal A_0$ and $\mathcal A_2$ is completely contained in the KW self-dual line and ends at the tricritical point as in figure \ref{fig:A5 mod 2}.
The $\mathcal A_0$ phase is centered at the point $(y=0,z=0)$ and the $\mathcal A_2$ phase is centered at the point $(y=\sqrt{2}, z=1)$. The interpolating line connecting the two points intersects the KW self-dual line at the point $(y=\frac{\sqrt{2}}{3}, z=\frac{1}{3})$. This point is exactly the critical point constructed by the equilibrium state of $\mathcal A_0$ and $\mathcal A_2$.

Next we show how the $A_5$ model is related to two copies Ising models and derive the critical points and the tricritical point in the phase diagram. 
The simple object $2$ in $A_5$ is the two dimensional irrep of $\textrm{Rep}(S_3)$. This fact motivates us to split the \(2\)-dimensional internal Hilbert space of simple object $2$ into direct sum of two internal degrees of freedom $2_1$ and $2_2$ with dimension equal to $1$. They satisfy the fractionalized fusion rules\cite{zhao2024,zhao2025}:
\begin{align}\label{eq:frac}
\qquad 2 = 2_1 \oplus 2_2, \qquad 2_1\otimes 2_1 = 2_2\otimes 2_2= \frac{1}{2}(0 + 2_2),\nonumber\\ 2_1\otimes 2_2 = 2_2\otimes 2_1 = \frac{1}{2} (2_1 \oplus 4),\qquad 4\otimes 2_1=2_2,\qquad 4\otimes& 2_2=2_1.
\end{align}
The fractionalized fusion coefficient \(N_{a_ib_j}^{c_k}\) (for anyon types $0$ and $4$ the second index is trivial) denotes the probability of obtaining the \emph{dyon} \(c_k\) when fusing dyons \(a_i\) and \(b_j\). They are exactly how the fusions of anyons are displayed in our finer structure---the HGW string-net model---where internal degrees of freedom of anyons are explicitly taken into account. Summing over the internal indices \(i,j\) for fixed anyon types reproduces the integer-valued fusion rules of the underlying anyons. Now that each edge degree of freedom in the string-net model corresponds one-to-one with a pure chargeon of the \(D(S_3)\) phase, it follows that only when three such edge labels—each including its internal index—can meet at a vertex precisely when their fractionalized fusion coefficient is nonzero. These fractionalized fusion rules are computable from the Clebsch–Gordan coefficients of the group \(S_3 \;=\;\bigl\langle r,\,s \mid r^3 = s^2 = (rs)^2 = e\bigr\rangle\). Notably, the fusion coefficients are not gauge invariant and depend on the basis we choose for dyon $2$. Here, we choose the real two-dimensional representation:
\[
D^2(r) = \frac{1}{2}
\begin{pmatrix}
-1 & -\sqrt{3} \\
\sqrt{3} & -1
\end{pmatrix},
\qquad
D^2(s) =
\begin{pmatrix}
1 & 0 \\
0 & -1
\end{pmatrix}.
\]
Then internal degrees of freedom $2_1$ and $2_2$ exactly labels the two dimension-indices of representation $2$ of $S_3$ group. The fractionalized fusion coefficients \eqref{eq:frac} are then given by the squared modulus of the corresponding Clebsch–Gordan coefficients:
\[
N_{2_i2_j}^{\,a}
=\left|\middle[\begin{array}{cc|c}
2 & 2 & a \\
i & j & 1
\end{array}\middle]\right|^2,\qquad N_{2_i2_j}^{\,2_k}
=\left|\middle[\begin{array}{cc|c}
2 & 2 & 2 \\
i & j & k
\end{array}\middle]\right|^2,\qquad N_{2_i a}^{\,2_j}
=\left|\middle[\begin{array}{cc|c}
2 & a & 2 \\
i & 1 & j
\end{array}\middle]\right|^2,
\]
where \(a\in\{0,4\}\).

We further assume that the internal degree of freedom corresponds to the state $\ket{2_1}, \ \ket{2_2}$ in the local Hilbert space and satisfy
\begin{equation*}
    \frac{1}{\sqrt{2}}\left(\left|2_1\right\rangle+\left|2_2\right\rangle\right)=|2\rangle, \quad \frac{1}{\sqrt{2}}\left(\left|2_1\right\rangle-\left|2_2\right\rangle\right)=\frac{1}{\sqrt{2}}(|0\rangle+|4\rangle).
\end{equation*}
Under this assumption we show that we can decompose the system into two Ising models and calculate precisely the critical points and the tricritical point.

For the octagon edge we temporarily use a new parametrization ($a$ and $b$) and later we will show how to map to the previous one ($y$ and $z$). The state on the octagon edge is set to be
$\left(\ket{0}+a\ket{2_2}\right)\left(\ket{0}+b\ket{4}\right) = \ket{0}+b\ket{4}+a\ket{2_2}+ab\ket{2_1}$.

\begin{figure}
    \centering
    \includegraphics[width=\linewidth]{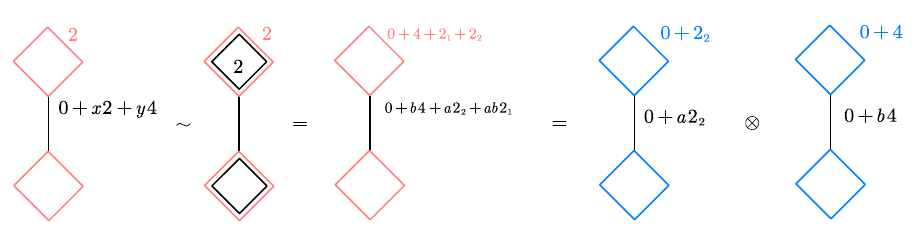}
    \caption{Double layer of Ising models.}
    \label{fig:double_ising_layer}
\end{figure}

In this way we have rewritten the system into two layers of Ising models, see figure \ref{fig:double_ising_layer}. 
We already know that the Ising model reaches the critical point when the octagon edge is tuned to  
$\ket{1_{\text{Ising}}} + (\sqrt{2}-1) \ket{\psi_{\text{Ising}}}$
, by the presciption (\ref{isingomeg}). In the current double layer system we identify $0$ as $1_{\text{Ising}}$ and identify $2_2$ or $4$ as $\psi_{\text{Ising}}$ in each layer. The first layer represents the competition between $0$ and $2_2$, while the second layer represents the competition between $0$ and $4$. 
Using KW analysis we can calculate the critical points of the double layer:
\begin{itemize}
    \item When $a=0$ and $b=\sqrt{2}-1$, the second layer is at the critical point while the first layer consists of simply local $0+2_2$ loops and contribute only a global constant factor to the strange correlator. Thus at this point the system is the critical Ising CFT. In the original parametrization this point corresponds to $y=0$ and $z=\sqrt{2}-1$, which is exactly the equilibrium state between $\mathcal A_0$ and $\mathcal A_1$.
    \item When $a=\sqrt{2}-1$ and $b=1$, the first layer is at the critical point while the second layer is at a topological fixed point described by the Frobenius algebra $\mathcal A = 1_{\text{Ising}} \oplus \psi_{\text{Ising}}$ as in equation \ref{fixed_point}. The fixed point state is invariant under F-move, so the second layer is simply transparent to the KW duality. Hence the double layer system is at its self-dual point and its critical behaviour is described by the Ising CFT. In the original parametrization this point corresponds to $y=2-\sqrt{2}$ and $z=1$, which is exactly the equilibrium state between $\mathcal A_1$ and $\mathcal A_2$. 
    \item When $a=b=\sqrt{2}-1$, the two layers are both at the critical point and the system is self-dual. The system is simply a stacking of two critical Ising models, so the central charge is $c=1/2+1/2=1$. In the original parametrization this point corresponds to $y=\frac{\sqrt{2}}{3}$ and $z=\frac{1}{3}$, which is exactly the tricritical point at which the two Ising phase transition lines coalesce.
    \item When $a=1$ and $b=\sqrt{2}-1$ or when $a=\sqrt{2}-1$ and $b=0$, these points lie outside the valid parameter range when mapped back to the original parametrization and represent unphysical states.
\end{itemize}

\subsection{Controlling the Other Topological Boundary and Ground State Degeneracies}  \label{sec:topbc}

As was discussed in the literature \cite{Bhardwaj:2023idu,Bhardwaj:2023bbf,Bhardwaj:2024qrf}, the ground state degeneracy is related to the number of topological lines that can end on both boundaries of the sandwich. To compute that in the current construction, one has to first recover the knowledge of both boundary conditions of the sandwich. 

In previous discussions, we have focussed on only one of the boundaries of the sandwich, described by the state $\langle \Omega |$. The other boundary condition is a topological boundary, and it is hidden in the precise form of $|\Psi\rangle_{\text{SN}({\Fus})}$. 

Recall that $|\Psi\rangle$ is expressed as a tensor network in terms of quantum 6j symbols \cite{Aasen:2020jwb, Vanhove:2018wlb}, as reviewed in Appendix \ref{app:symRG}.  
It was shown that there are other constructions of $|\Psi\rangle_{\text{SN}({\Fus})}$ \cite{Lootens:2020mso,Vanhove:2021nav}. The idea is to make use of a different type of quantum 6j symbols that are determined by the module category $\mathcal{M}_{\Fus}$. This $\mathcal{M}_{\Fus}$ can be understood as the module category of a Frobenius algebra $\A \in {\Fus}$. To pick appropriate topological boundary condition in the strange correlator, one simply needs to identify a $\A$ , and then obtain these 6j symbols associated to the corresponding $\mathcal{M}_{\Fus}$.
The canonical form of $|\Psi\rangle_{\text{SN}({\Fus})}$ that is constructed from the usual quantum 6j symbols correspond to taking $\mathcal{M}_{\Fus} = {\Fus}$, or equivalently the associated Frobenius algebra is $\A = \textrm{identity object}$. This Frobenius algebra exists in all fusion tensor category, corresponding to the fact that ${\Fus}$ is always a module of itself.  

The ground state degeneracy is then determined by similar considerations \cite{Bhardwaj:2023idu,Bhardwaj:2023bbf}, by counting common bulk lines that can end on both boundaries of the sandwich. i.e. for each boundary there are anyons that are already condensed. The number of shared condensed anyons between the two sides of the sandwich gives the number of degeneracy.

\subsubsection*{Illustration with $A_5$}
$A_5$ is the example we have discussed in detail in the previous sub-sections, with the phase diagram of three competing phases $\A_{0,2,3}$ shown explicitly in figure \ref{fig:A5 mod 2 tri}. The gapless phases at the phase boundaries and their ground state degeneracies were discussed previously. They can be explained using the counting explained above. 

\begin{enumerate}
    \item The ground state degeneracy $N_{02}$ for the $\A_0$ vs $\A_2$ critical line and the degeneracy $N_{21}$ for the $\A_3$ vs $\A_2$ critical line satisfies 
    \begin{equation}
        N_{02} = N_{32} = 3.
    \end{equation}

To explain that, we note that by picking the usual quantum 6j symbols to construct $|\Psi\rangle_{\text{SN}({\Fus})}$ corresponds to having a topological boundary charcterised by $\A_0$, the latter of which corresponds to the diagonal Lagrangian algebra $\oplus_{i=0}^4 i\bar i$. The critical line between $\A_0$ and $\A_2$ preserves symmetries given by their immediate shared ancestor, which is the $\Cent({\Z}_2)$ state, where $A\oplus F$ have condensed. Using table \ref{tab:S3vsA5}, we can translate this condensate back to their $\Cent(A_{5})$ labels, which is given by $0\bar 0 \oplus 4\bar 4 \oplus 2\bar  2$. The number of shared condensed anyons between the two boundaries is thus 3, matching the actual degeneracy observed. 
The same reasoning applies to the critical line between $\A_3$ and $\A_2$. That is because $\A_3$ is Morita equivalent to $\A_0$ as we have discused above. i.e. $\A_3$ corresponds to the same set of condensed anyons in the symTO as $\A_0$. Therefore, $N_{32}=N_{02}$.

\item The ground state degeneracy in the critical line between $\A_0$ and $\A_3$ is \begin{equation}
N_{03} = 2.
\end{equation}
We note that in the competition between $\A_0$ and $\A_3$, their closest common ancestor is the intermediate condensed state $\Cent({\tt Rep}S_3)$. The set of condensed anyons consists of $0\bar 0 \oplus 4\bar 4$. The number of shared condensed anyons with the other topological boundary is thus 2. 
\end{enumerate}

\section{Haagerup - and Novel CFTs}
\label{sec:Haagerup}
To show the power of our methods, let us apply them to the $H_3$ Haagerup {string-net} model to search for critical points and construct global phase diagrams. 
We will report in the following 5 critical points that came from different competing condensates. Two of which agrees with previously reported ones, but with the actual phase transition and global phase diagram clarified in our framework.  Three of them seems to be reported in the literature for the first time. 

The Haagerup fusion category $H_3$ arises from the Haagerup subfactor \cite{grossman2012quantum, asaeda1999exotic}. It has 6 simple objects given by $\{1,\alpha,\alpha^2,\rho, \alpha \rho, \alpha^2 \rho\}$.
The non-trivial fusion rules are given by 

\begin{align}
    &\alpha \otimes \alpha = \alpha^2, \qquad \alpha^3 = 1 \\

    & \alpha \otimes \rho = \alpha \rho = \rho \otimes \alpha^2, \qquad \alpha^2 \otimes \rho = \alpha^2 \rho = \rho \otimes \alpha \\
    & \rho \otimes \rho = 1 \oplus \rho \oplus \alpha \rho \oplus \alpha^2 \rho 
\end{align}
The objects $\{1,\alpha,\alpha^2\}$ form a ${\Z}_3$ subcategory. The nontrivial fusion rules among $\rho, \alpha\rho, \alpha^2 \rho$ can all be deduced from that of $\rho \otimes \rho $ by further fusing with $\alpha$. The quantum dimensions of the simple objects are given by 
\begin{equation}
    d_{\alpha} = d_{\alpha^2} =1, \qquad d_{\rho} = d_{\alpha \rho} = d_{\alpha^2 \rho} = d \equiv \frac{3+ \sqrt{13}}{2}.
\end{equation}

The Drinfeld center $\Cent(H_3)$ contains 12 different anyons,
\begin{equation}
    \{1,\pi_1,\pi_2,\sigma_1,\sigma_2,\sigma_3, \mu_1,\mu_2,\mu_3,\mu_4,\mu_5,\mu_6\},
\end{equation}
and has three Lagrangian condensates 
\begin{subequations}
\begin{align}
&L_0 = 1\oplus \pi_1 \oplus 2 \pi_2 \\
&L_1 = 1\oplus \pi_1 \oplus 2\sigma_1\\
&L_2 = 1\oplus \pi_1 \oplus \pi_2\oplus \sigma_1.
\end{align}
   
\end{subequations}

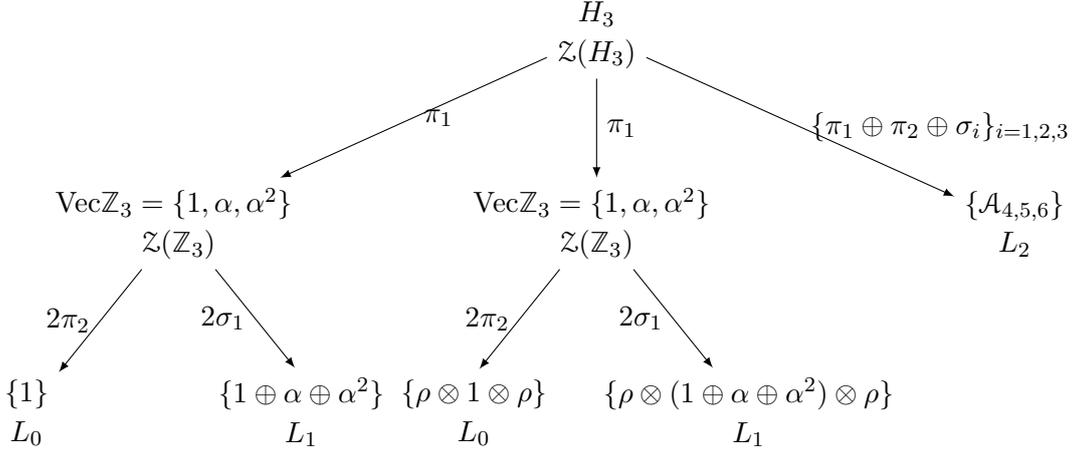
\begin{figure}
    \centering
   \begin{tikzpicture}
[grow = down,
    level distance = 25mm,
    edge from parent/.style = {draw, -latex},
    every node/.style = {align=center},
    level 1/.style={sibling distance=55mm, every node/.style = {align=center}},
    level 2/.style={sibling distance=40mm, every node/.style = {align=center}},
    level 3/.style={sibling distance=23mm, every node/.style = {align=center}},
    level 4/.style={sibling distance=23mm, every node/.style = {align=center}}
]
	\node{$H_3$ \\ $\Cent(H_3)$ }
		child {
            node {$ \textrm{Vec}{\Z}_3 = \{1,\alpha,\alpha^2\}$\,\,  \\
            $\Cent({\Z}_3)$ }
            child{node {$ \{1\}$ \\$L_0 $} edge from parent node[left]{$2\pi_2 $}}
            child{node{$\{1\oplus \alpha \oplus\alpha^2\}\qquad$ \\$L_1\qquad$}
                edge from parent node[left]{$2\sigma_1$}}  
            edge from parent node [right] {$\pi_1$}
            }
        child {
            node {$ \textrm{Vec}{\Z}_3 = \{1,\alpha,\alpha^2\}$\,\,  \\
                $\Cent({\Z}_3)$ }
            child{node {$\qquad \{\rho \otimes 1 \otimes \rho\}$ \\ $\qquad L_0$}
                edge from parent node[left]{$2\pi_2 $}}
            child{node{$\{\rho \otimes(1\oplus \alpha \oplus\alpha^2) \otimes \rho\}$ \\ $L_1$}
                edge from parent node[left]{$2\sigma_1$}}  
            edge from parent node [right] {$\pi_1$}
            } 
        child {
            node{$\{\mathcal{A}_{4,5,6} \}$\\$L_2$}
            edge from parent node[right]{$\{\pi_1 \oplus \pi_2\oplus \sigma_i\}_{i=1,2,3}$}
            };
\end{tikzpicture}
 \caption{Condensation tree for the doubled $H_3$ symTO. The right most branch abbreviates the 3 Morita equivalent branches, with the Frobenius algebras involved given in (\ref{A4}- \ref{A6}).  }
    \label{fig:Haagerup _tree}
\end{figure}

By nailing down the Frobenius algebras in $H_3$, as the input UFC of the HGW string-net model, the refined condensation diagram of $\Cent(H_3)$ is obtained and shown in figure \ref{fig:Haagerup _tree}. 

The UFC $H_3$ has three Morita equivalence classes \cite{grossman2012quantum} of Frobenius algebras, and by (\ref{eq:XAX}), it has only $7$ connected Frobenius algebras as follows.
\begin{subequations}
\begin{align}
& \mathcal{A}_0 = 1 \,\, \\
&\A_1 = \rho\otimes \A_0 \otimes \rho = 1\oplus \rho \oplus \alpha \rho \oplus \alpha^2 \rho\\
&\A_2 = 1 \oplus \alpha \oplus \alpha^2 \,\,\\
&\A_3 = \rho \otimes \A_2 \otimes \rho = (1\oplus \rho \oplus \alpha \rho \oplus \alpha^2 \rho) \otimes (1 \oplus \alpha \oplus \alpha^2). \\
& \A_4 = 1 \oplus \rho \oplus \alpha \rho, \,\,  \label{A4} \\
&\A_5 = \alpha \otimes \A_4 \otimes \alpha^2 = 1 \oplus \alpha \rho \oplus \alpha^2 \rho, \,\,  \label{A5}\\
&\A_6 = \alpha^2 \otimes \A_4 \otimes \alpha = 1 \oplus  \rho \oplus \alpha^2 \rho. 
 \label{A6}
\end{align}
\end{subequations}
The three Morita equivalence classes, corresponding to the three condensates in the output category $L_{1,2,3}$, are as follows: 
\begin{subequations}
\begin{align}
&L_0: \,\, \A_0,\,\A_1\\
&L_1:\,\, \A_2,\, \A_3 \\
&L_2: \,\, \A_4,\,\A_5,\, \A_6.
\end{align}    
\end{subequations}

We note that $L_2$ and $L_{0,1}$ has only the common parent $\Cent(H_3)$, even though $L_2$ and $L_{0,1} $ all contain $1\oplus\pi_1$. Condensing $\pi_1$ in  $\Cent(H_3)$ produces $\Cent({\Z}_3)$, which has only two Lagrangian algebras correspondinng to electric and magnetic condensations but does not have $L_2$. 

The seven Frobenius algebras have their modules tabulated as follows (details in Appendix \ref{appendix:data}.): 
\begin{center}\begin{tabular}{|c|c|}
\hline
$\A$ & Right-$\A$ modules \\ \hline
$1$ & \text{Every simple object is an independent module} \\ \hline
$\rho\otimes\rho$ & $i\otimes \rho$ where $i$ is any simple object  \\ \hline
$1\oplus\alpha\oplus\alpha^2$ & $1\oplus\alpha\oplus\alpha^2$, \  $\rho\otimes(1\oplus\alpha\oplus\alpha^2)$  \\ \hline
$\rho\otimes(1\oplus\alpha\oplus\alpha^2)\otimes\rho$ & $(1\oplus\alpha\oplus\alpha^2)\otimes\rho$, \ $\rho\otimes(1\oplus\alpha\oplus\alpha^2)\otimes\rho$    \\ \hline
$1\oplus\rho\oplus\alpha\rho$ & $1\oplus\rho\oplus\alpha\rho$, \ $\alpha\otimes(1\oplus\rho\oplus\alpha\rho)$, \ $\alpha^2\otimes(1\oplus\rho\oplus\alpha\rho)$, \ $\rho\oplus\alpha\rho\oplus\alpha^2\rho$ \\ \hline
$\alpha\otimes(1\oplus\rho\oplus\alpha\rho)\otimes\alpha^2$ & $M\otimes\alpha^2$ where $M$ is any right module of $1\oplus\rho\oplus\alpha\rho$ \\ \hline
$\alpha^2\otimes(1\oplus\rho\oplus\alpha\rho)\otimes\alpha$ & $M\otimes\alpha$ where $M$ is any right module of $1\oplus\rho\oplus\alpha\rho$ \\ \hline
\end{tabular}\end{center}

A module is a pair of module object and module function. Here, we only list the module objects for each algebra. In principle, two different modules may have the same module object and differ only in the module function. However, for the algebras in $H_3$, it turns out that each module has a distinct module object.

According to this table, one can construct critical points from pairs of Frobneius algebras readily. If possible, one picks a common module shared by two Frobenius algberas to remove other condensates from the reduced phase space, and finally constructs the equal weight combination of the normalised condensates, precisely as in (\ref{pair_compete}). 

\subsection{Competition of Algebras in the \eqs{H_3} Category}
Let us discuss several cases explicitly, combined with detailed numerical checks. 
Similar to the case of $A_{5}$ discussed earlier, numerics enjoys enhanced accuracy if we first perform symmetric RG flow of the boundary conditions. In this current example involving the Haagerup UFC, each object is self-dual. Consequently, each edge must be oriented, giving rise to a branching structure of the lattice. The orientation data needs to be included carefully when performing symmetric preserving renormalisation group flow, generalising the procedure studied in \cite{Vanhove:2018wlb, Chen:2022wvy}.
We thus detail this generalised symmetric RG method in Appendix \ref{app:symRG}. It's often suggested in the literature that the 6j symbol of  $H_3$ does not possess tetrahedral symmetry, for example in \cite{Hahn:2020cgf}; however, this is in fact not true \cite{Huang:2020lox}. We have found plenty of gauge choices for the Haagerup 6j symbols to be tetrahedral symmetric, and we picked an arbitrary one in our numerics. 
While the Turaev-Viro/string-net model may be defined with generic spherical fusion category ${\Fus}$, our RG algorithm is at present designed for those ${\Fus}$ where there are explicit tetrahedral symmetries. Generalisations to include models without tetrahedral symmetries would be discussed elsewhere.

\begin{figure}
    \centering
    \includegraphics[width=0.7\linewidth]{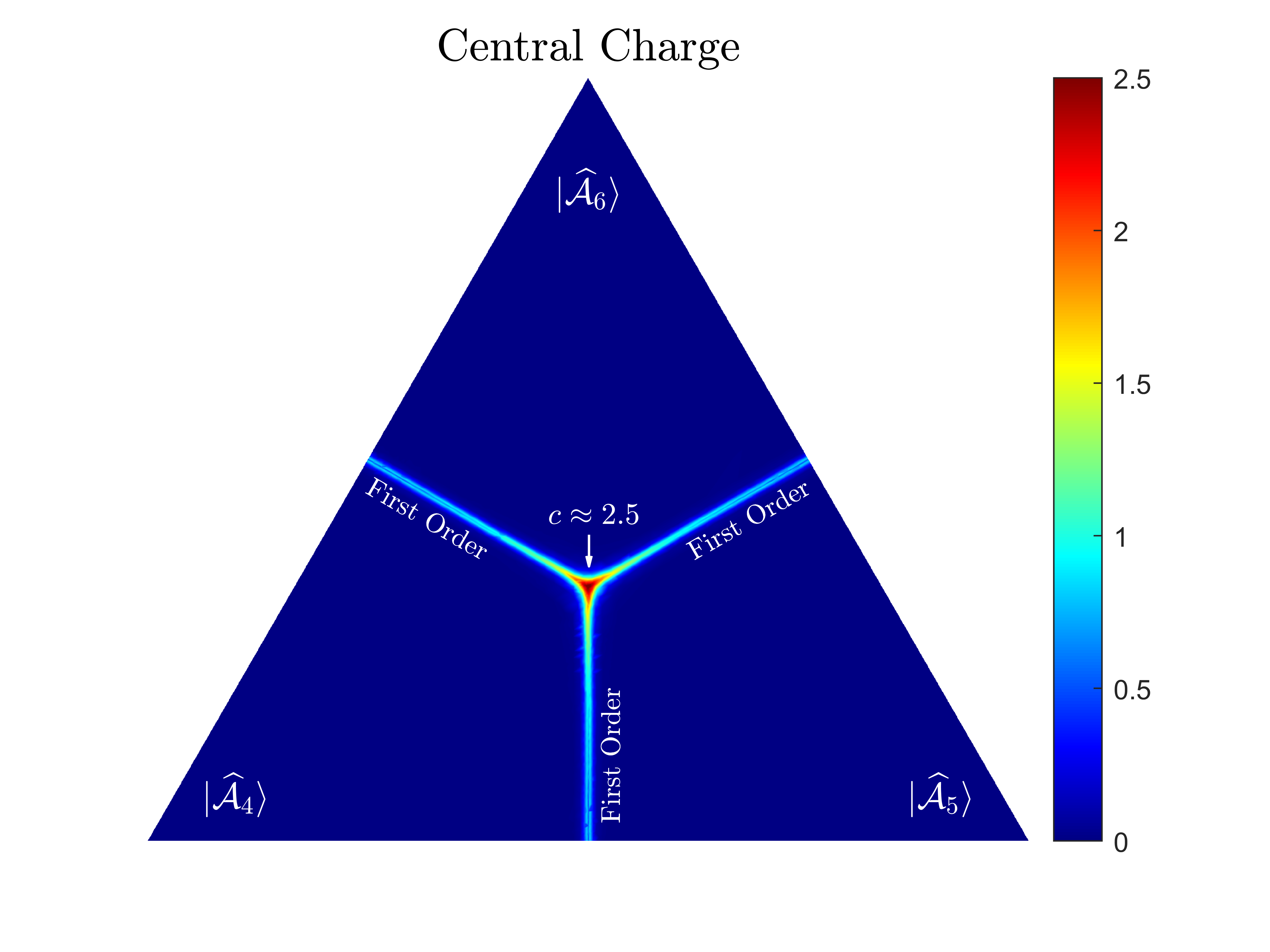}
    \caption{The ternary phase diagram of interpolating Frobenius algebras $\A_{4,5,6}$ in the $H_3$ UFC. The slanted edges are colored by modules containing object $\rho \oplus \alpha\rho \oplus \alpha^2\rho$. Each phase is marked by one of the nomalised states $\A_i$. The tri-critical point is a CFT with $c\approx 2.5$. Equal weight summation of any pair of condensates does not produce a CFT. There are two traces of non-zero central charges for each of the first-order phase transitions.  The actual phase transition lines sit at the center of these double lines. This may suggest a possible complex CFT at the vicinity of the first order transition as mentioned in section \ref{sec:first_order} with the 5-state Potts model following from the 3D $\Z_5$ quantum double as an exmaple . Improved accuracy and further simulations of the novel CFT will be reported elsewhere. 
    }
    \label{fig:H3 fro 456}
\end{figure}

\begin{figure}
    \centering
    \includegraphics[width=0.7\linewidth]{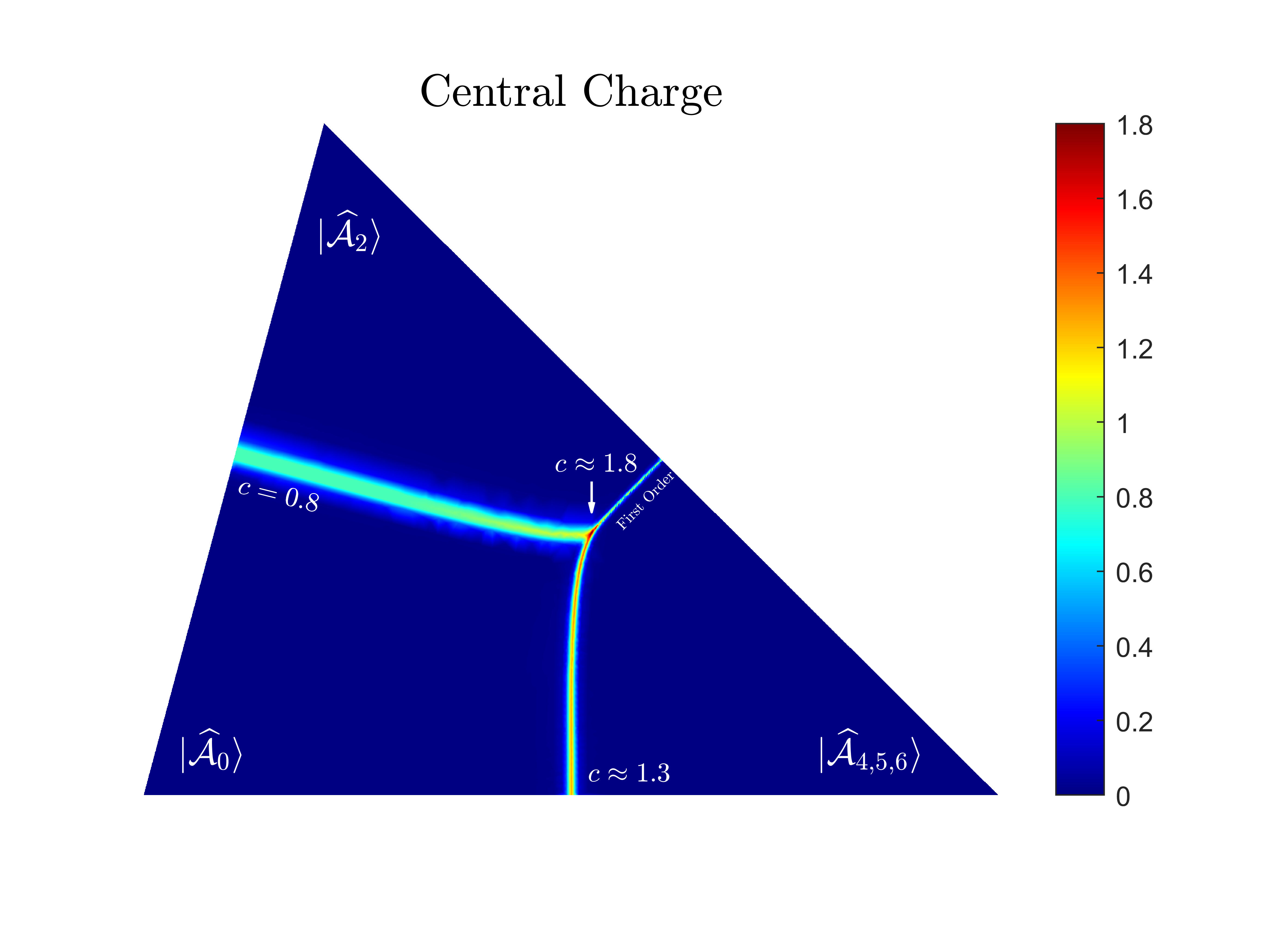}
    \caption{The ternary phase diagram of interpolating Frobenius algebras $\A_{0,2,(4,5,6)}$ in the $H_3$ UFC. The slanted edges are colored by modules consisting of object $\rho \oplus \alpha\rho \oplus \alpha^2\rho$. Each phase is marked by a nomalised states $\A_i$. The three phases are seperated by the critical lines with central charge $c>0$. CFT between $\A_0$ and $\A_2$ is the critical $3$ state Potts model with $c=0.8$. CFT between $\A_0$ and any of $\A_{4,5,6}$ has $c\approx 1.3$. Tri-critical point has $c\approx 1.8$. Phase transition point between $\A_2$ and any of the $\A_{4,5,6}$ is first-order. Appearance of non-zero central charges on the first order line are numerical fluctuations. Improved accuracy at the first order phase boundary and details of the novel CFTs will be reported elsewhere. }
    \label{fig:H3 fro 02456}
\end{figure}

\begin{figure}
    \centering
    \includegraphics[width=0.7\linewidth]{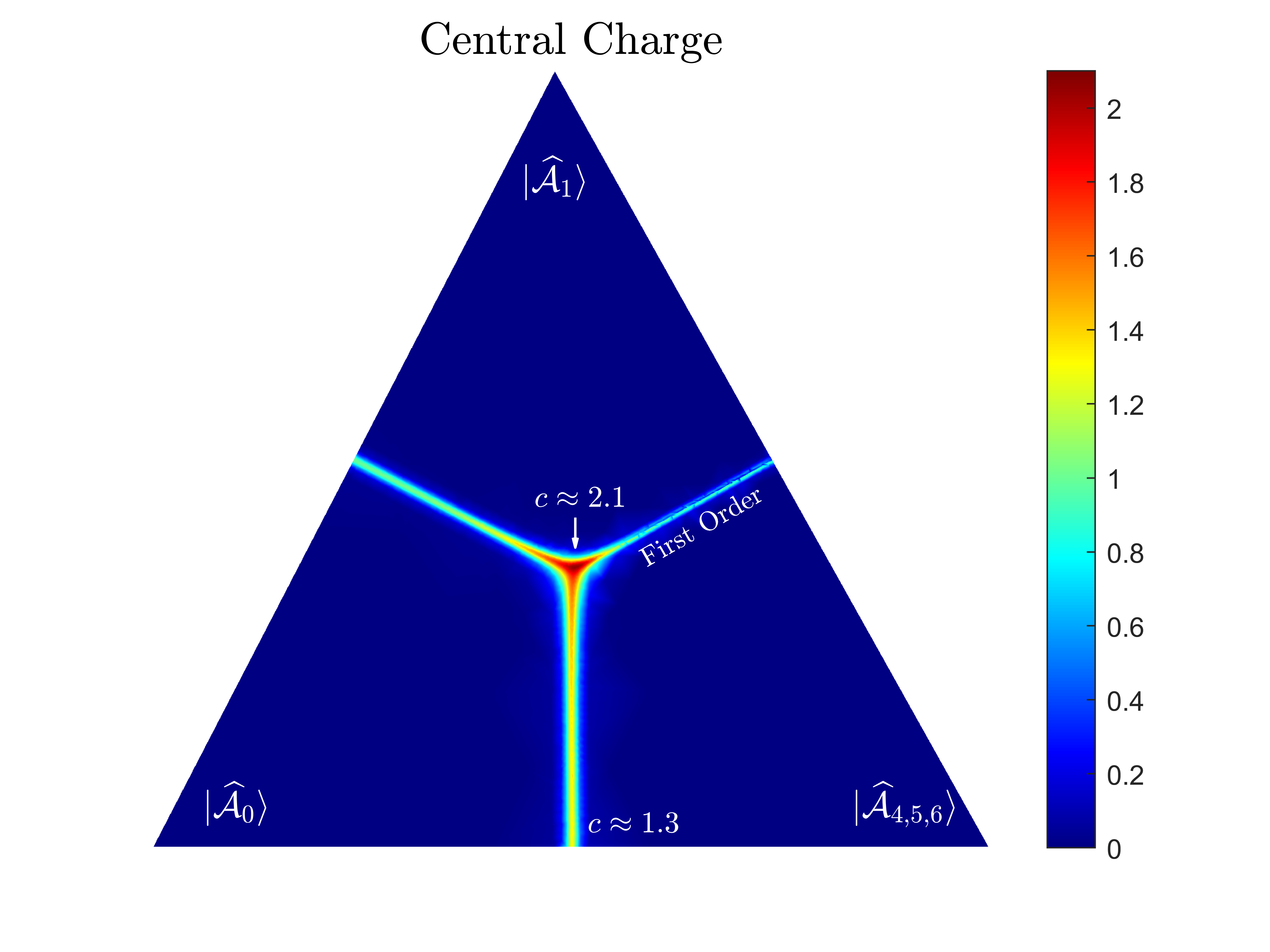}
    \caption{The ternary phase diagram of interpolating Frobenius algebras $\A_{0,1,(4,5,6)}$ in the $H_3$ UFC. The slanted edges are colored by modules containing object $\rho \oplus \alpha\rho \oplus \alpha^2\rho$. Each phase is marked by one of the nomalised states $\A_i$. It is very close to a equilateral triangle but not. CFT between $\A_0$ and any of $\A_{4,5,6}$ has $c\approx 1.3$. The tri-critical point has $c\approx 2.1$. The phase transition between $\A_0$ and $\A_1$ is debatable; please see main text for details. Phase transition between $\A_1$ and any of the $\A_{4,5,6}$ is first-order. There are two traces of non-zero central charges. The vague one conincides with the actual phase transition line. The more clear one may suggest a possible complex CFT at the vicinity of the first order as mentioned in section \ref{sec:first_order} with the $\Z_5$ quantum double as an exmaple. Improved accuracy and further simulations of the novel CFT will be reported elsewhere.
    }
    \label{fig:H3 fro 01456}
\end{figure}

\subsection*{Competition between $\A_0$ vs $\A_1$}
This is an intriguing example. The common ancestor goes all the way up to $H_3$, and so it is expected to produce a non-trivial realisation of the $\Cent(H_3)$ symmetry. 

$\A_0$ and $\A_1$ are Morita equivalent and supposedly would produce the same output condensate $L_0$. By Eq. \eqref{eq:XAX}, Morita equivalent Frobenius algebras are related to each other by conjugation by a simple object in the input UFC. Here, the conjugation is by $\rho$. Below we will also discuss their relation via the generalised KW duality. 

Here, it is also evident that $\rho$ as a simple object seems to be the most convenient shared module between $\A_0$ and $\A_1$. Choosing $\rho$ also has the virtue of excluding all other Frobenius algebras from the reduced phase space. Even then, the maximal rank of Frobenius algebra sharing the module $\rho$ is $N_\A =4$, and the number of gapped Frobenius algebras sharing $M=\rho$ is only $n_M=2$. Since $n_M < D_c = N_\A-1$, violating the sufficient condition \eqref{eq:2ndorder}, it is not \textit{a priori} obvious that this phase transition must be first order from our previous discussion.  


Substituting the data of the Frobenius algebra $\A_{0,1}$ into (\ref{pair_compete}) produces a fluctuating central charge under the symmetric RG algorithm. The entanglement scaling method however suggests a continuous phase transition of $c\approx 2$. We strongly suspect that this is the same model found in \cite{Huang:2021nvb}, and perhaps also the model in \cite{Vanhove:2021zop}. The choice of anyon in a generalised golden-chain model is essentially equivalent to picking a module $M$ in our model. In \cite{Huang:2021nvb}, the anyon chain is made up of $\rho$, which is equivalent to the choice made here. Therefore the critical point is precisely following from the competition between $\A_0$ and $\A_1$. We should thus land on the same critical point, which is supported by the matching central charge $c$. The hexagon type model in \cite{Vanhove:2021zop} is constructed with an $\bra{\Omega}$ made up mostly of $\rho$. This would be related to our square-octagon model with a larger unit cell. These enlarged kind of unit cell is discussed briefly in our discussion section. They are however beyond the scope of the current paper. Therefore, at the moment we are less certain about the interpretation of \cite{Vanhove:2021zop}. More data, such as determining the competing anyon condensates that protect this critical point, and also better numerics of the spectra, are needed to resolve this issue. These will be reported elsewhere.  

\subsection*{Competition between $\A_0$ vs $\A_2$}
In this case, we can pick a common module given by $\A_2$ itself. In this case, one can readily check numerically that the transfer matrix from (\ref{pair_compete}) yields the 3-state Potts model, which is obviously the case because their common ancestor is $\Cent({\Z}_3)$. This critical point is identical to the one we discussed in the previous section, and this is the same model considered in \cite{Bottini:2024eyv}. 

\subsection*{Competitions among $\A_{4,5,6}$ - New CFT alert }
We first look at the three Frobenius algebras $\A_{4,5,6}$, related by conjugation by objects $\{1,\alpha, \alpha^2\}$ that forms the $\Z_3 \in H_3$ subcategory. These three algebras have no common module. Each algebra $\A_{i=4,5,6}$ has a module $M_{\A_{i}}$ with the same object  $\rho\oplus\alpha\rho\oplus\alpha^2\rho$ but equipped with different module functions. To cut down the phase space as much as we could, we will pick each of these $M_{\A_i}$ to color the slanted edges in the interpolation between condensates. That is. the interpolation takes the form of (\ref{many_condensates}). The three algebras are mutually conjugated by $\Z_3$ generators. This $\Z_3$ symmetry is manifest in the phase diagram, as shown in figure \ref{fig:H3 fro 456}. Since the modules $M_{\A_{4,5,6}}$ are different, we have essentially turned on many different couplings, and so we do not expect their pair competitions to generate continuous phase transitions. This is confirmed by our numerics. Nonetheless,  it is pleasantly surprising that the {\bf tri-critical point, which corresponds precisely to the equal weight summation of the three condensates, seems to be a novel CFT of $c\approx 2.5$.} A more accurate determination of the central charge of this CFT and its spectrum will be reported elsewhere. The location of the phase boundaries match accurately with our general prediction (\ref{eq: crit line}).

\subsection*{Competitions between $\A_0$ vs $\A_{4,5,6}$ - New CFT alert}
As mentioned above, $\A_{4,5,6}$ are $\Z_3$ permutated. The pair competition between $\A_0$ and $\A_i$ for $i=4,5,6$ produces the same critical point. For each $i$, as always possible since $\A_0 = 1 \in \A_i$, we can pick the algebra $\A_i$ itself as the common module with $\A_0$.  The common ancestor between $\A_0$ and any of the $\A_{4,5,6}$ goes all the way up to $\Cent(H_3)$.  We find that the pair-competition indeed produces a second-order phase transition that respects the full $\Cent(H_3)$ symmetry. This is checked numerically, and {\bf we found a novel critical point at $c\approx 1.3$ }{\bf This is certainly a novel critical model, and potentially a new CFT in the IR. Details of this model will be reported elsewhere. }

We can also include any number of these six algebras $\A_{0,1,2,4,5,6}$ \footnote{Interpolation concerning $\A_3$ will be reported elsewhere.} in the interpolation (\ref{many_condensates}) , where we pick for each of them the module $\rho+\alpha\rho+\alpha^2\rho$. As expected, the phase diagram retains the permutation symmetry between $\A_{4,5,6}$. If we keep three phases in the interpolation, we can again plot a ternary phase diagram. We plot some of the ternary phase diagrams in figure \ref{fig:H3 fro 02456} and \ref{fig:H3 fro 01456}, with vertices corresponding to the RG fixed point condensed states. The algebras $\A_{1,2}$/$\A_{4,5,6}$ do not share any module function, and they appear to produce indeed a first-order phase transition. The CFTs between other pair-competetions are already given above. {\bf  The tri-critical points provide various new critical points that potentially corresponds to new CFTs.} There are three tri-critical points with $c\approx 1.8$, one of which is plotted in figure \ref{fig:H3 fro 02456}. There are two tri-critical points with $c\approx 2.1$, one of which is plotted in figure \ref{fig:H3 fro 01456}. All potential CFTs are summarized in the tables at appendix \ref{app:summary tables}.
Results with better accuracy will be reported elsewhere. Again, the location of the phase boundaries agrees accurately with the general prediction (\ref{eq: crit line}).

\subsection{Generalised KW Duality}

\begin{figure}
    \centering
    \includegraphics[width=0.95\linewidth]{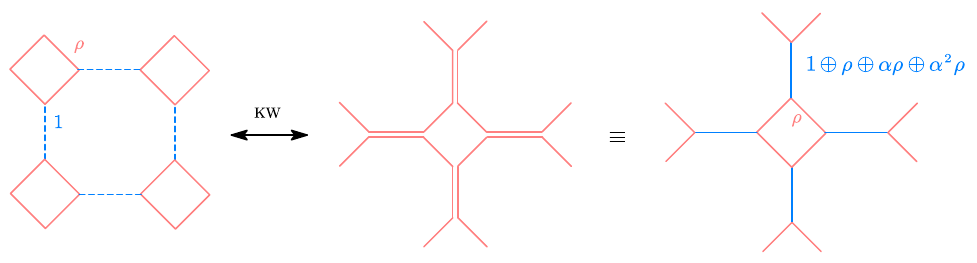}
    \caption{Generalised KW duality for the case where the module is $\rho$ and the competing algebras are $\mathcal{A}_0 = 1$ and $\mathcal{A}_1 = \rho\otimes\rho$.}
    \label{fig:Haagerup_KW_1}
\end{figure}
We also apply KW duality to analyze several intriguing competitions in $H_3$ above.
\subsubsection*{Competitions between $\A_0$ vs $\A_1$}
For the competition between $\mathcal A_0$ and $\mathcal A_1$ with $\rho$ as a shared module, we can also perform a generalised KW duality as shown in figure \ref{fig:Haagerup_KW_1}. The duality transformation maps $\rho$-squares to $\rho$-octagons, and the double $\rho$-lines on an octagon edge is equivalent to $\sum_{i\in\rho\otimes\rho} \frac{\sqrt{d_i}}{d}i$ lines. Then we can verify that the system is self-dual when the octagon edge takes the state 
\begin{equation*}
    \ket{1} + \frac{1}{\sqrt{3d+1}} \left(\ket{1} + \sqrt{d}\ket{\rho} + \sqrt{d}\ket{\alpha\rho} + \sqrt{d}\ket{\alpha^2\rho}\right)
\end{equation*}
This self-dual point coincides precisely with the point calculated by (\ref{pair_compete}).

\subsubsection*{Pair competitions within $\{\mathcal{A}_4, \mathcal{A}_5, \mathcal{A}_6\}$}
\begin{figure}
    \centering
    \includegraphics[width=0.5\linewidth]{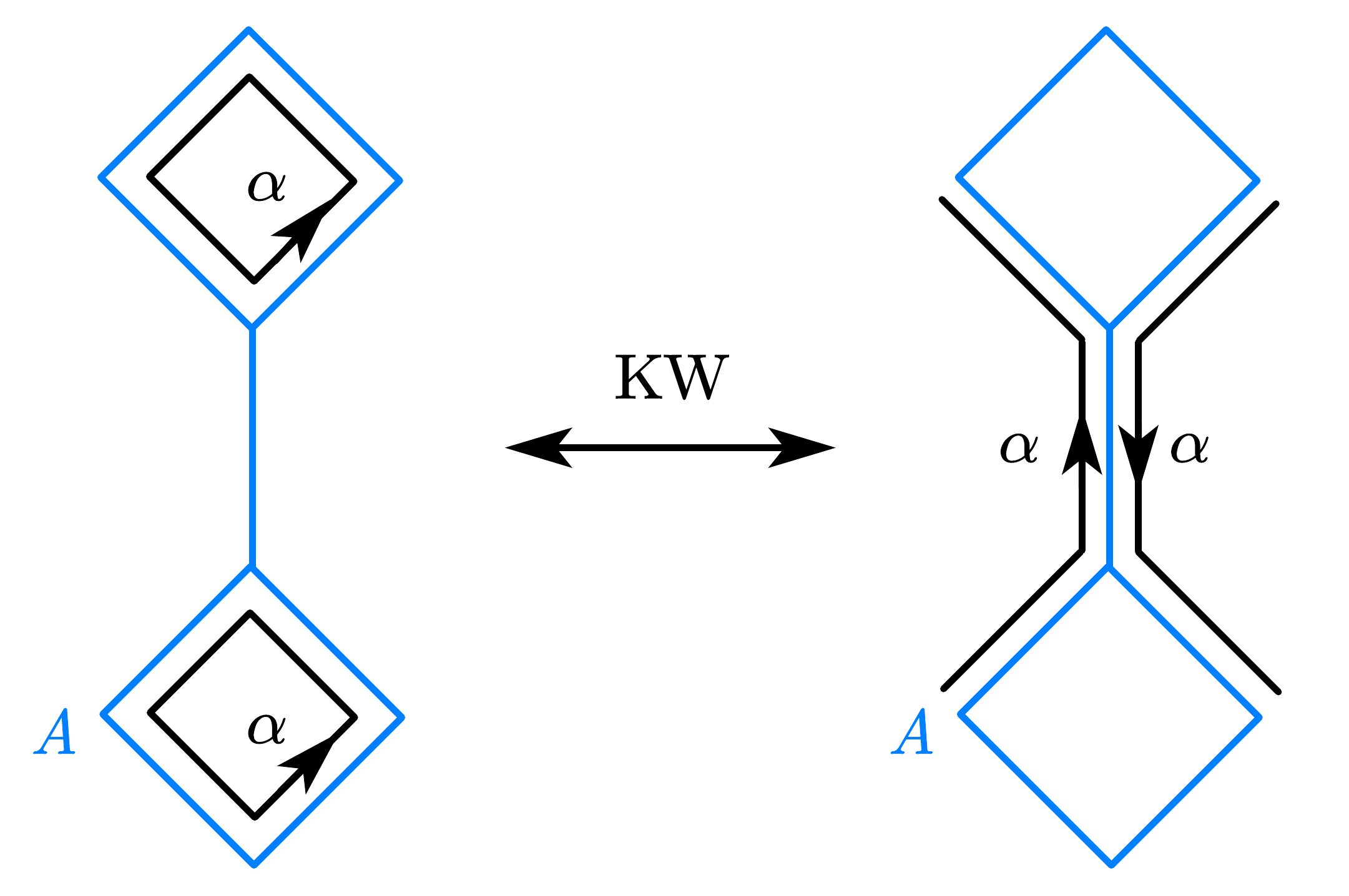}
    \caption{Generalised KW duality for the case where the competing Frobenius algebras are $\mathcal{A}$ and $\alpha\otimes\mathcal{A}\otimes\alpha^2$, and the modules are respectively $\alpha\mathcal{A}$ and $\mathcal{A}\alpha^2$ for any $\A\in\{\mathcal{A}_4, \mathcal{A}_5, \mathcal{A}_6\}$. Note that we have implicitly shifted the lattice by one unit cell after the KW dual, thereby moving our focus to the vertical edges, compared to KW in figure \ref{fig:Ak_KW} and figure \ref{fig:Haagerup_KW_1}.}
    \label{fig:Haagerup_KW_2}
\end{figure}

For the pair competitions within $\{\mathcal{A}_4, \mathcal{A}_5, \mathcal{A}_6\}$, we observe that each pair can be written as $\mathcal A$ and $\alpha\otimes\mathcal{A}\otimes\alpha^2$ for some $\mathcal{A}$ in the set. This observation allows us to reduce the problem to a double-layer structure, where the one layer is the algebra $\mathcal{A}$ and the other layer consists of $\alpha$ loops. The double-layer structure is illustrated in figure \ref{fig:Haagerup_KW_2}. In this setup, the competition between $\mathcal{A}$ and $\alpha\otimes\mathcal{A}\otimes\alpha^2$ is captured by the competition between the identity $1$ and $\alpha\otimes{\alpha^2}$ because the $\mathcal{A}$ layer is invariant under KW duality. It is analogous to the situation in figure \ref{fig:Haagerup_KW_1}, with the double $\rho$-lines replaced by $\alpha$-line and $\alpha^2$-line (or $\alpha$-line with the direction reversed). Nevertheless, $\alpha\otimes{\alpha^2}$ is the identity object, and hence the competition in this layer is trivial. As such, we expect the phase transition point between the pair to be a first-order phase transition. The numerical evidence for this argument has already been given above.

\section{Discussion}
\label{sec:Discussion}
In this paper, we have proposed a general method to construct critical lattice models using non-invertible symmetries and the strange correlator. The crux of designing critical lattice models using the strange correlator is to design appropriate state $\langle \Omega |$. We find that working with Frobenius algebras of the input category ${\Fus}$ defining the {string-net} model contains extra details not visible in the output condensate content. Our full strategy can be summarised as follows. 
\begin{enumerate}
    \item Choose an appropriate lattice that resembles a square lattice. 
    \item Reduce the problem to a low dimensional one by picking a unit cell surounded by ``module'' of condensates as a bath. An appropriate choice could help one further surpress the phase space and exclude other competing condensates, sieving out second order phase transitions from first order ones. 
    
    \item By making use of the definition of traces in UFCs, we create an equal amount of non-commuting anyon condensates in this bath, pushing the boundary state to criticality.

    \item We introduce a refined condensation tree to read off the symmetries preserved by the condensate. The diagram has to include condensates corresponding to Morita equivalent Frobenius algebras that are shown to be physically different.  Location of phase boundaries can also be deduced. 

    \item We generalise the notion of {KW} duality in this context and find an efficient and powerful way to locate many of the commensurate condensates. Combining with a novel observation that connects Frobenius algebras in the same Morita class, we find that much of the details of critical points, and natures of tri-critical points, can also be read-off, in addition to critical couplings.  

    \item We also make significant improvement to our numerical implementations of the symmetric RG method, allowing us to check our predictions with precision numerics. Unlike isolated models that were checked in the literature, where bond dimension is known to be high in more exotic situations, our method produces complete phase diagrams in multiple novel examples, including the long sought Haagerup model. 

\end{enumerate}

We demonstrate the power of our methodologies by first reproducing the $A$-series critical lattice models that would flow to the A-series minimal models, including the Ising model. 

Then we apply our methods to studying the phase diagram of the $A_5$ model, and found that we can analytically predict the location of critical points, and also a large part of the phase boundaries -- these are all checked by numerics using our improved symmetric RG method.

Finally we apply our methods to the Haagerup model. Not only do we reproduce known lattice models, we found at least three new critical points. We also obtain the first phase diagrams (whose phase boundaries are again located almost completely analytically) of the Haggerup model, which are confirmed by numerical results. 

Our program made significant progress in several ways:

\subsection{Landau Paradigm for Symmetric Lattice Models}
Our work lays the foundation for the Landau paradigm for lattice models. While the Landau paradigm has been proposed to extend to include non-invertible symmetries \cite{Ji:2019jhk,Bhardwaj:2024wlr},  it remains very difficult to write down an effective theory that incorporates given non-invertible symmetries which at the same time also carries detailed infrared dynamics of phase transitions. 
Such an effective theory description lies at the heart of the Landau paradigm because it allows a systematic method of constructing explicit and computable models, at least approximately. While the strange correlator, or its predecessor, the golden chain, is essentially equivalent to the sandwich realisation of symmetric theories, it carries extra significance because it provides a tangible bridge between generic non-invertible symmetries describable by fusion tensor categories and explicit lattice models carrying the symmetries.  The latter, even if not solved analytically, can at least be explored numerically. Therefore, the dynamics of the symmetric phases can in principle be extracted from it. For the last decade, however, the main bottle-neck of the strange-correlator strategy lies in the gigantic size of the phase space of states $\langle \Omega |$ forming the boundaries of the 3D TQFT. The strange correlator advances the golden-chain in that it separates the 1+1 D problem into a known 2+1 D topological wavefunction and a state $\langle \Omega|$ that encodes the dynamics. Symmetric RG shows that properties of $\langle \Omega|$ can be analysed locally, and it doesn't take solving the spectrum of a global Hamiltonian to know which phase we are in. This promises the possibility of significantly reducing the phase space of theories, by focusing on a few lattice sites in $\langle \Omega|$. Even then, unfortunately, there are too many possibilities for $\langle \Omega|$, and it is not clear where to start systematically. Contrast that with the case of effective theories. There, one can focus on the order parameters, which are generically limited to only a handful of fields and whose effective action can be written down systematically according to symmetries and scaling dimensions, that ultimately get reduced to a countable number of terms. Therefore, we need to find the counterpart of order-paremeter field in the context of the strange correlator. {\bf Our answer to this question is essentially the anyon creation operator in a unit cell.} Moving in phase space now corresponds simply to moving the vev of the order parameter in terms of the anyon condensate, thus providing physical handles in introducing different perturbations, and we can move in phase space in a physical and controlled manner. 

This Landau paradigm is related to a generalised Landau paradigm of 2+1 D topological orders that is proposed in a related paper by two of us \cite{zhao2025}, where Morita equivalent Frobenius algebras in the input UFC of the HGW string-net model (in fact an enlarged version of the HGW model) comprise the Goldstone mode parameter, allowing the order parameter fields corresponding to creation operators of condensing anyons to be defined. This encompasses topological orders (which have long be regarded beyond) back in a symmetry-breaking paradigm.

\subsection{CFT Factory and an Alternative to the Bootstrap Program}
A by-product of constructing a lattice model is that the model is by construction UV complete as well; hence, the model is defined completely non-perturbatively. This in a way makes it even more powerful than effective theory. The non-perturbative theory, as a well-defined UV complete theory, would automatically satisfy all the physical consistency conditions, such as modular invariance. Since non-commuting condensates could force the boundary into criticality, we actually use our construction to systematically generate novel CFTs that can be read off by taking the thermodynamic limit. This could offer a powerful alternative route to the bootstrap program.

\subsection{Comments on More Complicated Boundary Conditions and Enlarging the Unit Cell}

Our construction yet has many open problems for future investigation. A salient one is why we have chosen a square-like lattice and how to generalise the constructions to other lattices.

As explained in the main-text, our choice of square lattice is to ensure that there is an equal number of plaquettes and vertices, making the equal weight summation of competing condensates a natural location of criticality because we need not to distinguish fluxons and chargeons which supposedly reside on plaquettes and vertices respectively. For more generic lattices, we suspect that the equal mixture of competing condensates may not always generate a critical state globally. The correct relative weights may depend on the ratio of plaquettes and vertices in the given lattice. This is certainly worth exploring in the future. 

Another important question is that even as we work with the square-like lattice (or the square-octagon lattice chosen here), whether changing the choice of unit cells make a difference. This is an interesting and important question that deserves discourse in further detail in the following. 

For any given (non)-invertible symmetries, one expects that there are an infinite number of different systems sharing the same symmetries. In this paper, we have chosen a minimal unit cell, and explored the phase space of theories by varying couplings within the unit cell. Therefore, we are obviously exploring only a small corner of possible theories that realise the given symmetries. Hence, two paramount questions are 1) what the possible systems that can be realised within this restricted phase space are, and 2) how to systematically enlarge the space of theories that one explores. 

To the first question, we observe that when the subspace gets smaller, the central charge of the resultant critical point also appears to be smaller. For example, the $A$-series were realised by picking a module that essentially reduce the phase space in the unit cell to 2-dimensions. In that case, the central charge of the resultant critical point seems to be smaller than that with the same choice of unit cell but a large phase space. The number of degrees of freedom involved in the unit cell seems to be related to the central charge of any realised critical point. We suspect that the ``canonical'' CFT for a given symTO proposed in \cite{Kong:2019byq} can always be realised from models constructed in the unit cell. 

\begin{figure}
    \centering
    \includegraphics[width=0.7\linewidth]{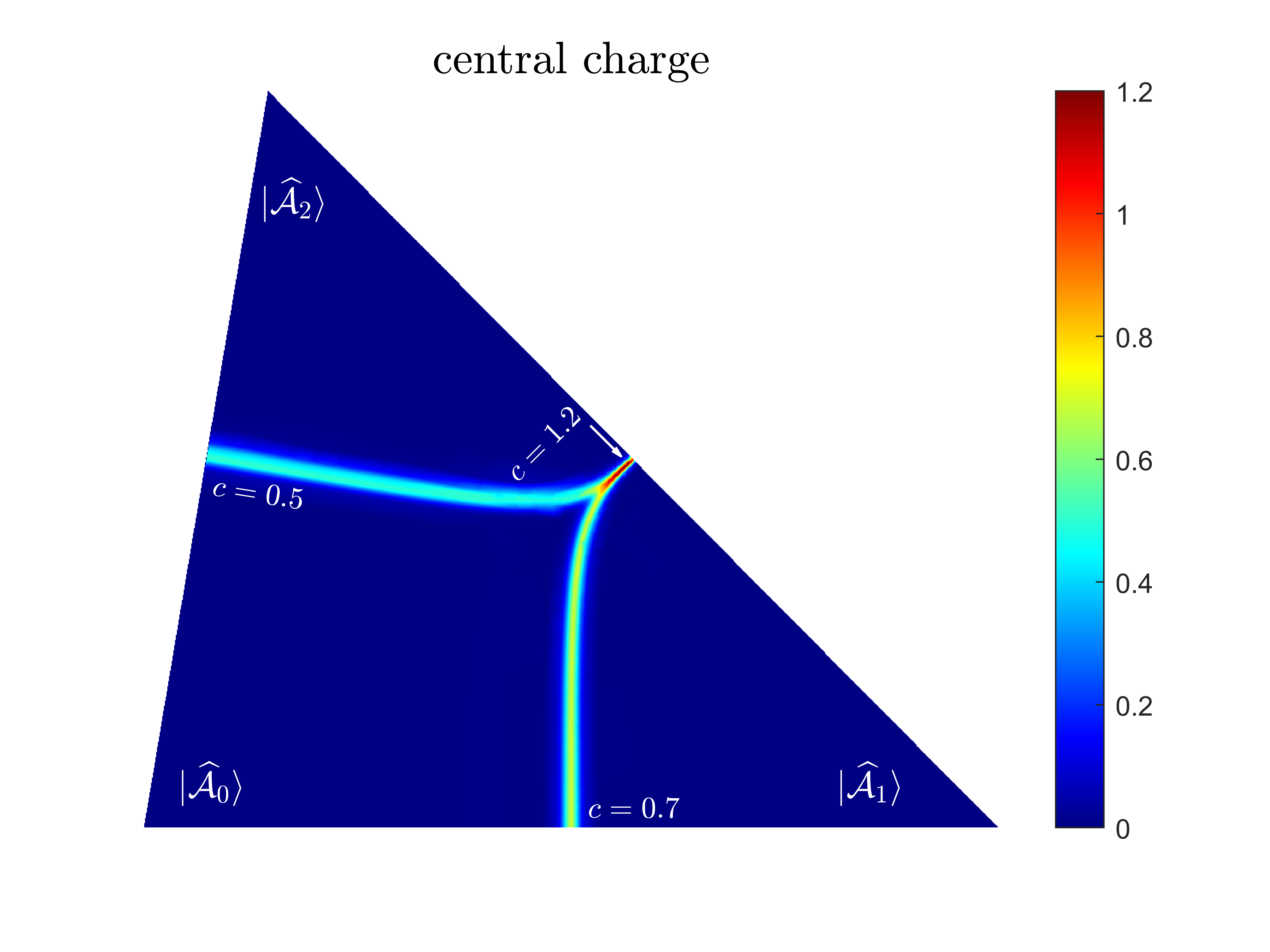}
    \caption{The ternary phase diagram of interpolating the Frobenius algebras $\A_0=0,\A_1=0\oplus2,\A_2=0\oplus3$ with module $1\oplus2$ in category $A_4$. They are all subalgebras of $\A_3=0\oplus1\oplus2\oplus3$. Each phase is marked by one of the three nomalised states $\A_i$. The three phases are seperated by the critical lines with central charge $c>0$. This is another illustration of the double-layer structure, see main text for details.}
    \label{fig:A3 mod 12 tri}
\end{figure}

This observation can be qualitatively explained by the generalised KW duality. As one increases the dimension of the subspace, the degrees of freedom often fall into decoupled multi-layer system, a direct generalisation of the double-layer structure described in the text. The CFT at the critical point of the multi-layer system is a tensor product of the CFTs of each critical layer; hemce, the central charge is larger. Apart from the exmaples above, a $c=1.2$ CFT is observed in category $A_4$ as plotted in figure \ref{fig:A3 mod 12 tri}. This CFT is realised in a subspace of $4$ dimensions. From the generalised KW duality, this system can be decomposed into a double layer, each bearing a 2-dimensional subspace. The two layers contribute a $c=0.5$ CFT and a $c=0.7$ CFT when both layers are tuned to its respective critical point. The observed $c=1.2$ CFT should simply be the stacking of the two single-layer CFTs.

With increasing size of the chosen unit cell, one expects that the strange correlator should be able to describe more complicated CFTs carrying the symmetries with larger central charges. We considered a larger unit cell as shown in figure \ref{fig:larger_unit_cell} below. 

\begin{figure}
    \centering
    \includegraphics[width=0.43\linewidth]{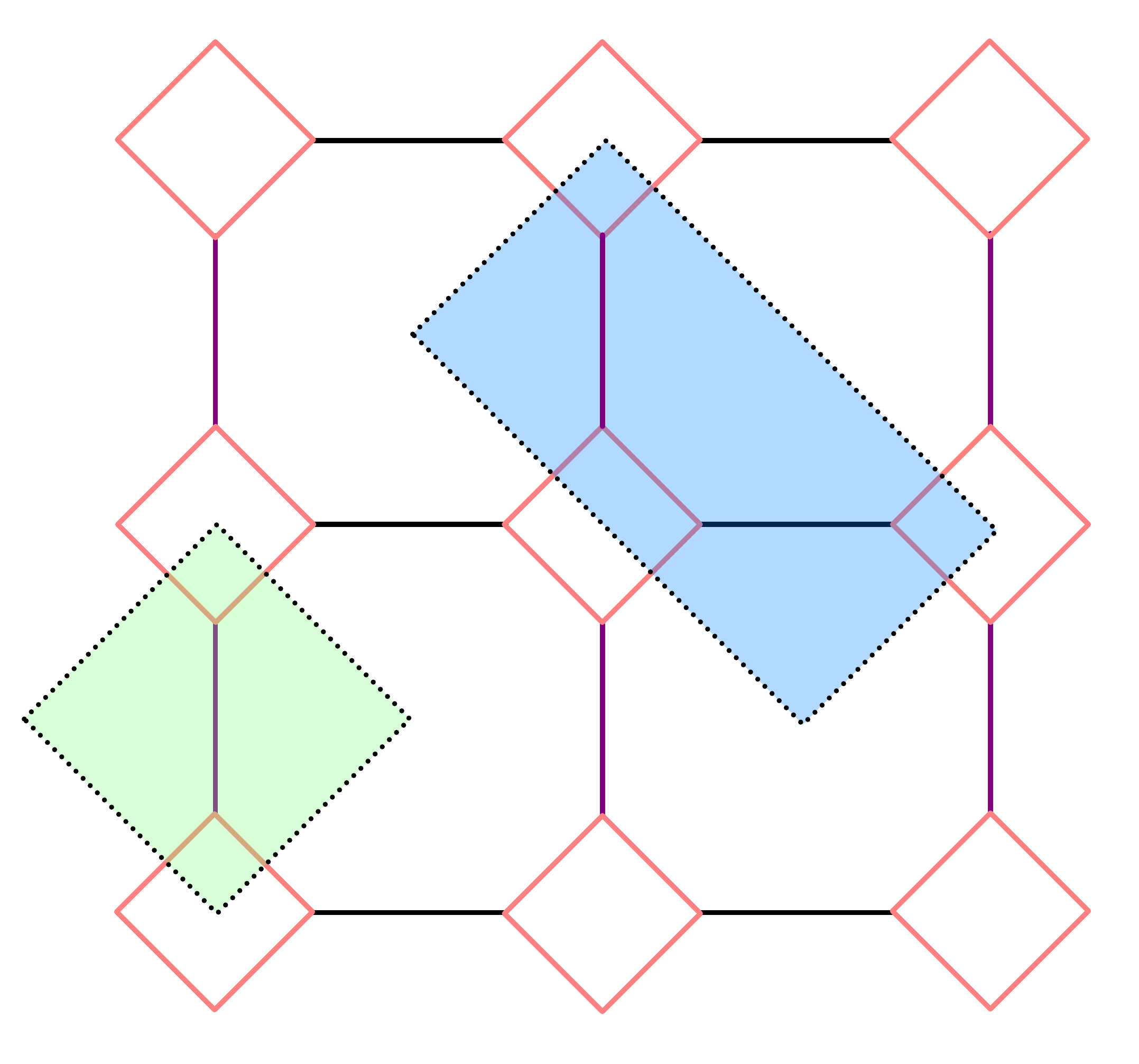}
    \caption{Example of larger unit cell. The blue region is the unit cell of the anisotropic state $\bra{\Omega}$, of which the horizontal and vertical edges are in general different. If $\bra{\Omega}$ is isotropic, then the unit cell is the green region, which is discussed extensively in the text above.}
    \label{fig:larger_unit_cell}
\end{figure}

We observe that generalised KW duality can effectively search for critical points here. The KW duality exchanges squares with octagons and swaps horizontal edges with vertical ones. On enlarging the unit cell to accomodate the anisotropic $\bra{\Omega}$, the $n$-dimensional phase space is promoted to a $2n$-dimensional phase space. In general, the critical self-dual manifold in the $n$-dimensional phase diagram is promoted to a critical self-dual manifold with higher dimensions in the $2n$-dimensional phase diagram. The extra dimensions of the critical manifold are often irrelevant to the CFT. We have checked this in the case of Ising, Fibonacci, and $A_k$ categories and the results will be presented elsewhere.

Larger unit cells can be converted to an entangled $\langle \Omega |$ state and computed by the same numerical RG process. We have not yet looked into a more general ansatz to systematically produce critical models for $\langle \Omega |$ that carries entanglement. This and other important questions will be discussed elsewhere. 

\begin{acknowledgments}
The authors thank Rui-Zhen Huang, Yunfeng Jiang, Zhengwei Liu, Shuang Ming, Sébastien Palcoux, Yilong Wang, Jinsong Wu, Zhihao Zhang for inspiring and helpful discussions. LYH is supported by NSFC (Grant No. 11922502, 11875111). CS is supported by the postdoctoral fund of Beijing Institute of Mathematical Sciences and Applications. 
Part of this work was done during the KITP Program “Generalized Symmetries in Quantum Field Theory: High Energy Physics, Condensed Matter, and Quantum
Gravity”, and supported in part by grant NSF PHY-2309135 to the Kavli Institute for Theoretical
Physics (KITP). YW is supported by NSFC Grant No. KRH1512711, the Shanghai Municipal Science and Technology Major Project (Grant No. 2019SHZDZX01), Science and Technology Commission of Shanghai Municipality (Grant No. 24LZ1400100), and the Innovation Program for Quantum Science and Technology (No. 2024ZD0300101).  
YW is grateful for the hospitality of the Perimeter Institute during his visit, where the main part of this work is done. This research was supported in part by the Perimeter Institute for Theoretical Physics. Research at Perimeter Institute is supported by the Government of Canada through the Department of Innovation, Science and Economic Development and by the Province of Ontario through the Ministry of Research, Innovation and Science. LYH acknowledges Lin Chen, Yikun Jiang, Bingxin Lao for collaboration and discussions on related problems.
\end{acknowledgments}

\newpage
\appendix
\section{The HGW String-Net Model}
\label{sec:HGW}

\begin{figure}
\centering
\includegraphics[height=0.3\linewidth]{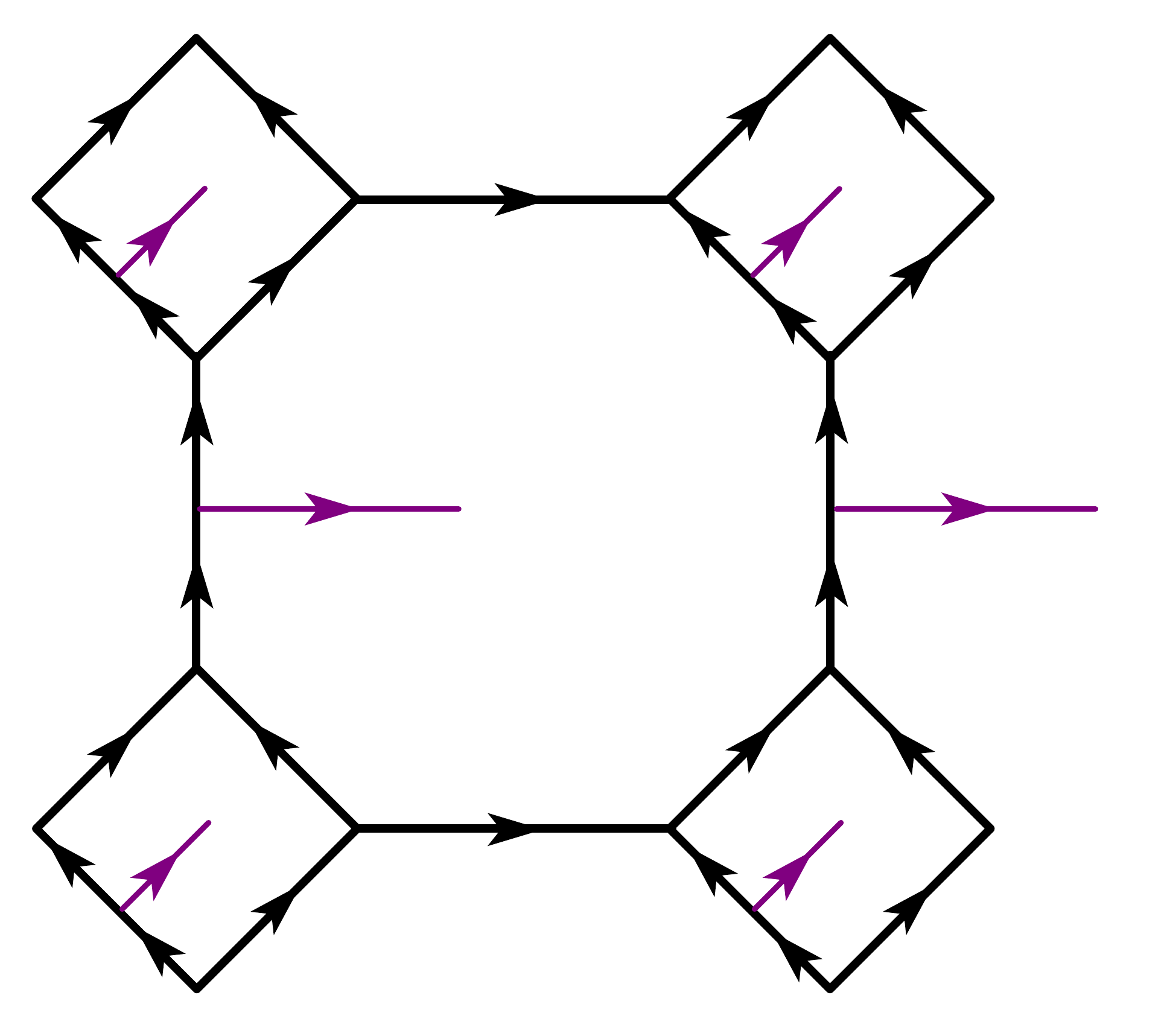}
\caption{The HGW model.}
\label{fig:square_octagon_with_tail}
\end{figure}

The HGW model\cite{hu2018full, zhao2024noninvertible, Zhao:2024ilc} can be defined on a trivalent lattice, where a tail (purple lines in Fig. \ref{fig:square_octagon_with_tail}) is attached to a chosen edge of a plaquette, while the choice is topologically irrelevant . The lattice shape is irrelevant in the string-net model describing topological orders, but it plays an important role in conformal field theories. In this article, for convenience, we only define the HGW model on the truncated square lattice as shown in Figure \ref{fig:square_octagon_with_tail}, which facilitates the subsequent renormalization group procedure. The basic configuration is established by labeling each edge and tail with a simple object from the input unitary fusion category \(\Fus\) of the HGW model, subject to the constraint on all vertices that \(\delta_{ijk} = 1\) for the three incident edges or tails meeting at this vertex. The Hilbert space \(\Hil\) of the model is spanned by all possible configurations of these labels on the edges and tails. Edges and tails are oriented, but the choice of orientation does not affect the physics. 

In the HGW model, anyons reside on the plaquettes of the lattice, and their types are labeled by the simple objects of the Drinfeld center \(\Cent(\Fus)\) of the input UFC \(\Fus\). A key pro of the HGW construction is that it explicitly manifests the internal gauge degrees of freedom of non-Abelian anyons. Specifically, anyons are realized in the model by \emph{dyons}---a pair consisting of an anyon type \(J\) and its internal gauge degree of freedom \(p\), where \(p\) is the degree of freedom on the tail of the plaquette where the anyon resides. A given anyon type \(J\), as a simple object in \(\Cent(\Fus)\), may carry multiple types of internal degree of freedom \(p\), thereby enlarging its internal gauge space. Although these \(p\) are gauge degrees of freedom and hence unobservable in the topological phase, they play a central role in the construction of the CFT states, where topological invariance is broken to expose these internal degrees of freedom, which become physical local degrees of freedom determine the physical phenomena that may be captured by a critical CFT.

In the HGW model, anyons are created in pairs by acting a \emph{creation operator} on the ground state \(\ket{\Psi}\). It suffices to define the action of the shortest creation operator \(W_E^{J; pq}\), which creates a pair of dyons \((J^\ast, p^\ast)\) and \((J, q)\) in the two adjacent plaquettes separated by an edge \(E\):

\eqn{
W_E^{J; pq} \ExcitedA\quad :=\quad \sum_{k \in L_\Fus} \sqrt{\frac{d_k}{d_j}} \ \overline{z_{pqj}^{J;k}} \ \ \ExcitedB \ ,
}
where \(J\) is a simple object in \(\Cent(\Fus)\), \(p, q\) are internal gauge degrees of freedom on tails of anyons, \(j\) is the degree of freedom on edge \(E\), and \(z_{pqj}^{J;k}\) is the coefficients called \emph{half-braiding tensors}. All other creation operators in the model can be generated by composing such shortest creation operators. .

In a companion paper\cite{Zhao:2024ilc}, it is shown that the edges of the lattice carry a flat gauge field valued in $\Fus$, and the open ends of the tails host anyon excitations coupled with the gauge field via the Gauss law and gauge connection. This finding allows one to recast topological phases back in the Landau-Ginzburg paradigm, though in a more general sense, where phase transitions are triggered by anyon condensation. The HGW model is a convenient framework for constructing any possible kind of anyon condensation in a doubled topological phase and study the consequent phenomena\cite{Zhao:2024ilc}.

\eqn[eq:Frob]{\FrobProdA\quad =\quad \sum_{abc\in L_\A}f_{abc^\ast}\quad\FrobProdB\ ,\qquad\FrobProdC\quad =\quad \sum_{abc\in L_\A}f_{cb^\ast a^\ast}\quad\FrobProdD\ ,}

$$\ContractA\qquad\Longrightarrow\qquad \ContractB\ ,$$

\subsection{Definitions of Frobenius Algebras and Modules}\label{app:alg_mod}

In this appendix, we quickly review the mathematical definitions of Frobenius algebras and modules. One who wants to construct their own CFTs via our method can take use of the definition of Frobenius algebras and modules in this section.

\begin{figure}
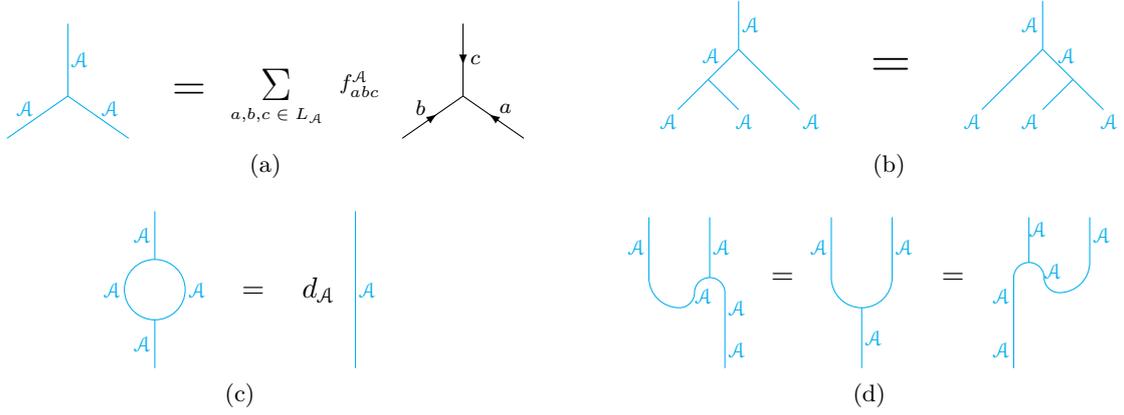
\centering
\subfloat[]{\FrobeniusalgebraA}\hspace{40pt}\subfloat[]{\FrobeniusalgebraB}\\
\hspace{30pt}\subfloat[]{\FrobeniusalgebraC}\hspace{80pt}\subfloat[]{\FrobeniusalgebraD}\\
\caption{(a) Frobenius algebras in basis form. (b) The Associativity: Eq. \eqref{eq:FrobA}. (c) The unitarity: Eq. \eqref{eq:FrobB}. (d) Equations \eqref{eq:FrobC}. Taking different graphical bases yields different equations in Eq. \eqref{eq:FrobC}. This equation is equivalent to the Frobenius condition\cite{hu2018boundary}.}
    \label{fig:Frobalg}
\end{figure}

A unitary Frobenius algebra in ${\Fus}$ is a (possibly composite) objects in ${\Fus}$ i.e. 
\begin{equation}
    \A = \bigoplus_an_aa, \qquad a \in {\Fus},\qquad n_a \in \mathbb{N},
\end{equation}
equipped with a product \(\A\otimes\A\to\A\) and coproduct \(\A\to\A\otimes\A\), satisfying certain properties. Here, $n_a$ are non-negative integers describing the multiplicities of object $a\in {\Fus}$ participating in $\mathcal{A}$. To avoid clutter, the discussion in this section is limited to $n_a \le 1$, although it can be readily generalized by introducing extra indices $i$ for each anyon label $a \in \A$ whose $n_a > 1$. For the case where all \(n_a \leq 1\), the products and co-products of \(\A\) encoded in a cyclically symmetric function \(f^\A: L_\A^3\to\CC\), satisfying
\begin{align}
\sum_{t \in L_\A} f^\A_{rst} f^\A_{abt^\ast} G^{rst}_{abc} \sqrt{d_c d_t} &= f^\A_{acs} f^\A_{rc^\ast b}\ ,\label{eq:FrobA}\\
\sum_{ab\in L_\A} f^\A_{abc} f^\A_{c^\ast b^\ast a^\ast} \sqrt{d_a d_b} &= d_\A \sqrt{d_c},\label{eq:FrobB}\\ 
f^\A_{abc} = f^\A_{bca}, \quad f^\A_{1ab} = \delta_{ab^\ast},\ \ &\ \ f^\A_{abc} = (f^\A_{c^\ast b^\ast a^\ast})^\ast .\label{eq:FrobC}
\end{align}
where \(L_\A = \{a\in \Fus: n_a = 1\}\) is the set of simple objects appearing in \(\A\), and \(d_\A = \sum_{a\in L_\A}d_a\) is the quantum dimension of \(\A\). One can express this map in basis form, which is depicted in figure \ref{fig:Frobalg}.

A right module of Frobenius algebra \(\A\) in a UFC \(\Fus\) is a (possibly composite) object
\[
M_\A := \bigoplus_{x\in L_\Fus} m_x x,
\]
equipped with an algebra action on \(M_\A\): \(M_\A\otimes \A \to M_\A\). Here, $x$ is a collection of simple objects of $\Fus$ that appear in the module, and \(m_x\in\NN\) is the multiplicity of \(x\) in \(M\). For simplicity, in this paper, we restrict to the case \(m_x \le 1\), and define \(L_{M_\A} = \{x\in L_\Fus\mid m_x  =  1\}\). The algebra action is recorded in a \emph{module function} $\rho_{M_\A}: L_\A\otimes L_{M_\A}^2\to\CC$, which satisfies an associativity constraint:
\begin{equation}\label{eq:rightmodule}
[\rho_{M_\A}]^a_{xy}[\rho_{M_\A}]^b_{yz}G^{bz^\ast by}_{xac} = f^\A_{abc^\ast}[\rho_{M_\A}]^c_{xz},\qquad [\rho_{M_\A}]^a_{xy} = ([\rho_{M_\A}]^{a^\ast}_{x^\ast y^\ast})^\ast.
\end{equation}
In particular, for UFC \(\Vec(G)\) with a finite group \(G\), the modules over a Frobenius algebra \(\A = \bigoplus_{g \in H} g\) (with multiplication \(f^\A_{fgh} = \delta_{e, fgh}\)) for a subgroup \(H \leq G\) correspond to the representations of \(H\),  and function \(\rho\) encode the representation matrix entries.
 
One can express the module function in basis form, which is depicted in figure \ref{fig:right_A_module}: The vertex connecting the edges colored by $\mathcal{A}$ and the chosen module $M_\A$ are weighted by the coefficients $[\rho_{M_\A}]_{xy}^a$ that defines the action of $\mathcal{A}$ on the module.
\begin{figure}\centering
\subfloat[]{\RModuleA}\hspace{60pt}\subfloat[]{\RModuleB}
\caption{(a) The basis representation of right-$A$ module \(M\). (b) The right-module condition: Eq. \eqref{eq:rightmodule}.}
\label{fig:right_A_module}
\end{figure}

\subsection{Anyon Condensation in the HGW Model}

The HGW model provides a finer description of anyon-condensation-induced topological phase transitions.

To describe the procedure of condensing a Lagrangian set $L$ of anyons in the HGW model, we can add to the HGW Hamiltonian a condensation term:
\[
H = H_\Fus - \lim_{\Lambda\to\infty}\sum_E P_E^\A,
\]
where $H_\Fus$ is HGW Hamiltonian of the parent topological phase, and $P^\A_E$ is a local projector acting on an edge $E$:
\[
P_E^\A = \sum_{J\in L}\quad\sum_{p,q}\pi_J^{p,q}W_E^{J;pq}\ = {}_E\ket{\A}\bra{\A}_E ,
\]
where \(W_E^{J;pq}\) are creation operators of condensed dyons, \(\pi_J^{p, q}\in\mathbb{C}\) are coefficients to make the sum a projector. In the limit \(\Lambda \to \infty\), projectors \(P_E^\A\) ensures that the new ground states $\ket{\A}_E$ are \(+1\) eigenstates of all \(P_E^\A\), which are coherent states filled with arbitrarily many condensed anyons throughout the lattice. This state can be locally represented as:
$$\ket{A}_E =\quad \ExcitedC\quad =\quad \sum_{j,k,p,q\in L_\A}\ f_{jk^\ast p^\ast}f_{kq^\ast j^\ast}\quad \ExcitedD\ ,$$
where \(E\) is an edge of the lattice, while the violet lines refer to auxiliary tails in plaquettes, which will then be contracted via topological moves and result in only one tail in each plaquette, representing the entanglements between different edges. The global ground state of the trivial topological order is the state where each edge and tail in the original lattice \ref{fig:4-8-lattice} carries Frobenius algebra object \(A\).

\begin{figure}
    \centering
    \includegraphics[height=0.3\linewidth]{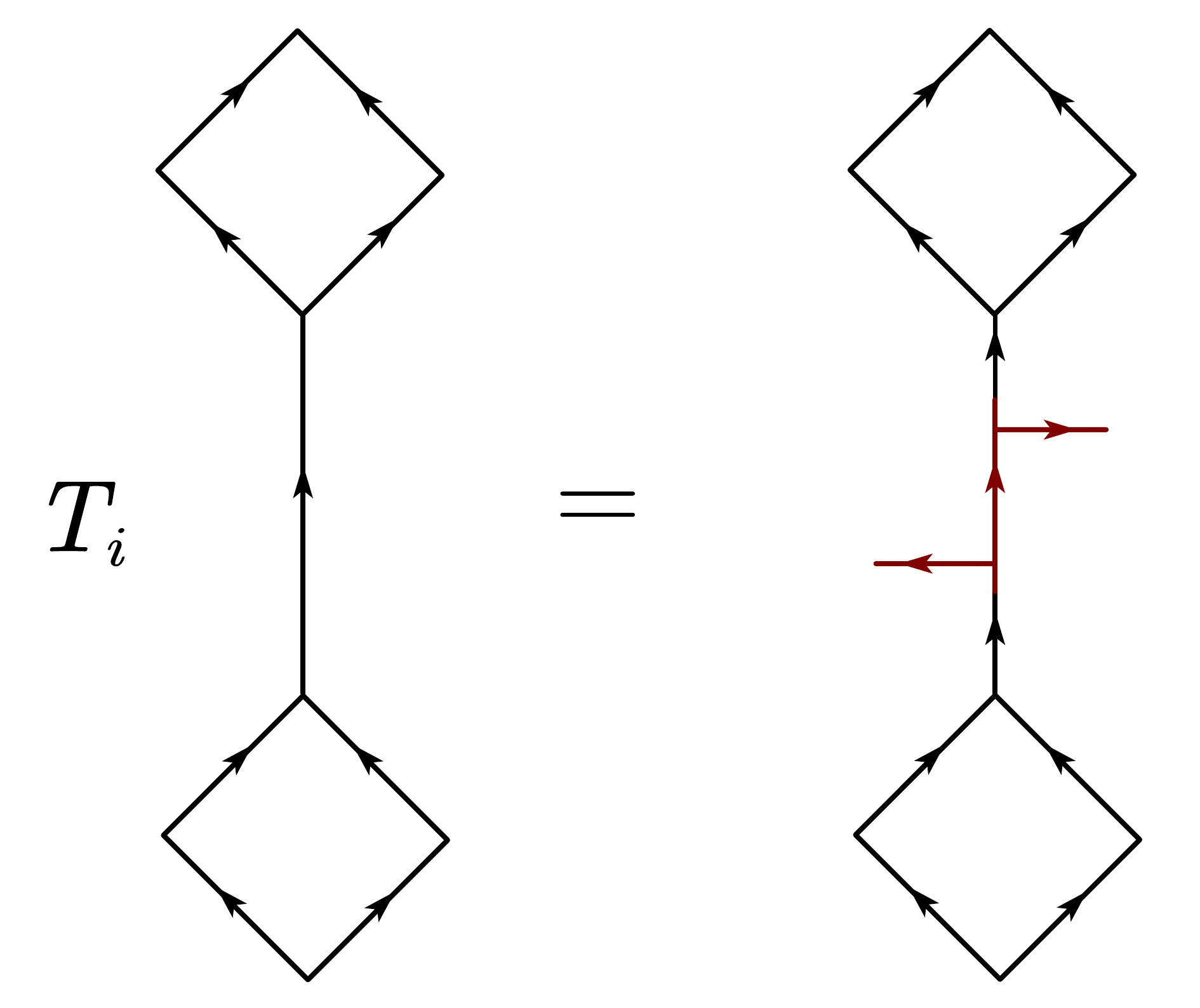}
    \caption{Anyon creation operator}
    \label{fig:anyon_creation}
\end{figure}

\begin{figure}
    \centering
    \includegraphics[height=0.3\linewidth]{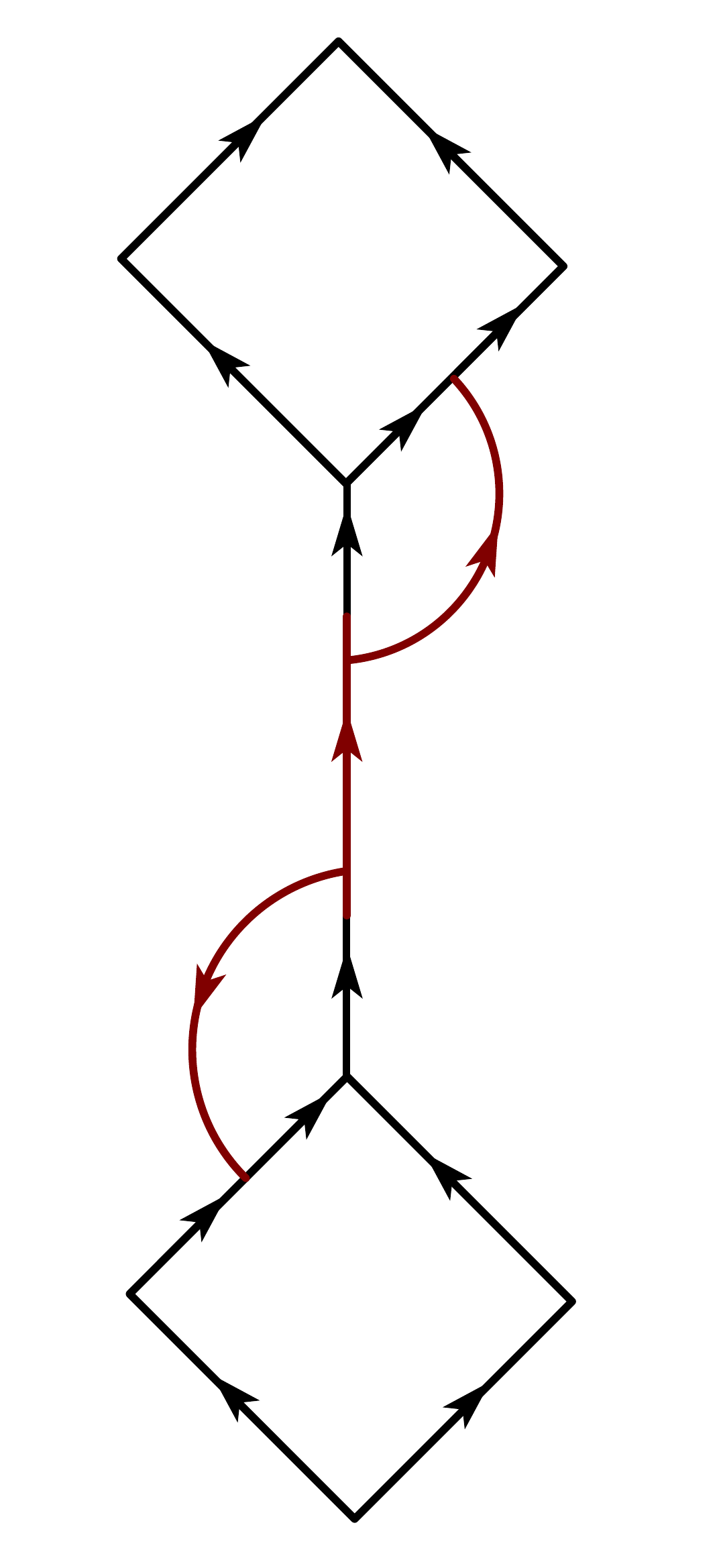}
    \caption{Anyon condensation}
    \label{fig:enter-label}
\end{figure}

In the case of the Doubled Ising model, there are two Frobenius algebras of the input Ising UFC:
\[
\A_1 = 1, \qquad [f_1]_{111} = 1,\qquad \A_2 = 1 \oplus \psi, \qquad [f_2]_{111} = [f_2]_{1\psi\psi} = 1.
\]
These two input Frobenius algebras are Morita equivalent because they have the same full center \(L = 1\bar 1\oplus\psi\bar\psi\oplus\sigma\bar\sigma.\)
The corresponding projectors are
\begin{equation}
P_E^{\A_1} = \frac{W_E^{1\bar 1;1,1} + W_E^{\psi\bar\psi;1,1} + 2W_E^{\sigma\bar\sigma,1,1}}{4} = \delta_{j_E, 0},\qquad P_E^{\A_2} = \frac{W_E^{1\bar 1;1,1} + W_E^{\psi\bar\psi;1,1} + 2W_E^{\sigma\bar\sigma,\psi,\psi}}{4}.
\label{projector}
\end{equation}

In our square-octagon lattice as shown in the main text, the anyon operator, the pair creation operator should act as shown in figure \ref{fig:anyon_creation}.  The HGW lattice is different from the {HGW} string net with these extra tails. To make contact with the original version of the strange correlator which follows from the {HGW} ground state wave-function, we would like to remove these tails. One very natural way of achieving this is to connect these lines to the square, as shown in figure \ref{fig:enter-label}, and then removing the extra bubble by F-moves.  

We note that the critical state (\ref{isingomeg}) is the eigenstate of $P_E^{\A_1} + P_E^{\A_2}$ with the largest eigenvalue.

This story can be repeated in all pair-competition in which $\A_i \subset \A_j$ in other fusion category. One can show that 
(\ref{pair_compete}) is always the eigenstate with the largest eigenvalue of the sum of projection operators, analogously defined as the Ising case above, irrespective of the choice of the common module between $\A_{i,j}$.

\section{Some Details of Various Categories}\label{appendix:data}
\subsection{The Modular Tensor Category \eqs{\mathcal{Z}(S_3)}}\label{appendix:DS3}
We would like to review here some basic data of the $\mathcal{Z}(S_3)$ category. The symmetry group $S_3=\langle x,y|x^3=y^2=e,yx=x^2y \rangle$.
The anyons of $\mathcal{Z}(S_3)$ are labeled by a pair $(W,\gamma_W)$, where $W$ is a conjugacy class of the group $S_3$, and $\gamma_W$ an irreducible representation of the centralizer of $W$. There're eight kinds of anyons in $\mathcal{Z}(S_3)$, conventionally labelled by letters $A$ through $H$. A summary of all the anyons are listed below. 

\begin{center}
\begin{tabular}{c|ccc|cc|ccc}
    \hline\hline
     & $A$ & $B$ & $C$ & $D$ & $E$ & $F$ & $G$ & $H$ \\
    \hline
    conjugacy class $W$ & \multicolumn{3}{|c|}{$\{e\}$} & \multicolumn{2}{c|}{$\{y,xy,x^2y\}$} & \multicolumn{3}{c}{$\{x,x^2\}$} \\
    \hline
    centralizer $\cong$ & \multicolumn{3}{|c|}{$S_3$} & \multicolumn{2}{c|}{$\Z_2$} & \multicolumn{3}{c}{$\Z_3$} \\
    \hline
    irrep $\gamma_W$ of centralizer  & $\bf 1$ & sign & $\bf \pi$ & $\bf 1$ & $\bf -1$ & $\bf 1$ & $\bf \omega$ & $\bf \omega^*$ \\
    \hline
    dim($\gamma_W$) & 1 & 1 & 2 & 1 & 1 & 1 & 1 & 1 \\
    \hline
    quantum dimension $d=|W|\times$ dim$(\gamma_W)$ & 1 & 1 & 2 & 3 & 3 & 2 & 2 & 2\\
    \hline
    twist $\theta$ & 1 & 1 & 1 & 1 & -1 & 1 & $e^{2\pi i/3}$ & $e^{-2\pi i/3}$\\
    \hline\hline
\end{tabular} 
\end{center}

Their fusion rules are given by

\begin{center}
\scalebox{0.7}{
\begin{tabular}{c|ccc|cc|ccc}
    \hline\hline
    $\otimes$ & $A$ & $B$ & $C$ & $D$ & $E$ & $F$ & $G$ & $H$ \\ \hline
    $A$ & $A$ & $B$ & $C$ & $D$ & $E$ & $F$ & $G$ & $H$ \\ 
    $B$ & $B$ & $A$ & $C$ & $E$ & $D$ & $F$ & $G$ & $H$ \\ 
    $C$ & $C$ & $C$ & $A\oplus B\oplus C$ & $D\oplus E$ & $D\oplus E$ & $G\oplus H$ & $F\oplus H$ & $F\oplus G$ \\ \hline
    $D$ & $D$ & $E$ & $D\oplus E$ & $A\oplus C\oplus F\oplus G\oplus H$ & $B\oplus C\oplus F\oplus G\oplus H$ & $D\oplus E$ & $D\oplus E$ & $D\oplus E$ \\ 
    $E$ & $E$ & $D$ & $D\oplus E$ & $B\oplus C\oplus F\oplus G\oplus H$ & $A\oplus C\oplus F\oplus G\oplus H$ & $D\oplus E$ & $D\oplus E$ & $D\oplus E$ \\ \hline
    $F$ & $F$ & $F$ & $G\oplus H$ & $D\oplus E$ & $D\oplus E$ & $A\oplus B\oplus F$ & $C\oplus H$ & $C\oplus G$ \\ 
    $G$ & $G$ & $G$ & $F\oplus H$ & $D\oplus E$ & $D\oplus E$ & $C\oplus H$ & $A\oplus B\oplus G$ & $C\oplus F$ \\
    $H$ & $H$ & $H$ & $F\oplus G$ & $D\oplus E$ & $D\oplus E$ & $C\oplus G$ & $C\oplus F$ & $A\oplus B\oplus H$ \\
    \hline\hline
\end{tabular}
}
\end{center}
    
There're $4$ distinct Lagrangian algebras in $\mathcal{Z}(S_3)$, labeled by the $4$ different subgroups of $S_3$. The Lagrangian algebras corresponding to each subgroup are listed in the following table.
\begin{center}
\begin{tabular}{c|c}
    \hline\hline
    subgroup $K$ & Lagrangian algebra $L$ \\
    \hline
    $\mathbb{1} $ & $A\oplus B\oplus 2C$ \\
    $\Z_2$ & $A\oplus C\oplus D$ \\
    $\Z_3$ & $A\oplus B\oplus 2F$ \\
    $S_3$ & $A\oplus D\oplus F$ \\
    \hline\hline
\end{tabular}
\end{center}

\subsection{Haagerup Fusion Category \eqs{H_3}}

The \emph{Haagerup} fusion category is a notably special category. It contains six types of simple objects, labeled by
$$1,\qquad \alpha,\qquad \alpha^2,\qquad \rho,\qquad \alpha\rho,\qquad\alpha^2\rho,$$
with the following quantum dimensions
$$d_1 = d_\alpha = d_{\alpha^2} = 1,\qquad d_\rho = d_{\alpha\rho} = d_{\alpha^2\rho} = \frac{3 + \sqrt{13}}{2}.$$
The fusion rules are
\begin{center}\begin{tabular}{||c|c|c|c|c|c||}\hline\hline
$1$ & $\alpha$ & $\alpha^2$ & $\rho$ & $\alpha\rho$ & $\alpha^2\rho$ \\ \hline
$\alpha$ & $\alpha^2$ & $1$ & $\alpha\rho$ & $\alpha^2\rho$ & $\rho$ \\ \hline
$\alpha^2$ & $1$ & $\alpha$ & $\alpha^2\rho$ & $\rho$ & $\alpha\rho$ \\ \hline
$\rho$ & $\alpha^2\rho$ & $\alpha\rho$ & $1\oplus\rho\oplus\alpha\rho\oplus\alpha^2\rho$ & $\alpha^2\oplus\rho\oplus\alpha\rho\oplus\alpha^2\rho$ & $\alpha\oplus\rho\oplus\alpha\rho\oplus\alpha^2\rho$ \\ \hline
$\alpha\rho$ & $\rho$ & $\alpha^2\rho$ & $\alpha\oplus\rho\oplus\alpha\rho\oplus\alpha^2\rho$ & $1\oplus\rho\oplus\alpha\rho\oplus\alpha^2\rho$ & $\alpha^2\oplus\rho\oplus\alpha\rho\oplus\alpha^2\rho$ \\ \hline
$\alpha^2\rho$ & $\alpha\rho$ & $\rho$ & $\alpha^2\oplus\rho\oplus\alpha\rho\oplus\alpha^2\rho$ & $\alpha\oplus\rho\oplus\alpha\rho\oplus\alpha^2\rho$ & $1\oplus\rho\oplus\alpha\rho\oplus\alpha^2\rho$ \\ \hline\hline
\end{tabular}\end{center}

\subsubsection{Frobenius Algebras and Modules}

The Haagerup category has seven Frobenius algebras divided into three Morita classes. The Haagerup fusion rules are not commutative, so the left and right modules are slightly different. We only consider the right modules here. The module function tensor component $[\rho_M^{\A}]^a_{xy}$ represents the algebra object $a\in\A$ fuses from right to module object $x\in M$ and transform it to $y\in M$, i.e., $y\in x\otimes a$. The three minimal algebras along with their modules are listed below:

\begin{enumerate}
\item The trivial Frobenius algebra $\A_0 = 1$, such that $f_1^{11}$ = 1. It has six \emph{right} modules: $M_x = x, \rho^1_{xx} = 1$, where $x$ is a Haagerup simple object.
\item The $\Z_3$ Frobenius algebra $\A_2 = 1\oplus\alpha\oplus\alpha^2$, such that $f_{a}^{bc} = N_a^{bc}$, the fusion rules. It has two simple modules:
$$M_1 = 1\oplus\alpha\oplus\alpha^2,\qquad [\rho_1]^a_{xy} = N_y^{xa},$$
$$M_{\rho} = \rho\oplus\alpha\rho\oplus\alpha^2\rho,\qquad [\rho_\rho]^a_{xy} = N_y^{xa}(-1)^{\delta_{a,\alpha^2}\delta_{x,\alpha\rho}\delta_{y,\alpha^2\rho}},$$
\item The special Frobenius algebra $\A_4 = 1\oplus\rho\oplus\alpha\rho$. The algebra multiplication $f_a^{bc}$ is too cumbersome to write here. It has four simple right modules:
$$M_1 = 1\oplus\rho\oplus\alpha\rho,\qquad [\rho_1]^a_{xy} = f_y^{xa}.$$
$$M_\alpha = \alpha\otimes M_1 = \alpha\oplus\alpha\rho\oplus\alpha^2\rho,\qquad [\rho_\alpha]^a_{xy} = \frac{f_y^{xa}}{[F^{\alpha, \alpha^2\otimes x, a}_{y}]_{x, \alpha^2\otimes y}}.$$
$$M_{\alpha^2} = \alpha^2\otimes M_1 = \alpha^2\oplus\rho\oplus\alpha^2\rho,\qquad [\rho_{\alpha^2}]^a_{xy} = \frac{f_y^{xa}}{[F^{\alpha^2, \alpha\otimes x, a}_{y}]_{x, \alpha\otimes y}},$$
$$M_\rho = \rho\oplus\alpha^2\rho\oplus\alpha\rho. \qquad {}$$

\end{enumerate}
The module function of $M_\rho$ is somewhat messy and will not be presented in the text.

The other connected Frobenius algebras can be obtained by conjugating these minimal algebras with some simple objects. The modules of the other non-minimal Frobenius algebras can be easily calculated from those of the minimal ones. We take $\A_5 = \alpha\otimes\A_4\alpha^2$ as an illustrating example. For $M_i$ any right-$\A_4$ module, $M_i\otimes\alpha^2$ must be a right-$\A_5$ module. And the module function of $M_i\otimes\alpha^2$ (on $\A_5$) differ from $M_i$ (on $\A_4$) only by some F-symbols, which can be determined immediately.

\section{Symmetry Preserving RG of the Strange Correlator}
\label{app:symRG}
\subsection{The Strange Correlator }
\begin{figure}
    \centering
    \includegraphics[width=0.9\linewidth]{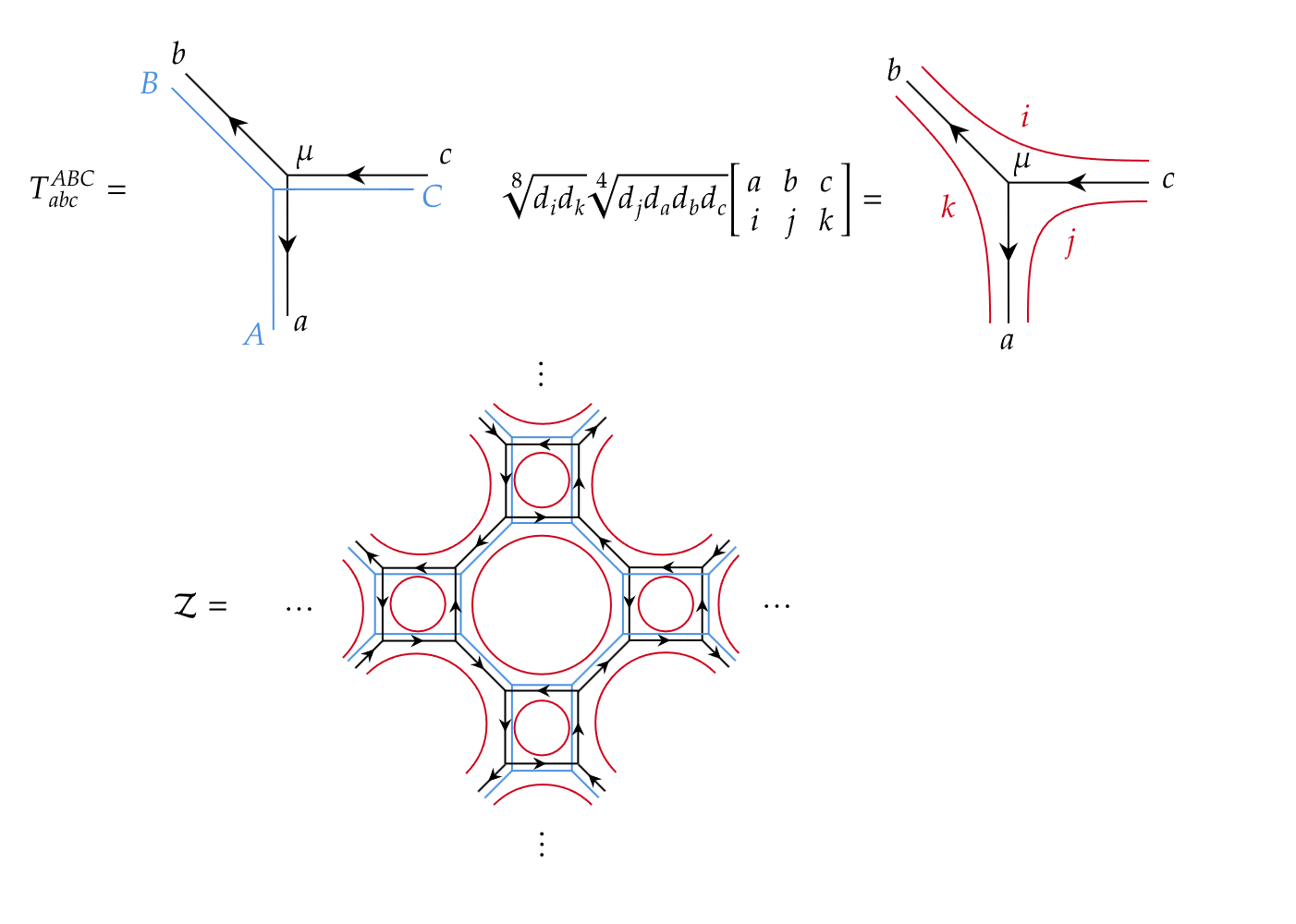}
    \caption{The strange correlator $\langle\Omega|\Psi\rangle$ is constructed from unit cells composed of tensors $T$, the tetrahedron symbols and weights involving quantum dimensions. Blue indices are summed within $\langle\Omega|$, red indices within $|\Psi\rangle$, while the black indices are shared and summed between $\langle\Omega|$ and $|\Psi\rangle$.}
    \label{fig:strange correlator}
\end{figure}

A strange correlator\footnote{There are other recent applications of the strange correlator in constructing 1+1 D phases. See for example \cite{cenjie,xu2020tensor,Zhang:2023ldq,Lu:2025rwd}.} partition fucntion $Z$ is the overlap between a string-net ground state $|\Psi\rangle$ and a tensor network state $|\Omega\rangle$ (see Figure \ref{fig:strange correlator}), 
\begin{equation}
     Z = \sum_{\{a\}} \langle \Omega|\Psi\rangle.
\end{equation}
$|\Omega\rangle$ is composed of local tensors $T_{abc}^{ABC}$ . The black lines (indices $a,b,c$) are labeled by simple objects that contract with $|\Psi\rangle$. The fusion rule at each vertex is $a\otimes b=c$, as indicated by the arrows.\footnote{For simplicity, we assume all tensor elements are real numbers, which holds for all tensors used in this work.  If complex numbers were used, neighboring tensors would be complex conjugates of each other, arranging the tensor network in a checkerboard pattern.  This arrangement stems from the unitarity of the module tensors and ensures the partition function remains real.  Generalization to arbitrary F-symbols is possible, though considerably more complex, and is left for future work.}  We denote these open legs by a set of orthonormal basis state $|\{a\} \rangle$. The blue lines (indices $A,B,C$) are auxilary legs contracted within $|\Omega\rangle$. Each auxiliary leg has a bond dimension $d_\chi$, so $A,B,C\in \{1,2,\dots, d_\chi\}$. The symbol $\mu$ denote different local tensors.  The state is formally written as
\begin{equation}
    |\Omega\rangle =\sum _{\{A\}}\prod _{\mu } T_{abc}^{ABC }( \mu ) |\{a\} \rangle.
\end{equation}

The state $|\Psi\rangle$ can also be represented as a tensor network (see figure \eqref{fig:strange correlator}). Its ingredients are tetrahedron symbols, which are related to the $6j$- symbols of the fusion category:
\[\Tsym{a}{b}{c}{i}{j}{k} = \frac{1}{\sqrt{d_ed_f}}[F^{abi}_j]_{ck} = G^{b^\ast a^\ast c}_{ji^\ast k}.\]
The red lines (indices $i,j,k$), connecting across two tensor legs, are summed over all simple objects in the category. The black lines are, again, open legs that contract with $|\Omega\rangle$. The state is thus 
\begin{equation}
    |\Psi \rangle =\sum _{\{i\}}\sqrt{d_{a}} d_{i}\prod _{\mu } \Tsym{a}{b}{c}{i}{j}{k}( \mu ) |\{a\} \rangle.
\end{equation}

In total, the partition function is 
\begin{equation}\label{eq:ZinTot}
     Z =\sum _{\{a\},\{A\},\{i\}}\prod _{\mu}z_{abc,ijk}^{ABC}(\mu).
\end{equation}
where we have split the weight of quantum dimensions into local parts as
\begin{equation}\label{eq:zinTot}
    z_{abc,ijk}^{ABC}(\mu)=\sqrt[4]{d^2_id_jd^2_kd_ad_bd_c} T_{abc}^{ABC\ }( \mu ) \Tsym{a}{b}{c}{i}{j}{k}( \mu )
\end{equation}

In this paper, a unitary and tetrahedron-symmetric gauge is chosen for the tetrahedron symbol\footnote{We have made the generalization to arbitrary $F$ symbols possible but much more complicated. We leave this to future work.}. The unitarity implies:
\begin{equation}
\label{eq:unitarity}
    \sum_f d_ed_f\Tsym{a}{b}{e}{c}{d}{f}{\Tsym{a}{b}{e'}{c}{d}{f}}^\ast =\delta_{ee'}N_{ab}^eN_{ec}^d.
\end{equation}

The tetrahedron symmetry implies
\begin{equation}
    \Tsym{a}{b}{e}{c}{d}{f}=\Tsym{a^*}{e}{b}{c}{f}{d}=\Tsym{e}{b^*}{a}{f}{d}{c}=\Tsym{a}{f}{d}{c^*}{e}{b}.
\end{equation}

As a consequence of the tetrahedron symmetry, the pentagon equation of the $F$ symbol can be written in the following form that will be used in the RG,
\begin{equation}
\label{eq:tet sym pen}
    \Tsym{a}{b}{e}{c}{d}{f}\Tsym{a}{b'}{e'}{c'}{d}{f}=\sum_x d_x \Tsym{a}{b}{e}{x}{e'}{b'}\Tsym{e'}{x}{e}{c}{d}{c'}\Tsym{b'}{x}{b}{c}{f}{c'}.
\end{equation}

\subsection{Categorical Symmetric RG Procedure} \label{sec:symRG}
\begin{figure}
    \centering
    \includegraphics[width=0.8\linewidth]{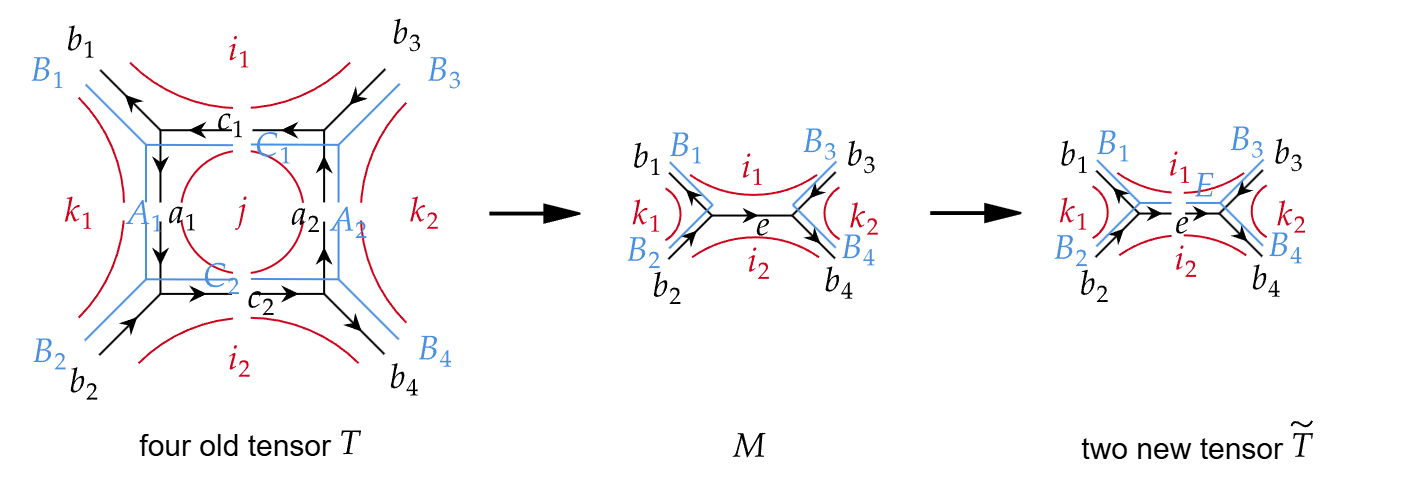}
    \caption{In each RG step, four old tensors $T$ are combined into an intermediate tensor $M$. $M$ is then decomposed into two new tensors $\tilde T$ via SVD, practically implemented using an algorithm like loop-TNR. The red legs suggest the underlying tetrahedron symbols and are not included as a part of the tensors $T$ or $M$.}
    \label{fig:tnr}
\end{figure}

A key virtue of the strange correlator construction is that $F$-moves and within the tensor network state $|\Omega\rangle$ leave the partition function $Z$ invariant provided the underlying tetrahedron symbols are deformed accordingly. Consequently, different lattice discretizations are related by Pachner moves.  This feature enables the following RG procedure.  In each RG step, a block of four local tensors is transformed into two new ones (see figure \ref{fig:tnr})

We begin by rewriting the product of four tetrahedron symbols as shown below. (\ref{eq:tet sym pen}) is used to go from the first to the second line, and (\ref{eq:unitarity}) is used from the third to the fourth line, where a sum over $\delta_{e_1e_2}$ is performed.

\begin{align}\label{four f sym}
    &\sum_{j} d_j \Tsym{a_1}{b_1}{c_1}{i_1}{j}{k_1}\Tsym{a_1}{b_2}{c_2}{i_2}{j}{k_1}\Tsym{a_2}{b_3}{c_1}{i_1}{j}{k_2}\Tsym{a_2}{b_4}{c_2}{i_2}{j}{k_2} \notag\\
    =&\sum_{j} d_j \sum_{e_1}d_{e_1}\Tsym{a_1}{b_2}{c_2}{e_1}{c_1}{b_1} \Tsym{c_1}{e_1}{c_2}{i_2}{j}{i_1}\Tsym{b_1}{e_1}{b_2}{i_2}{k_1}{i_1} \sum_{e_2}d_{e_2}\Tsym{a_2}{b_4}{c_2}{e_2}{c_1}{b_3} \Tsym{c_1}{e_2}{c_2}{i_2}{j}{i_1}\Tsym{b_3}{e_2}{b_4}{i_2}{k_2}{i_1} \notag\\
    =&\left ( \sum_{j} d_j \Tsym{c_1}{e_1}{c_2}{i_2}{j}{i_1}\Tsym{c_1}{e_2}{c_2}{i_2}{j}{i_1} \right ) \sum_{e_1,e_2}d_{e_1}d_{e_2}\Tsym{a_1}{b_2}{c_2}{e_1}{c_1}{b_1}\Tsym{b_1}{e_1}{b_2}{i_2}{k_1}{i_1} \Tsym{a_2}{b_4}{c_2}{e_2}{c_1}{b_3} \Tsym{b_3}{e_2}{b_4}{i_2}{k_2}{i_1} \notag\\
    =&\sum_{e}\sqrt{d_e} \Tsym{b_1}{e}{b_2}{i_2}{k_1}{i_1}  \Tsym{b_3}{e}{b_4}{i_2}{k_2}{i_1}\left (\sqrt{d_e}\Tsym{a_1}{b_2}{c_2}{e}{c_1}{b_1}\Tsym{a_2}{b_4}{c_2}{e}{c_1}{b_3}\right ).
\end{align}

We combine the part in brackets with $T$ to get an intermediate tensor $M$,

\begin{align}
\label{eq:M tensor}
M^{B_1B_2B_3B_4}_{b_1b_2b_3b_4e}=\hspace{-0.36cm}\sum_{a_1,a_2,c_1,c_2}^{A_1,A_2,C_1,C_2}\hspace{-0.33cm} \sqrt{d_{a_1}d_{a_2}d_{c_1}d_{c_2}d_e} \hspace{-0.1cm}\Tsym{a_1}{b_2}{c_2}{e}{c_1}{b_1}\Tsym{a_2}{b_4}{c_2}{e}{c_1}{b_3} T^{A_1B_1C_1}_{a_1b_1c_1}T^{A_1B_2C_2}_{a_1b_2c_2}T^{A_2B_3C_1}_{a_2b_3c_1}T^{A_2B_4C_2}_{a_2b_4c_2} .
\end{align}

A new tensor $\tilde T$ is then obtained by applying a singular value decomposition (SVD) to the tensor $M$,

\begin{equation}\label{eq:svd M}
   M^{B_1B_2B_3B_4}_{b_1b_2b_3b_4e}= \sum_E \tilde T_{b_1eb_2}^{B_1EB_2} \tilde T_{b_3eb_4}^{B_3EB_4}.
\end{equation}
In practice, the bond dimension of the new index $E$ must be truncated below a certain $d_\chi$. We employ the loop-TNR procedure \cite{Yang_2017} with gradient descents to search for an optimal approximation for this decomposition. In total, we have
\begin{align}
&\sum_{e}^E\sqrt{d_e} \Tsym{b_1}{e}{b_2}{i_2}{k_1}{i_1}  \Tsym{b_3}{e}{b_4}{i_2}{k_2}{i_1}  \tilde T_{b_1eb_2}^{B_1EB_2} \tilde T_{b_3eb_4}^{B_3EB_4}=\hspace{-0.33cm}\sum_{a_1,a_2,c_1,c_2,j}^{A_1,A_2,C_1,C_2}\hspace{-0.33cm} \notag\sqrt{d_{a_1}d_{a_2}d_{c_1}d_{c_2}}d_j \\
     &\Tsym{a_1}{b_1}{c_1}{i_1}{j}{k_1}\Tsym{a_1}{b_2}{c_2}{i_2}{j}{k_1}\Tsym{a_2}{b_3}{c_1}{i_1}{j}{k_2}\Tsym{a_2}{b_4}{c_2}{i_2}{j}{k_2}T^{A_1B_1C_1}_{a_1b_1c_1}T^{A_1B_2C_2}_{a_1b_2c_2}T^{A_2B_3C_1}_{a_2b_3c_1}T^{A_2B_4C_2}_{a_2b_4c_2}.
\end{align}

Now we fully recover the original form of the strange correlator. Thus, the categorical symmetry, embedded in the underlying tetrahedron symbols, is preserved throughout the RG step. The fusion rules at the new vertices are $b_1\otimes e=b_2$ and $b_3\otimes e=b_4$. One can alternatively obtain $b_2\otimes e'=b_1$ and $b_4\otimes e'=b_3$ by a different application of the pentagon equation in \ref{four f sym}.

In our typical setup, a unit cell comprises two local $T$ tensor. The initial tensor for the RG flow can be considered as the tensor $M$ in (\ref{eq:M tensor}) and (\ref{eq:svd M}). Consequently, an initial decomposition step of (\ref{eq:svd M}) is performed. For unit cells corresponding to (\ref{many_condensates}) where each $\langle\unitcellColorAlgMod{\A_i}{M_{\A_i}}|$ is weighted by parameter $r_i$, the initial tensor $M$ is
\begin{equation}
\label{eq:ini tensor}
   M_{b_1b_2b_3b_4e}=\sum_i r_i [\rho_{M_{\A_i}}]_{b_1b_2}^e ([\rho_{M_{\A_i}}]_{b_3b_4}^e)^\ast
\end{equation}

We omit the auxilary legs $B_i$ in $M$. For general initial tensors, $B_i$ and possibly also $E$ account for multiplicities if a simple object $b_i$ or $e$ appear multiple times in  $M_\A$ or $\A$. In present cases, the multiplicity is no more than $1$, so the initial tensor has trivial auxiliary legs. During the RG procedure, their effective bond dimensions can increase, reflecting the growing entanglement among the local tensors. 

\subsection{Critical Strange Correlator of the Ising Category}
\label{sec:critIsing}
As an example, we explicitly write down the local transfer matrix for competing Ising anyons. The critical unit cell (\ref{isingomeg}) reads

\[M_{\sigma\sigma\sigma\sigma 1}=(\sqrt{2}+1)/2,M_{\sigma\sigma\sigma\sigma \psi}=1/2. \]

From equation (\ref{eq:zinTot}), the local transfer matrix for a unit cell is 
\begin{equation}
    z_{b_1b_2b_3b_4e,i_1i_2k_1k_2}=\sqrt[4]{d_e^2d_{b_1}d_{b_2}d_{b_3}d_{b_4}d_{i_1}d_{i_2}d_{k_1}d_{k_2}}M_{b_1b_2b_3b_4e}\Tsym{b_1}{e}{b_2}{i_2}{k_1}{i_1}  \Tsym{b_3}{e}{b_4}{i_2}{k_2}{i_1}
\end{equation}
Here $b_1=b_2=b_3=b_4=\sigma$. We omit $b_i$ and sum over $e$ for $z_{i_1i_2k_1k_2}=\sqrt{2}\sum_e z_{\sigma\sigma\sigma\sigma e,i_1i_2k_1k_2}$:
\begin{subequations}\label{eq:zofIsing}
\begin{align} 
    z_{\sigma\sigma11}=\sqrt 2+1, z_{\sigma\sigma\psi\psi}=\sqrt 2+1, z_{\sigma\sigma1\psi}=1, z_{\sigma\sigma\psi1}=1,\\
    z_{11\sigma\sigma}=\sqrt 2+1, z_{\psi\psi\sigma\sigma}=\sqrt 2+1, z_{1\psi\sigma\sigma}=1, z_{\psi1\sigma\sigma}=1.
\end{align}
\end{subequations}

If we interpret $1$ and $\psi$ as spin up and down, each line in (\ref{eq:zofIsing}) gives one independent copy of the local weights $e^{2\beta}=\sqrt 2+1$ for a classical Ising model with the critical inverse temperature $\beta=\frac{1}{2}\textrm{ln}(\sqrt2+1)$. The $\sigma$ acts as a placeholder to keep these two copies disentangled.

We thus conclude that the strange correaltor with the critical unit cell (\ref{isingomeg}) is a partition functions of two copies of critical Ising.

\subsection{Computing the Phase Diagram}
\begin{figure}
    \centering
    \includegraphics[width=0.7\linewidth]{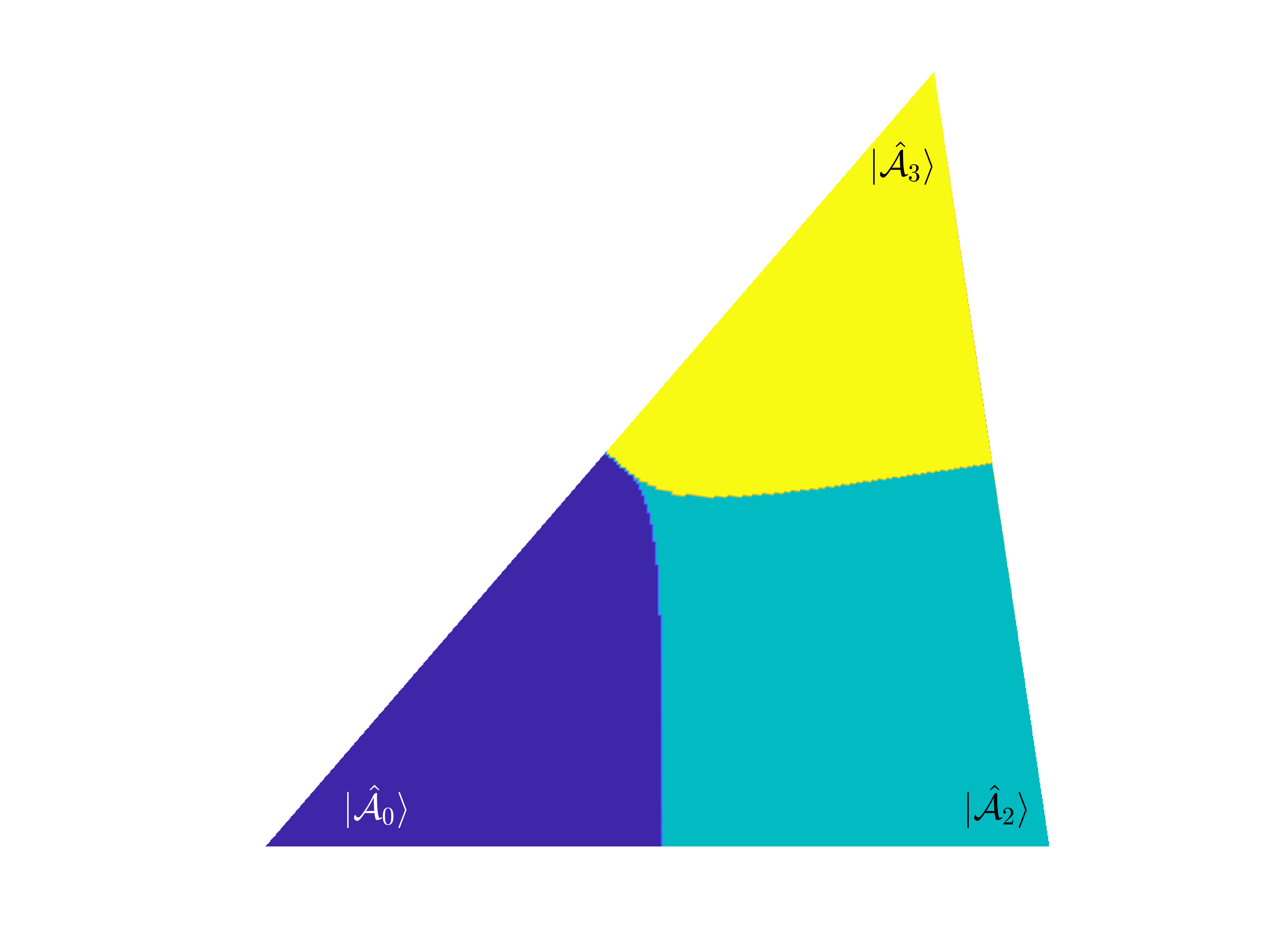}
    \caption{Ternary phase diagram for the interpolation of algebras $\A_{0,2,3}$ under module $2$ of the $A_5$ category. Different colors correspond to different converged fixed-point tensors, computed with $d_\chi=1$. This illustrates the role of the fixed-point tensor as an order parameter and demonstrates the precision of our categorical symmetric RG framework even at a small bond dimension.  }
    \label{fig:A5 order}
\end{figure}

It can be verified that if the initial tensor (\ref{eq:ini tensor})corresponds to a single algebra, the local tensor $T$ flows under RG to the corresponding Frobenius tensor $f^{\A}$ in one RG step, and that $f^{\A}$ is then a fixed-point of the RG equations. In general cases, the local tensor of any gapped state will gradually flow to one of the $f^{\A}$, which is different for each phase. The converged fixed-point tensor thus acts as an order parameter distinguishing different phases. Phase diagrams constructed using these fixed-point tensors, even with a minimal bond dimension $d_\chi=1$, show remarkable accuracy.  This highlights the efficiency of our algorithm in mapping phase diagrams.  For instance, figure \ref{fig:A5 order} shows a ternary phase diagram generated with $d_\chi=1$ demonstrating excellent precision. 

The central charge and operator spectrum can be extracted from the transfer matrix constructed from the converged RG tensors.  Among various available techniques, we adopt a straightforward method described in  for the results presented in \cite{gu2009tensor}.  Significant scope remains for enhancing the precision of central charge calculations.  In this work, the bond dimension $d_\chi$ used for central charge calculations ranges from $4$ to $10$, depending on the specific underlying category.  All computations were performed on a Tesla V100 GPU with $32$ GB of memory.  Typically, the computed central charge stabilizes within $4$ to $6$ RG steps, requiring approximately $5$ to $15$ seconds per run.  

The central charge can also be plotted as phase diagram if the phase transitions are all CFTs. For first-order phase transitions,  the central charge is expected to be zero.  However, our current method for calculating the central charge may yield small, non-zero residual values at first-order transitions after a moderate number of RG steps.  Beyond this number of steps, the calculated central charge for genuine CFTs also tends to degrade from its stable value.  To provide the most reliable estimates for the CFTs, we present results obtained before this degradation, which may leave some numerical artifacts (small non-zero $c$) at the first-order phase boundaries in figures such as \ref{fig:H3 fro 456} and \ref{fig:H3 fro 02456}. This residual artifact at first-order transitions relates to the algorithm's slow convergence near a boundary between basins of attraction of different fixed-point tensors; it typically vanishes with further RG steps.  We are currently implementing an improved algorithm for central charge calculation, and updated results will be reported in a future publication. 

\section{Summary Tables of Phase Boundaries}
\label{app:summary tables}
We list all phase transitions constructed in this paper in the tables below. The central charges for (possibly) new CFT candidates are numerical values, suggested by an $\approx $ sign. We also color them red for clarity.

For the competetion between $\A_i \subset \A_j$ with all module funcion in $\A_i$ shared, the resulting lattice model is independent of the specific module we choose. The listed modules $M$ are the ones used in numerics and many of them are changeable.

\begin{center}\begin{tabular}{|c|c|c|c|c|c|c|}
\hline
Cat & $Z_2$ & $Z_3$ &  \multicolumn{3}{c|}{$Z_4$} & {$Z_5$} \\ \hline
$\A_i$ & $0$ & $0$ & \multicolumn{2}{c|}{$0$} & {$0 \oplus 2$} & $0$ \\ \hline
$\A_j$ & $0\oplus 1$ & $0\oplus 1\oplus 2$ & $0\oplus 2$  & \multicolumn{2}{c|}{$0\oplus 1\oplus 2\oplus 3$} & $0\oplus 1\oplus 2\oplus 3\oplus 4$ \\ \hline
$M$ & $0\oplus 1$ & $0\oplus 1\oplus 2$ & $0\oplus 2$  & \multicolumn{2}{c|}{$0\oplus 1\oplus 2\oplus 3$}& $0\oplus 1\oplus 2\oplus 3\oplus 4$ \\ \hline
$D_\textrm{unit}$ &$2 $&$ 3 $&$ 2 $& $4$ & $4$ & $5$ \\ \hline
Type & \multicolumn{5}{c|}{Second order} &First order  \\ \hline  
CFT & \makecell{Ising\\ $c=0.5$} & \makecell{3-Potts \\ $c=0.8$}& \makecell{Ising\\ $c=0.5$} & \makecell{4-Potts \\ $c=1$}  & \makecell{Ising\\ $c=0.5$} &$-$  \\ \hline  
\end{tabular}
\end{center}
\vspace{0.4cm}

\begin{center}
\begin{tabular}{|c|c|c|c|c|c|c|c|c|}
\hline
Cat & $A_2$ & $A_3$ &  \multicolumn{6}{c|}{$A_4$}  \\ \hline
$\A_i$ & $0$ & $0$ & \multicolumn{3}{c|}{$0$} & \multicolumn{2}{c|}{$0 \oplus 2$} & $0 \oplus 3$ \\ \hline
$\A_j$ & $0\oplus 1 $ & $0\oplus 2$ & $0\oplus 2$ & $0\oplus 3$ & $0\oplus 1\oplus 2\oplus 3$ & $0\oplus 3$& \multicolumn{2}{c|}{$0\oplus 1\oplus 2\oplus 3$} \\ \hline
$M$ & $0\oplus 1 $ & $1$ & $1$ & $0\oplus 3$ & \multicolumn{4}{c|}{$1\oplus 2$}  \\ \hline
$D_\textrm{unit}$ &$2 $&$ 2 $&$ 2 $& $2$ & $4$ & $3$ & $4$ &$4$ \\ \hline
Type & \multicolumn{8}{c|}{Second order}  \\ \hline 
CFT & \multicolumn{2}{c|}{\makecell{Ising\\ $c=0.5$}} & \makecell{tri-Ising\\ $c=0.7$}  & \makecell{Ising\\ $c=0.5$} & \multicolumn{2}{c|}{\makecell{Ising $\otimes$ tri-Ising\\ $c=1.2$}} & \makecell{Ising\\ $c=0.5$} & \makecell{tri-Ising\\ $c=0.7$} \\ \hline  
\end{tabular}
\end{center}
\vspace{0.4cm}

\begin{center}
{
\begin{tabular}{|c|c|c|c|c|c|c|c|}
\hline
Cat & \multicolumn{7}{c|}{$A_5$}  \\ \hline
$\A_i$ & \multicolumn{3}{c|}{$0$} & \multicolumn{2}{c|}{$0 \oplus 4$} & \multicolumn{2}{c|}{$0 \oplus 2$} \\ \hline
$\A_j$ & $0\oplus 2$ & $0\oplus 4$  & $0\oplus 2\oplus 4$  & $0\oplus 2$ &  \multicolumn{3}{c|}{$0\oplus 2\oplus 4$} \\ \hline
$M$ & $1$ & $2$  & $2$  & $1\oplus 3$ & $2$& $1\oplus 3^*$ & $1\oplus 3^{**}$ \\ \hline
$D_\textrm{unit}$ &$2 $&$ 2 $&$ 3 $& $3$ & $3$ & $4$ & $4$ \\ \hline
Type & \multicolumn{6}{c|}{Second order} & First order \\ \hline  
CFT & \makecell{3-Potts \\ $c=0.8$} & \makecell{Ising\\ $c=0.5$} & \makecell{4-Potts\\ $c=1$} & \makecell{Ising $\otimes$ 3-Potts\\ $c=1.3$} & \makecell{Ising\\ $c=0.5$} & \makecell{3-Potts \\ $c=0.8$}  & $-$   \\ \hline  
\multicolumn{8}{l}{\small$^*$ uses a connected module function while $^{**}$ uses a disconnected one for the algebra $0\oplus2$.}\\ 
\end{tabular}
}\end{center}
\vspace{0.4cm}

\begin{center}\begin{tabular}{|c|c|c|c|c|c|c|c|}
\hline
Cat &   \multicolumn{7}{c|}{$H_3$}  \\ \hline
$\A_i$ & \multicolumn{3}{c|}{$\A_0$} & \multicolumn{2}{c|}{$\A_1$} & {$\A_2$}& {$\A_{4,5,6}$} \\ \hline
$\A_j$ & {$\A_1$} & {$\A_2$}& {$\A_{4,5,6}$} & {$\A_2$}& {$\A_{4,5,6}$} & {$\A_{4,5,6}$} & {$\A_{(5,6),(4,6),(4,5)}$} \\ \hline
$M$ & \multicolumn{7}{c|}{$\rho \oplus \alpha \rho \oplus \alpha^2 \rho$} \\ \hline
$D_\textrm{unit}$ & $4$ & $3$ & $3$ & $6$ & 6 & 5 & 5 \\ \hline
Type & \multicolumn{3}{c|}{Second order}& \multicolumn{4}{c|}{First order}  \\ \hline 
CFT & ${\color{red}c\approx2.0}^*$ & \makecell{3-Potts \\ $c=0.8$} & ${\color{red}c\approx1.3}$ & \multicolumn{4}{c|}{$-$ }  \\ \hline  
\multicolumn{7}{l}{\small$^*$ is possibly the CFT that has been observed in \cite{Vanhove:2021zop,Huang:2021nvb}. See section \ref{sec:Haagerup} for details.}\\ 
\end{tabular}
\end{center}

\vspace{0.4cm}
\begin{center}\begin{tabular}{|c|c|c|c|c|c|c|c|c|}
\hline
Cat &   \multicolumn{8}{c|}{$H_3$}  \\ \hline
$\A_i$ & {$\A_0$} & {$\A_0$} &{$\A_0$} & {$\A_1$}& {$\A_{1}$} & {$\A_{0}$} & {$\A_{2}$} & {$\A_4$} \\ \hline
$\A_j$ & {$\A_1$} & {$\A_2$}& {$\A_1$}& {$\A_2$} &\multicolumn{3}{c|}{{$\A_{4,5,6}$}}& {$\A_5$}\\ \hline
$\A_k$ & {$\A_2$} & {$\A_{4,5,6}$}& {$\A_{4,5,6}$}& {$\A_{4,5,6}$}&\multicolumn{3}{c|}{{$\A_{(5,6),(4,6),(4,5)}$}} & {$\A_6$}\\ \hline
$M$ & \multicolumn{8}{c|}{$\rho \oplus \alpha \rho \oplus \alpha^2 \rho$} \\ \hline
$D_\textrm{unit}$ & $6$ & $5$ & $6$ & $8$ & 8 & 5 & 7 & $7$\\ \hline
Type & \multicolumn{3}{c|}{Second order} & \multicolumn{2}{c|}{First order} & \multicolumn{3}{c|}{Second order}  \\ \hline 
CFT & ${\color{red}c\approx1.8}$  & ${\color{red}c\approx1.8}$ & ${\color{red}c\approx2.1}$& $-$ & $-$& ${\color{red}c\approx1.8}$ & ${\color{red}c\approx2.1}$ & ${\color{red}c\approx2.5}$\\ \hline  
\end{tabular}
\end{center}
The algebras of $H_3$ are listed below for reference:
\begin{subequations}\nonumber
\begin{align} 
& \mathcal{A}_0 = 1, \,\, \A_1  = 1\oplus \rho \oplus \alpha \rho \oplus \alpha^2 \rho, \,\,\A_2 = 1 \oplus \alpha \oplus \alpha^2 \,\,\\
& \A_4 = 1 \oplus \rho \oplus \alpha \rho, \,\, \A_5 = 1 \oplus \alpha \rho \oplus \alpha^2 \rho, \,\,  \A_6 = 1 \oplus  \rho \oplus \alpha^2 \rho. 
\end{align}
\end{subequations}

\section{The Effect of the Collection of Objects \eqs{M_{\A_i} \neq M_{\A_j}} in the interpolation in a unit cell}
\label{app:entangle}
In section \ref{sec:puddle}, we emphasized that to deconstruct the competition of condensates in the global lattice into competitions within the unit cells, it is crucial to color the slanted edges in the unit cell by module $M_{\A_{i,j}}$ such that these modules contain exactly the same set of objects, even though generically module functions could differ. 

Here, we illustrate with examples that when $M_{\A_i}$ and $M_{\A_j}$ contain different objects, condensates in neighbouring unit cells would be entangled non-trivially. The equilibrium constructed within one unit cell in (\ref{pair_compete}) would be disrupted by its neighbors. 

As an illustration, let us consider the Ising model as in section \ref{sec:ising}. We again consider the competition between $\A_1= 1$ and $\A_2 = 1 \oplus \psi$ with however the following interpolation
\begin{equation}
     \big\langle \unitcell_{(\mathcal{A}_1, \sigma),(\A_2, 1\oplus 
     \psi)}\big| =\langle \widehat \A_1 |_{\sigma}+ r \langle \widehat \A_2 |_{1\oplus \psi},
\end{equation}
where we have chosen $M_{\A_1} = \sigma$ and $M_{\A_2} = 1\oplus \psi$.
Now consider assembling the global $\ket{\Omega}$ from the unit cell, as indicated in figure \ref{fig:4-8-lattice}.
It is clear that for two neighbouring unit cells connected by a slanted edge in which the module runs, when $\A_1$ appears in one unit cell, all of its neighbours must also be $\A_1$, and as a result all unit cells are forced into $\A_1$. Similarly when $\A_2$ appears in one unit cell, all unit cells in the global state are simultaneousely $\A_2$. Each of these corresponds to a global RG fixed point state of a globally condensed phase expressible in terms of (\ref{fixed_point}) 

There is thus complete correlation between all unit cells, and the equilibrium intended for a single unit cell in equation (\ref{pair_compete}) did not correspond to the phase transition point as $r$ is varied. Instead, it is found that the critical coupling occurs at $r_c = 2^{1/4}=\sqrt{\frac{d_{M_{\A_2}}}{d_{M_{\A_1}}}}$. 
One can readily check that this is equivalent to the equal weight summation of the $\A_1$ and $\A_2$ condensate, with normalisation defined by the {\it global lattice} instead of the unit cell. Also, given that the entire global lattice now plays the role of the unit cell, the dimension of phase space is huge and not surprisingly the phase transition point is first order. 

When the module objects are common between $M_{\A_i}$ and $M_{\A_j}$, one can readily show that every module object in one unit cell to be fed to the next neighbouring cell would be acted upon by every member of the condensate in the linear combination of condensates in the next cell. Therefore each unit cell forms a common background to all the condensates appearing in its neighbour cell. As a result equilibrium achieved in a unit cell is not disrupted by its neighbour. The unit cells are essentially factorised as desired. 

\section{The Phase Diagram of Ashkin-Teller Model}\label{sec:AT}
The isotropic Ashkin–Teller model on a two-dimensional square lattice is defined by the Hamiltonian
\[
H \;=\; -\,\beta \sum_{\langle i j\rangle}\bigl(\sigma_i \sigma_j + \tau_i \tau_j\bigr)
\;-\;\alpha \sum_{\langle i j\rangle}\sigma_i \sigma_j \,\tau_i \tau_j
\]
where 
\begin{itemize}
  \item \(\sigma_i,\tau_i=\pm1\) are two independent Ising spins on each site \(i\),
  \item \(\beta\) is the nearest-neighbour Ising coupling (same for both species),
  \item \(\alpha\) is the four-spin interaction coupling.
\end{itemize}

There exists a self-dual line
\begin{equation}
    e^{-2\alpha} = \sinh{2\beta}
\end{equation}
which extends from the decoupled critical Ising point $(\alpha=0, \beta=\beta_c^{\text{Ising}}\approx0.44)$ to the tricritical point $(\alpha=\beta=\frac{1}{4}\log 3\approx0.2747)$.  This line is the second order phase transition line between the paramagnetic phase and the ferromagnetic phase of Ashkin-Teller model.
Along this line the continuum limit is a compactified boson CFT with $c=1$ and continuously varying radius.  For $\alpha > \frac{1}{4}\log 3$ the phase transition line bifurcates into two distinct Ising-critical branches, although their precise separation remains analytically undetermined.

\begin{figure}
    \centering
    \includegraphics[width=\linewidth]{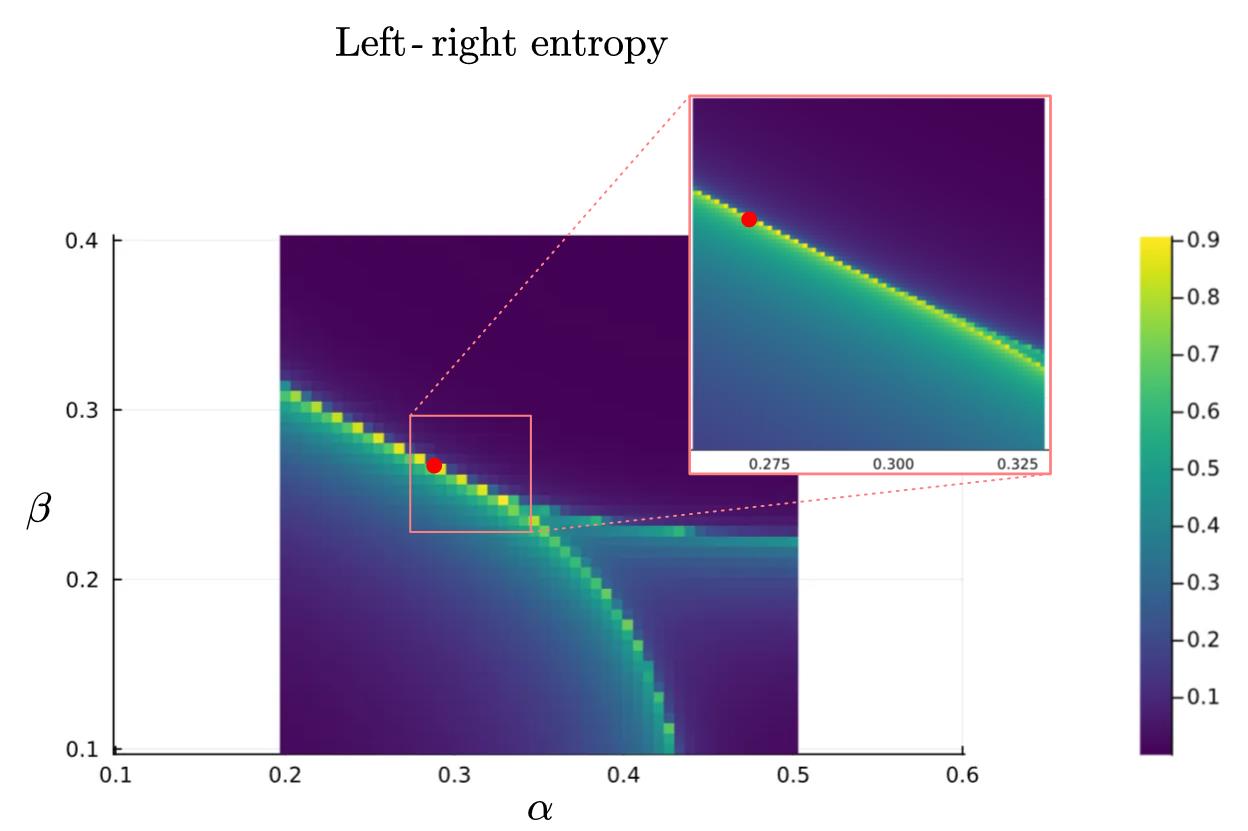}
    \caption{Heat-map of the “left–right” bipartite entropy $S_{LR}$ of the Ashkin–Teller model as a function of the two coupling parameters $\alpha$ (horizontal axis) and $\beta$ (vertical axis). Here $\alpha$ is the four-spin coupling and $\beta$ the nearest-neighbour Ising coupling. The inset at upper right zooms in on the boxed region around the tricritical point. The red point $(\alpha=\beta=\frac{1}{4}\log 3)$ is the analytic tricritical point. This phase diagram is obtained using a bond dimension $\chi=50$ in iMPS. As shown in the inset, in a small region beyond the tricritical point the two Ising branches almost coalesce and are numerically indistinguishable in the current precision.}
    \label{fig:AT_phase_diagram}
\end{figure}

We compute the phase diagram numerically (Fig.\,\ref{fig:AT_phase_diagram}) by evaluating the left-right bipartite entanglement entropy $(S_{LR})$ of the dominant eigenstate of the Ashkin–Teller transfer matrix on an infinite cylinder.  In the vicinity of the tricritical point ($\alpha \gtrsim \frac{1}{4}\log 3$), the two Ising lines merge within our resolution and thus appear as a single “apparent” critical line.  Notably, the numerically observed bifurcation point is shifted away from the analytic tricritical point; this discrepancy reflects the extreme sensitivity of the model to small numerical errors near the tricritical point.

To verify the true nature of this apparent coalescence, we examine the scaling of $S_{LR}$ with the correlation length $\xi$.  In a genuine critical (gapless) regime, the entropy satisfies $S_{LR} = \frac{c}{6}\log\xi$, 
whereas in a gapped phase \(S_{LR}\) eventually saturates (area law) as the bond dimension \(\chi\) is increased.  For a point on the self–dual line just beyond the tricritical value, our initial simulations (with modest \(\chi\)) yield a linear \(S_{LR}\)–\(\log\xi\) relation consistent with \(c\approx 1\).  However, upon increasing both the bond dimension and the precision, the same point shows a clear tendency toward entropy saturation, indicating it is actually gapped and only masquerades as critical at lower precision.

By contrast, at true second‐order transition points the linear \(S_{LR}\)–\(\log\xi\) behaviour persists and becomes more stable as \(\chi\) grows, reducing the relative error in the extracted central charge.

\newpage
\bibliography{ref.bib}
\bibliographystyle{utphys}
\end{document}

%% file: input.tex
\newcommand{\ExcitedA}{
\begin{tikzpicture}[baseline={([yshift=-.8ex]current bounding box.center)},scale=1.2]
\draw [->-] (3.2,5.6) -- (3.2,7.2);
\node [centered, rotate=0.] at (3.,6.4) {\scriptsize $j$};
\end{tikzpicture}
}

\newcommand{\ExcitedB}{
\begin{tikzpicture}[baseline={([yshift=-.8ex]current bounding box.center)},scale=1.2]
\draw [->-] (3.2,5.6) -- (3.2,6.1);
\draw [->-] (3.2,6.1) -- (3.2,6.7);
\draw [->-] (3.2,6.7) -- (3.2,7.2);
\draw [thick, violet, ->-] (3.2,6.7) -- (4.,6.7);
\draw [thick, violet, ->-] (3.2,6.1) -- (2.4,6.1);
\node [centered, rotate=0.] at (3.,5.8) {\scriptsize $j$};
\node [centered, rotate=0.] at (3.,6.4) {\scriptsize $k$};
\node [centered, rotate=0.] at (3.4,7.) {\scriptsize $j$};
\node [centered, rotate=0.] at (4.2,6.7) {\scriptsize $q$};
\node [centered, rotate=0.] at (2.2,6.1) {\scriptsize $p^\ast$};
\end{tikzpicture}
}

\newcommand{\ExcitedC}{
\begin{tikzpicture}[baseline={([yshift=-.8ex]current bounding box.center)},scale=1.2]
\draw [blue] (3.2,5.6) -- (3.2,6.1);
\draw [blue] (3.2,6.1) -- (3.2,6.7);
\draw [blue] (3.2,6.7) -- (3.2,7.2);
\draw [thick, violet] (3.2,6.7) -- (4.,6.7);
\draw [thick, violet] (3.2,6.1) -- (2.4,6.1);
\node [centered, rotate=0.] at (3.,5.8) {\scriptsize $A$};
\node [centered, rotate=0.] at (3.,6.4) {\scriptsize $A$};
\node [centered, rotate=0.] at (3.4,7.) {\scriptsize $A$};
\node [centered, rotate=0.] at (4.2,6.7) {\scriptsize $A$};
\node [centered, rotate=0.] at (2.2,6.1) {\scriptsize $A$};
\end{tikzpicture}
}

\newcommand{\ExcitedD}{
\begin{tikzpicture}[baseline={([yshift=-.8ex]current bounding box.center)},scale=1.2]
\draw [->-] (3.2,5.6) -- (3.2,6.1);
\draw [->-] (3.2,6.1) -- (3.2,6.7);
\draw [->-] (3.2,6.7) -- (3.2,7.2);
\draw [thick, violet, ->-] (3.2,6.7) -- (4.,6.7);
\draw [thick, violet, ->-] (3.2,6.1) -- (2.4,6.1);
\node [centered, rotate=0.] at (3.,5.8) {\scriptsize $j$};
\node [centered, rotate=0.] at (3.,6.4) {\scriptsize $k$};
\node [centered, rotate=0.] at (3.4,7.) {\scriptsize $j$};
\node [centered, rotate=0.] at (4.2,6.7) {\scriptsize $q$};
\node [centered, rotate=0.] at (2.2,6.1) {\scriptsize $p$};
\end{tikzpicture}
}

\newcommand{\FrobProdA}{
\begin{tikzpicture}[baseline={([yshift=-.8ex]current bounding box.center)},scale=.6]
\draw [blue] (-.1, 0) -- (1, .8);
\draw [blue] (2.1, 0) -- (1, .8);
\draw [blue] (1, .8) -- (1, 2.1);
\node [centered, rotate=0.] at (.2,.6) {\scriptsize $A$};
\node [centered, rotate=0.] at (1.8,.6) {\scriptsize $A$};
\node [centered, rotate=0.] at (1.2,1.5) {\scriptsize $A$};
\end{tikzpicture}
}

\newcommand{\FrobProdB}{
\begin{tikzpicture}[baseline={([yshift=-.8ex]current bounding box.center)},scale=.6]
\draw [->-] (-.1, 0) -- (1, .8);
\draw [->-] (2.1, 0) -- (1, .8);
\draw [->-] (1, .8) -- (1, 2.1);
\node [centered, rotate=0.] at (.1,.5) {\scriptsize $a$};
\node [centered, rotate=0.] at (1.9,.5) {\scriptsize $b$};
\node [centered, rotate=0.] at (1.2,1.5) {\scriptsize $c$};
\end{tikzpicture}
}

\newcommand{\FrobProdC}{
\begin{tikzpicture}[baseline={([yshift=-.8ex]current bounding box.center)},scale=.6]
\draw [blue] (-.2, 0) -- (1, -.8);
\draw [blue] (2.2, 0) -- (1, -.8);
\draw [blue] (1, -.8) -- (1, -2.1);
\node [centered, rotate=0.] at (.2,-.6) {\scriptsize $A$};
\node [centered, rotate=0.] at (1.7,-.6) {\scriptsize $A$};
\node [centered, rotate=0.] at (1.2,-1.5) {\scriptsize $A$};
\end{tikzpicture}
}

\newcommand{\FrobProdD}{
\begin{tikzpicture}[baseline={([yshift=-.8ex]current bounding box.center)},scale=.6]
\draw [-<-] (0, 0) -- (1, -.8);
\draw [-<-] (2, 0) -- (1, -.8);
\draw [-<-] (1, -.8) -- (1, -2.1);
\node [centered, rotate=0.] at (.3,-.6) {\scriptsize $a$};
\node [centered, rotate=0.] at (1.7,-.6) {\scriptsize $b$};
\node [centered, rotate=0.] at (1.2,-1.5) {\scriptsize $c$};
\end{tikzpicture}
}

\newcommand{\ContractA}{
\begin{tikzpicture}[baseline={([yshift=-.8ex]current bounding box.center)},scale=1]
\draw [violet] (3.7,8.) ..controls (3.77,8.152) and (3.84,8.303) .. (4.,8.4)
			 ..controls (4.16,8.497) and (4.411,8.538) .. (4.6,8.5)
			 ..controls (4.789,8.462) and (4.915,8.344) .. (5.,8.2)
			 ..controls (5.085,8.056) and (5.129,7.886) .. (5.1,7.7)
			 ..controls (5.071,7.514) and (4.967,7.313) .. (4.8,7.2)
			 ..controls (4.633,7.087) and (4.401,7.062) .. (4.2,7.1)
			 ..controls (3.999,7.138) and (3.828,7.239) .. (3.8,7.4)
			 ..controls (3.772,7.561) and (3.886,7.78) .. (4.,8.);
\draw [blue] (4.,8.8) -- (4.8,8.8);
\draw [blue] (4.8,8.8) -- (5.4,8.2);
\draw [blue] (5.4,8.2) -- (5.4,7.4);
\draw [blue] (5.4,7.4) -- (4.8,6.8);
\draw [blue] (4.8,6.8) -- (4.,6.8);
\draw [blue] (4.,6.8) -- (3.4,7.4);
\draw [blue] (3.4,7.4) -- (3.4,8.2);
\draw [blue] (3.4,8.2) -- (4.,8.8);
\draw [blue] (4.,8.8) -- (3.6,9.2);
\draw [blue] (3.4,8.2) -- (3.,8.6);
\draw [blue] (3.4,7.4) -- (3.,7.);
\draw [blue] (4.,6.8) -- (3.6,6.4);
\draw [blue] (4.8,6.8) -- (5.2,6.4);
\draw [blue] (5.4,7.4) -- (5.8,7.);
\draw [blue] (5.4,8.2) -- (5.8,8.6);
\draw [blue] (5.2,9.2) -- (4.8,8.8);
\draw [violet] (3.4,8.) -- (4.4,8.);
\draw [violet] (3.8,8.6) -- (4.,8.4);
\draw [violet] (4.6,8.8) -- (4.6,8.5);
\draw [violet] (5.,8.2) -- (5.2,8.4);
\draw [violet] (5.1,7.7) -- (5.4,7.7);
\draw [violet] (4.8,7.2) -- (5.,7.);
\draw [violet] (4.2,6.8) -- (4.2,7.1);
\draw [violet] (3.6,7.2) -- (3.8,7.4);
\draw [violet] (3.4,8.6) -- (3.6,8.4);
\draw [violet] (4.2,9.2) -- (4.2,8.8);
\draw [violet] (5.,8.6) -- (5.3,8.9);
\draw [violet] (5.4,8.) -- (5.8,8.);
\draw [violet] (5.2,7.2) -- (5.4,7.);
\draw [violet] (4.6,6.8) -- (4.6,6.5);
\draw [violet] (3.8,7.) -- (3.6,6.8);
\draw [violet] (3.,7.6) -- (3.4,7.6);
\node [violet, centered, rotate=0.] at (3.7,8.3) {\scriptsize $\times$};
\node [violet, centered, rotate=0.] at (4.3,8.6) {\scriptsize $\times$};
\node [violet, centered, rotate=0.] at (4.9,8.5) {\scriptsize $\times$};
\node [violet, centered, rotate=0.] at (5.2,8.) {\scriptsize $\times$};
\node [violet, centered, rotate=0.] at (5.1,7.4) {\scriptsize $\times$};
\node [violet, centered, rotate=0.] at (4.6,7.) {\scriptsize $\times$};
\node [violet, centered, rotate=0.] at (4.,7.) {\scriptsize $\times$};
\node [violet, centered, rotate=0.] at (3.6,7.8) {\scriptsize $\times$};
\node [violet, centered, rotate=0.] at (4.4,7.8) {\scriptsize $A$};
\end{tikzpicture}
}

\newcommand{\ContractB}{
\begin{tikzpicture}[baseline={([yshift=-.8ex]current bounding box.center)},scale=1]
\draw [blue] (4.,8.8) -- (4.8,8.8);
\draw [blue] (4.8,8.8) -- (5.4,8.2);
\draw [blue] (5.4,8.2) -- (5.4,7.4);
\draw [blue] (5.4,7.4) -- (4.8,6.8);
\draw [blue] (4.8,6.8) -- (4.,6.8);
\draw [blue] (4.,6.8) -- (3.4,7.4);
\draw [blue] (3.4,7.4) -- (3.4,8.2);
\draw [blue] (3.4,8.2) -- (4.,8.8);
\draw [blue] (4.,8.8) -- (3.6,9.2);
\draw [blue] (3.4,8.2) -- (3.,8.6);
\draw [blue] (3.4,7.4) -- (3.,7.);
\draw [blue] (4.,6.8) -- (3.6,6.4);
\draw [blue] (4.8,6.8) -- (5.2,6.4);
\draw [blue] (5.4,7.4) -- (5.8,7.);
\draw [blue] (5.4,8.2) -- (5.8,8.6);
\draw [blue] (5.2,9.2) -- (4.8,8.8);
\draw [violet] (3.4,8.) -- (4.4,8.);
\draw [violet] (3.4,7.55) -- (3.1, 7.55);
\draw [violet] (4.2,8.8) -- (4.2,9.1);
\draw [violet] (5.7,8.) -- (5.4,8.);
\draw [violet] (4.6,6.8) -- (4.6,6.5);
\draw [violet] (3.4,8.6) -- (3.6,8.4);
\draw [violet] (5.,8.6) -- (5.3,8.9);
\draw [violet] (5.2,7.2) -- (5.4,7.);
\draw [violet] (3.8,7.) -- (3.6,6.8);
\node [violet, centered, rotate=0.] at (4.4,7.8) {\scriptsize $A$};
\end{tikzpicture}
}

\newcommand{\FrobeniusalgebraA}{
\begin{tikzpicture}[scale=0.8]
\draw [cyan] (0, -.2) -- (1, 0.5) -- (2, -.2);
\draw [cyan] (1, 0.5) -- (1, 1.7);
\node [cyan] at (0.3, 0.3) {\scriptsize $\A$};
\node [cyan] at (1.7, 0.3) {\scriptsize $\A$};
\node [cyan] at (1.2, 1.1) {\scriptsize $\A$};
\node [] at (3.7, .7) {\LARGE $=\quad\Sigma$};
\node [] at (4.45, .18) {\tiny $a$,$b$,$c\in L_\A$};
\node [] at (5.8, .7) {\scriptsize $f_{abc}^\A$};
\draw [->-] (6.5, -.2) -- (7.5, 0.5);
\draw [-<-] (7.5, 0.5) -- (8.5, -.2);
\draw [-<-] (7.5, 0.5) -- (7.5, 1.7);
\node [] at (8.2, 0.3) {\scriptsize $a$};
\node [] at (6.8, 0.3) {\scriptsize $b$};
\node [] at (7.7, 1.1) {\scriptsize $c$};
\end{tikzpicture}
}

\newcommand{\FrobeniusalgebraB}{
\begin{tikzpicture}[scale = 0.8]
\draw [cyan] (0., 0.) -- (1., 1.) -- (1., 1.8);
\draw [cyan] (0.5, 0.5) -- (1., 0.);
\draw [cyan] (1., 1.) -- (2., 0.);
\node [cyan] at (-.15, -.2) {\scriptsize $\A$};
\node [cyan] at (1.1, -.2) {\scriptsize $\A$};
\node [cyan] at (2.2, -.2) {\scriptsize $\A$};
\node [cyan] at (.55, .9) {\scriptsize $\A$};
\node [cyan] at (1.2, 1.4) {\scriptsize $\A$};
\node [] at (3.5, .7) {\huge $=$};

\draw [cyan] (5., 0.) -- (6., 1.) -- (6., 1.8);
\draw [cyan] (6.5, 0.5) -- (6., 0.);
\draw [cyan] (6., 1.) -- (7., 0.);
\node [cyan] at (4.85, -.2) {\scriptsize $\A$};
\node [cyan] at (5.8, -.2) {\scriptsize $\A$};
\node [cyan] at (6.4, .9) {\scriptsize $\A$};
\node [cyan] at (5.8, 1.4) {\scriptsize $\A$};
\node [cyan] at (7.1, -.2) {\scriptsize $\A$};
\end{tikzpicture}
}

\newcommand{\FrobeniusalgebraC}{
\begin{tikzpicture}[scale = 0.8]
\draw [cyan] (0., 0.2) -- (0., 1.) arc (-90:270:0.5);
\draw [cyan] (0., 2.) -- (0., 2.8);
\node [cyan] at (-.2, .6) {\scriptsize $\A$};
\node [cyan] at (-.2, 2.4) {\scriptsize $\A$};
\node [cyan] at (-.7, 1.5) {\scriptsize $\A$};
\node [cyan] at (.7, 1.5) {\scriptsize $\A$};
\node [] at (2.2, 1.5) {$=\quad d_\A$};
\draw [cyan] (3.3, 0.2) -- (3.3, 2.8);
\node [cyan] at (3.5, 1.5) {\scriptsize $\A$};
\end{tikzpicture}
}

\newcommand{\FrobeniusalgebraD}{
\begin{tikzpicture}[scale = 0.8]
\draw [cyan] (0, 2) -- (0, 1) arc (180:270:0.5) arc (-90:0:.25) arc (180:0:0.25) -- (1.25, -.5);
\draw [cyan] (1., 1.) -- (1., 2);
\node [cyan] at (-.2, 1.5) {\scriptsize $\A$};
\node [cyan] at (1.2, 1.5) {\scriptsize $\A$};
\node [cyan] at (1.45, .5) {\scriptsize $\A$};
\node [cyan] at (1.45, -.2) {\scriptsize $\A$};
\node [cyan] at (.9, .7) {\scriptsize $\A$};

\node [] at (2.2, 1.) {$=$};

\draw [cyan] (3., 2.) -- (3., 1.) arc (180:360:0.5) -- (4., 2.);
\draw [cyan] (3.5, .5) -- (3.5, -.5); 
\node [cyan] at (2.8, 1.5) {\scriptsize $\A$};
\node [cyan] at (4.2, 1.5) {\scriptsize $\A$};
\node [cyan] at (3.7, 0.) {\scriptsize $\A$};

\node [] at (5., 1.) {$=$};

\draw [cyan] (6., -.5) -- (6., 1.) arc (180:0:0.25) arc (180:270:0.25) arc (-90:0:0.5) -- (7.25, 2.);
\draw [cyan] (6.25, 1.25) -- (6.25, 2); 
\node [cyan] at (5.8, .7) {\scriptsize $\A$};
\node [cyan] at (5.8, -.2) {\scriptsize $\A$};
\node [cyan] at (7.45, 1.7) {\scriptsize $\A$};
\node [cyan] at (6.65, 1.1) {\scriptsize $\A$};
\node [cyan] at (6.4, 1.8) {\scriptsize $\A$};

\end{tikzpicture}
}

\newcommand{\RModuleA}{
\begin{tikzpicture}[scale = 0.8]
\draw [red] (0., 0.) -- (0., 2.);
\draw [cyan] (0., 1.) -- (1., 0.);
\node [red] at (-.25, .5) {\scriptsize $M$};
\node [red] at (-.25, 1.5) {\scriptsize $M$};
\node [cyan] at (1.2, 0.) {\scriptsize $\A$};

\node [] at (3., 1.) {\LARGE $=\quad\Sigma$};
\node [] at (3.8, 0.5) {\tiny $a\in L_\A$};
\node [] at (3.8, 0.2) {\tiny $x, y\in L_M$};
\node [] at (5., 1.) {\scriptsize $[\rho_{M_\A}]^a_{xy}$};

\draw [->-] (6.5, 0.) -- (6.5, 1.);
\draw [->-] (6.5, 1.) -- (6.5, 2.);
\draw [-<-] (6.5, 1.) -- (7.5, 0.);
\node [] at (6.3, 0.5) {\scriptsize $x$};
\node [] at (6.3, 1.5) {\scriptsize $y$};
\node [] at (7.2, .65) {\scriptsize $a$};

\end{tikzpicture}
}

\newcommand{\RModuleB}{
\begin{tikzpicture}[scale = 0.8]
\draw [red] (0., 0.) -- (0., 2.);
\draw [cyan] (0., 0.7) -- (0.7, 0.);
\draw [cyan] (0., 1.3) -- (0.7, 0.6);
\node [red] at (-.25, .3) {\scriptsize $M$};
\node [red] at (-.25, 1.) {\scriptsize $M$};
\node [red] at (-.25, 1.7) {\scriptsize $M$};
\node [cyan] at (0.9, 0.) {\scriptsize $\A$};
\node [cyan] at (.9, .6) {\scriptsize $\A$};

\node [] at (1.8, 1.) {$=$};

\draw [red] (3., 0.) -- (3., 2.);
\draw [cyan] (3., 1.4) -- (4.4, 0.);
\draw [cyan] (3.9, .5) -- (3.5, 0.);
\node [cyan] at (3.35, .2) {\scriptsize $\A$};
\node [cyan] at (4.6, 0.) {\scriptsize $\A$};
\node [cyan] at (3.6, 1.1) {\scriptsize $\A$};
\node [red] at (2.75, 1.7) {\scriptsize $M$};
\node [red] at (2.75, .6) {\scriptsize $M$};
\end{tikzpicture}
}